\documentclass[12pt]{article}
\usepackage{amsmath,amsthm,amsfonts,amssymb} % graphicx
\usepackage{cite,faktor}
\usepackage{tikz,ytableau,subfig}
\usepackage{mathtools}
\usepackage{xcolor,colortbl}
\usepackage{arydshln}

\newlength{\xtrawidth}
\newlength{\xtraheight}
\numberwithin{equation}{section}
\numberwithin{table}{section}
\numberwithin{figure}{section}

\allowdisplaybreaks[2]

\DeclareMathOperator{\Tr}{Tr}

\def\ystd{\ytableausetup{smalltableaux,aligntableaux=center,boxsize=0.45em}}
\def\ystdnum{\ytableausetup{smalltableaux}}
\ystd

\begin{document}

\begin{titlepage}
\begin{center}
\hfill BONN-TH-2014-12\\
\vskip 0.75in

{\Large \bf A note on colored HOMFLY polynomials for}\\[2ex]
{\Large \bf hyperbolic knots from WZW models}\\[2ex]
\vskip 0.4in
{\bf{Jie Gu and Hans Jockers}}\\
\vskip 0.4in
{\em Bethe Center for Theoretical Physics}\\
{\em Physikalisches Institut, Universit\"at Bonn}\\
{\em 53115 Bonn, Germany}\\
\vskip 0.2in
{\tt jiegu@th.physik.uni-bonn.de}\\  
{\tt jockers@uni-bonn.de} \\ 
\end{center}

\vskip 0.35in

\begin{center} {\bf Abstract} \end{center}

Using the correspondence between Chern--Simons theories and Wess--Zumino--Witten models we present the necessary tools to calculate colored HOMFLY polynomials for hyperbolic knots. For two--bridge hyperbolic knots we derive the colored HOMFLY invariants in terms of crossing matrices of the underlying Wess--Zumino--Witten model. Our analysis extends previous works by incorporating non--trivial multiplicities for the primaries appearing in the crossing matrices, so as to describe colorings of HOMFLY invariants beyond the totally symmetric or anti--symmetric representations of $SU(N)$. The crossing matrices directly relate to 6j--symbols of the quantum group $\mathcal{U}_qsu(N)$. We present powerful methods to calculate such quantum 6j--symbols for general $N$. This allows us to determine previously unknown colored HOMFLY polynomials for two--bridge hyperbolic knots. We give explicitly the HOMFLY polynomials colored by the representation $\{2,1\}$ for two--bridge hyperbolic knots with up to eight crossings. Yet, the scope of application of our techniques goes beyond knot theory; e.g., our findings can be used to study correlators in Wess--Zumino--Witten conformal field theories or --- in the limit to classical groups --- to determine color factors for Yang Mills amplitudes.

\vfill

\noindent July, 2014

\end{titlepage}
%%%%%%%%%%%%%%%%%%%%%%%%%%%%%%%%%%%%%%%%%%%%%%%%%%%%%%%%%%%%%%%%%%%%
\tableofcontents
\newpage

%%%%%%%%%%%%%%%%%%%%%%%%%%%%%%
\section{Introduction}
%%%%%%%%%%%%%%%%%%%%%%%%%%%%%%
In the seminal work \cite{Witten:1988hf} Witten shows that Wilson loop observables of three-dimensional Chern--Simons theory naturally describe knot invariants on three manifolds. In particular, Wilson loop expectation values of $SU(N)$ Chern--Simons theory on $S^3$ determine the colored HOMFLY polynomials of knots on the three sphere.

Over the years, the connection between Chern--Simons and knot theory has resulted in a lot of progress in both research fields and has led to many surprising correspondences.\footnote{For a review, see for example refs.~\cite{Labastida:1998ud,MR2177747,Marino:2004uf,Gukov:2012jx} and references therein.} For instance, $SU(N)$ Chern--Simons theory on $S^3$ enjoys an interpretation as a topological string theory on the deformed conifold geometry $T^*S^3$ \cite{Witten:1992fb}. Based on the large $N$ transition by Gopakumar and Vafa \cite{Gopakumar:1998ki} --- realizing a topological version of AdS/CFT duality --- Wilson loop expectation values and hence knot invarants are computed by certain open topological string amplitudes on the resolved conifold geometry. The relationship to the topological string on the resolved conifold furnishes many non--trivial results and checks for the described chain of dualities. However, most of the explicit results concern simple knots such as the unknot and torus knots\cite{Labastida:2000zp,Labastida:2000yw,Ooguri:1999bv,Marino:2001re,Brini:2011wi,Diaconescu:2011xr,Jockers:2012pz}. But recent progress --- both in knot theory and in topological string theory --- has opened up new possibilities to study hyperbolic knots in terms of topological strings on the conifold as well \cite{MR2376818,Ng:2012qq,Aganagic:2012jb,Aganagic:2013jpa,Gu:2014yba}.

To further establish and to check the above described developments, the knowledge of colored HOMFLY polynomials for non--torus knots is crucial. For example, to test the assertions of our recent work \cite{Gu:2014yba}, it would be interesting to have HOMFLY polynomials for hyperbolic knots colored with Young diagrams with up to two rows at our disposal. There are some results on HOMFLY polynomials for certain hyperbolic knots colored with totally symmetric and/or anti--symmetric representations \cite{RamaDevi:1992np,Zodinmawia:2011oya,Nawata:2012pg,Itoyama:2012re,Itoyama:2012fq, Kawagoe:2012bt,Nawata:2013qpa}. More recently, for certain classes of knots HOMFLY invariants for colorings with more general representations have explicitly been obtained in refs.~\cite{Anokhina:2012rm, Anokhina:2013ica}. The aim of this note is to provide for the necessary tools to calculate HOMFLY invariants for representations beyond the totally symmetric/anti--symmetric cases for two--bridge hyperbolic knots in Chern--Simons theory directly. Following the interesting works~\cite{Ramadevi:2000gq,Borhade:2003cu, Zodinmawia:2011oya}, we perform our calculations in two steps. Firstly, we realize the three sphere $S^3$ --- with a Wilson line along the knot suitably embedded --- as a connected sum of two solid balls $B_2$. This topological surgery is then carried over to the partition function of $SU(N)$ Chern--Simons theory \cite{Witten:1988hf}. Secondly, we use the correspondence between $SU(N)$ Chern--Simons theory on the solid ball $B_2$ and the $\widehat{su(N)}_k$ Wess--Zumino--Witten (WZW) conformal field theory on the boundary $S^2\equiv\partial B_2$. 

In order to arrive at Wilson loop expectation values for two--bridge hyperbolic knots in $SU(N)$ Chern--Simons theory, it is necessary to consistently combine the $SU(N)$ partition functions of the two solid balls $B_2$. This step requires the knowledge of certain $\mathcal{U}_qsu(N)$ quantum 6j--symbols for general $N$ in terms of representations associated to the to--be--determined colored HOMFLY polynomial \cite{Ramadevi:2000gq,Borhade:2003cu}. Deriving the relevant quantum 6j--symbols for general $N$ is conceptually challenging and computationally expensive --- in particular if they involve representations arising with multiplicities. We employ and extend known bootstrap techniques \cite{Butler:1981,MR832771} to calculate $\mathcal{U}_qsu(N)$ quantum 6j--symbols for general $N$ involving multiplicities. In particular, using recoupling relations among s--, t-- and u--channels for conformal blocks in WZW conformal field theories, we develop a new method --- called the eigenvector method --- to derive new and non--trivial relations among quantum 6j--symbols. Combining all these techniques, we explicitly compute previously unknown quantum 6j--symbols with multiplicities, so as to determine new colored HOMFLY invariants for two--bridge hyperbolic knots.

Furthermore, we establish a new approach to compute classical 6j--symbols of $SU(N)$ for general $N$ using projectors. Our method is inspired by the projector approach for $U(N)$ groups of refs.~\cite{MR2418111,Elvang:2003ue}, but goes beyond the $U(N)$ case, as it is designed to realize projectors for representations of $SU(N)$ and their conjugates representations simultaneously. While this projector method may prove useful for applications in $SU(N)$ Yang Mills theory, it is used here as tool to calculate classical 6j--symbols of $SU(N)$. Since $\mathcal{U}_qsu(N)$ quantum 6j--symbols specialize in the limit $q\to 1$ to the classical 6j--symbols, the projector approach serves as an independent and non--trivial check on our derivation of quantum 6j--symbols. 

In this work the developed computational techniques to determine quantum 6j--symbols are employed to derive knot invariants. However, our approach has a much broader scope of application in the context of conformal field theories. The relationship between crossing matrices and quantum 6j--symbols combined with the bootstrap method in conformal field theory, allows us to express any correlator in WZW conformal field theories in terms of quantum 6j--symbols. Furthermore, quantum 6j--symbols arise in the context of boundary operator product expansions for WZW models with branes as well, see for instance the discussions in refs.~\cite{Behrend:1999bn,Felder:1999ka,Felder:1999mq}. While this work focuses on $\widehat{su(N)}_k$ WZW models, many of the presented techniques readily generalize to other affine Lie groups as well.

Apart from its relevance for knot theory and WZW models, the derivation of (classical) 6j--symbols has other interesting applications in physics. For instance, in quantum field theory 6j--symbols describe recoupling relations among $s$--, $t$-- and $u$--channels of scattering amplitudes involving matter fields transforming in non--trivial representations of the gauge group, see, e.g., refs.~\cite{MR2418111,Elvang:2003ue}. More generally, the need of 6j--symbols arises in quantum mechanics in the context of recoupling problems of tensor products of states transforming under a continuous symmetry group \cite{Wigner:1959}. Such recoupling problems appear in atomic physics for spin states transforming under $SU(2)$. More complicated recoupling problems are for instance accociated to quantum states of ultracold alkaline--earth atoms, which transform in representations of $SU(N)$~\cite{Gorshkov:2009jk}.

The outline of this paper is as follows: In Sec.~\ref{sec:curves} we review and extend the procedure presented in refs.~\cite{Ramadevi:2000gq, Zodinmawia:2011oya} to compute colored HOMFLY invariants in the framework of WZW models, including the necessary modifications so as to accommodate for non--trivial multiplicities of representations. Sec.~\ref{sec:toprec} is devoted to the detailed presentation of three methods for computing quantum and classical 6j--symbols: the bootstrap technique, the eigenvector method, and the projector approach.  Sec.~\ref{sec:knotinv} states the calculated quantum 6j--symbols relevant for the computation of the HOMFLY knot invariant colored by $\ydiagram{2,1}$.  This colored HOMFLY invariant is explicitly computed for two--bridge hyperbolic knots with up to eight crossings. We find agreement with the findings of refs.~\cite{Anokhina:2012rm,Anokhina:2013ica} for the subset of two--bridge hyperbolic knots that are analyzed there as well. The symmetry properties of the HOMFLY invariants are also discussed. Sec.~\ref{sec:conclusion} concludes and discusses the possible ways and potential computational problems in extending this method to compute higher--bridge knots colored by more complicated representations. In Appendix~\ref{sec:BootstrapExample} we present in detail a sample calculation of a quantum 6j--symbol. Finally, in Appendix~\ref{sec:HOMFLYFormulae} we list the formulae for all two--bridge knots with up to eight crossings that compute the colored HOMFLY invariants in terms of crossing matrices of the $\widehat{su(N)}_k$ WZW model.

%%%%%%%%%%%%%%%%%%%%%%%%%%%%%%
\section{WZW models for colored HOMFLY polynomials} \label{sec:curves}
%%%%%%%%%%%%%%%%%%%%%%%%%%%%%%
To set the stage for our calculations, we first review the interesting works \cite{Ramadevi:2000gq, Zodinmawia:2011oya}, in which HOMFLY invariants colored with certain representations are calculated using conformal field theory techniques applied to the $\widehat{su(N)}_k$~WZW~model. We extend their method in allowing for multiplicities in the appearing recoupling matrices acting on conformal blocks, which makes the framework (in principal) applicable for HOMFLY invariants colored with any representation of $SU(N)$.

We consider the Chern--Simons theory with gauge group $U(1) \times SU(N)$ and levels $k_1, k$ on $S^3$. The action reads
\begin{equation}
	S= \frac{k_1}{4\pi} \int_{S^3} B\wedge dB + \frac{k}{4\pi}\int_{S^3} \Tr_R \left(A \wedge dA + \frac{2}{3} A\wedge A \wedge A \right) \ .
\end{equation}
We pick the integer $n$ for the representation of $U(1)$ and the representation $R$ for $SU(N)$. Let $\mathcal{K}$ be a knot in $S^3$. The associated Wilson loop operator factorizes
\[
		W^{\mathcal{K}}_{(n,R)}[B,A] = \Tr_n U_{\mathcal{K}}[B] \Tr_R U_{\mathcal{K}}[A] \ ,
\]
and so does the expectation value of the Wilson loop operator
\begin{equation}
	\mathcal{W}^{U(1)\times SU(N)}_{(n,R)} (\mathcal{K}) = \langle W^{\mathcal{K}}_{(n,R)}[B,A] \rangle = \mathcal{W}^{U(1)}_n(\mathcal{K}) \, \mathcal{W}^{SU(N)}_R(\mathcal{K}) \ .
\end{equation}
These are called the quantum knot invariants with the respective gauge groups.

The Abelian quantum invariant $\mathcal{W}^{U(1)}_n$ here serves to regularize the framing transformation. It is know that when the framing of the knot $\mathcal{K}$ increases by $\Delta f$, the quantum knot invariant with gauge group $G$ transforms by \cite{Witten:1988hf}
\begin{equation}
	\mathcal{W}^G_R(\mathcal{K}) \mapsto \exp\left( 2\pi i h_R\cdot \Delta f \right) \mathcal{W}^G_R(\mathcal{K}) \ .
\end{equation}
Here $h_R$ is the conformal weight of the WZW primary field in the integrable representation $R$
\begin{equation}
	h_R = \sum_a \frac{(T^a)^2}{2(k+g)} = \frac{C_R}{k+g}	\ ,
\end{equation}
where the summation is performed over all the generators $T^a$ of the gauge group. So $C_R$ is the quadratic Casimir of the group $G$. $k$ is the level of the Chern--Simons theory, and $g$ is the dual coxeter number of the gauge group $G$. For the group $SU(N)$ with level $k$, the dual coxeter number $g$ is $N$. The quadratic Casimir $C_R$ for the representation $R$ is
\begin{equation}
	C_R = \frac{1}{2}\left[N\ell + \kappa_R - \frac{\ell^2}{N}\right], \quad \kappa_R = \ell + \sum_i(\ell^2_i -2i \ell_i) \ .
\end{equation}
$\ell_i$ is the number of boxes on the $i$-th row of the Young diagram associated to the representation $R$, and $\ell$ is the total number of boxes of the Young diagram. The framing transformation is then,
\begin{equation}\label{equ:SU(N)Framing}
		\mathcal{W}^{SU(N)}_R(\mathcal{K}) \mapsto  \lambda^{\frac{1}{2} \ell \Delta f } q^{\frac{1}{2} \kappa_R \Delta f} q^{-\frac{\ell^2}{2N} \Delta f } \mathcal{W}^{SU(N)}_R(\mathcal{K}) \ .
\end{equation}
which has the factor $q^{-\frac{\ell^2}{2N} \Delta f }$ with explicit $N$ dependence. Here the usual convention is used,
\begin{equation}
		q = \exp\left( \frac{2\pi i}{k+N}\right) ,\quad \lambda = q^N \ .
\end{equation}

The Abelian quantum knot invariant $\mathcal{W}^{U(1)}_n(\mathcal{K})$ with level $k_1$ is much simpler. It is known that in zero framing, the Abelian quantum knot invariant is always 1. So in framing $f$ the only contribution comes from the framing transformation,
\[
		\mathcal{W}^{U(1)}_n (\mathcal{K}) = \exp\left( 2\pi i h_n \cdot f \right) = \exp\left( 2\pi i \frac{n^2}{2 k_1}  f \right) \ .
\]
A further $\Delta f$ framing transformation is obtained by 
\begin{equation}\label{equ:U(1)Framing}
		\mathcal{W}^{U(1)}_n(\mathcal{K}) \mapsto \exp\left( 2\pi i \frac{n^2}{2 k_1} \Delta f \right) \mathcal{W}^{U(1)}_n(\mathcal{K}) \ .
\end{equation}

Combing the framing transformations \eqref{equ:SU(N)Framing} and \eqref{equ:U(1)Framing} for the factorized quantum knot invariant, we see that the explicit $N$--dependent factor $q^{-\frac{\ell^2}{2N} \Delta f }$ vanishes for
\begin{equation}
		k_1 = N(k+N), \quad n = \ell \ .
\end{equation}
Note that if we start with the action of the $U(N)$ Chern--Simons theory with level $k$ and extract the $U(1)$ sector, we would get exactly this choice for $k_1$ and $n$. It is then defined in ref.~\cite{Zodinmawia:2011oya} that $\mathcal{W}^{U(1)\times SU(N)}_{(n,R)}(\mathcal{K})$ with this regularization is the unnormalized HOMFLY invariant for three--manifolds colored with representation $R$ \footnote{The choice of $k_1$ and $n$ is slightly different from ref.~\cite{Zodinmawia:2011oya} to ensure that the charge $n$ is integral. The regularization of the unnormalized colored HOMFLY invariants is nonetheless the same.}
\begin{equation}\label{equ:U1Regularization}
	H_R(\mathcal{K} ) = \mathcal{W}^{U(1)\times SU(N)}_{(n,R)}(\mathcal{K})\Big|_{\substack{k_1 =N( k+N) \\ n = \ell } } = q^{\frac{\ell^2}{2 N} f } \mathcal{W}^{SU(N)}_R(\mathcal{K}) \ .
\end{equation}
It transforms when the framing increases by $\Delta f$ as,
\begin{equation}\label{equ:HOMFLYFramingTransform}
	H_R(\mathcal{K} ) \mapsto q^{\frac{1}{2}\kappa_R\cdot \Delta f} \lambda^{\frac{1}{2} \ell \cdot \Delta f} H_R(\mathcal{K}) \ .
\end{equation}
Clearly if the so-defined colored HOMFLY invariant in one framing is only a function of $q$ and $\lambda$ (i.e. the apparent dependence on $N$ drops off), it remains so in an arbitrary framing. Note that in some mathematical literature as well as for our results presented in Sec.~\ref{sec:knotinv} the colored HOMFLY invariants are normalized as
\[
		\bar{H}_R(\mathcal{K}) = \frac{H_R(\mathcal{K})}{H_R(\bigcirc)} \ .
\]

\begin{figure}[t]
\center
\includegraphics[width=0.3\textwidth]{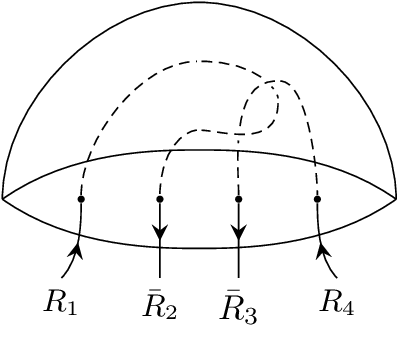}
\caption{A three--manifold with boundary and four punctures from the strands of a knot.}\label{fig:BoundaryState}
\end{figure}

The method to compute the quantum knot invariant $\mathcal{W}^{SU(N)}_R(K)$ is based on the original idea of Witten \cite{Witten:1988hf}. We cut $S^3$ into two three--manifolds $M_1, M_2$ with boundaries $\Sigma_1, \Sigma_2$. Each of the two shared boundaries has only four punctures by the strands of the knot. Then the path integral in either of the two three--manifolds $M_1, M_2$ produces a quantum state on the two boundaries $\Sigma_1$ and $\Sigma_2$, respectively. The inner product of the two quantum states is the quantum knot invariant. Witten has derived the Hilbert space for such boundary quantum states in ref.~\cite{Witten:1988hf}. Namely, associate each puncture with the representation $R$ or its conjugate $\bar{R}$ of the strand that goes through the puncture, depending on whether the oriented strand goes in or out of the three--manifold. For instance on the surface in Fig.~\ref{fig:BoundaryState}, the four punctures are associated with $R_1, \bar{R}_2, \bar{R}_3, R_4$, respectively. Then the Hilbert space of boundary quantum states is isomorphic to the space of conformal blocks of four point functions in the $\widehat{su(N)}_k$~WZW model, where the four fields in the conformal block become the WZW~primaries of the integral representations assigned to the four punctures.  

\begin{figure}[t]
\center
\subfloat[The basis state $|\phi^{(1)}_{t,r_3r_4}(R_1,R_2,R_3,R_4)\rangle$]{\includegraphics[width=0.38\textwidth]{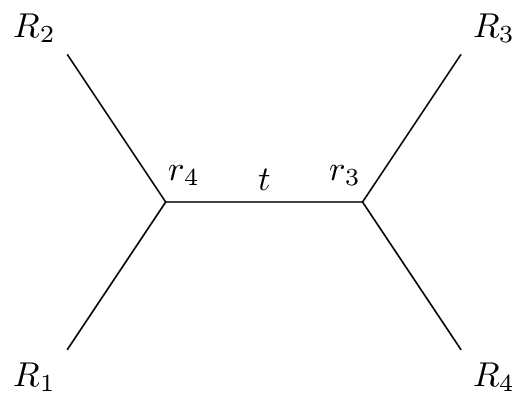}} \quad \quad \quad \quad
\subfloat[The basis state $|\phi^{(2)}_{s,r_1r_2}(R_1,R_2,R_3,R_4)\rangle$]{\includegraphics[width=0.25\textwidth]{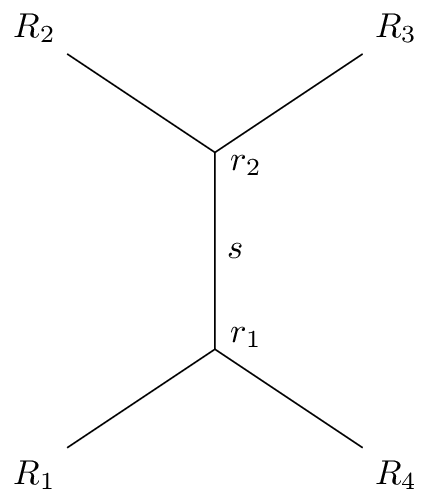}}
\caption{Two types of conformal blocks representing two bases of the Hilbert space of the four--punctured boundary of a three--manifold.}\label{fig:ConformalBlocks}
\end{figure}

\subsection{Braiding operators}

\begin{figure}
\center
\begin{minipage}{0.4\linewidth}
\centering
\includegraphics[width=0.4\linewidth]{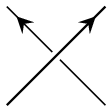}\\
over--crossing
\end{minipage}
\begin{minipage}{0.4\linewidth}
\centering
\includegraphics[width=0.4\linewidth]{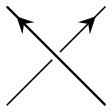}\\
under--crossing
\end{minipage}
\caption{The convention for an over--crossing and an under--crossing.}\label{fig:Crossings}
\end{figure}

\begin{figure}[t]
\center
\begin{minipage}{0.2\linewidth}
\center
\includegraphics[width=0.5\linewidth]{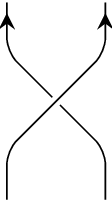}\\
$b^{(+)}_i$\\
\includegraphics[width=0.5\linewidth]{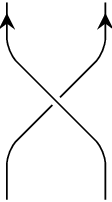}\\
$(b^{(+)}_i)^{-1}$
\end{minipage}
\begin{minipage}{0.2\linewidth}
\center
\includegraphics[width=0.5\linewidth]{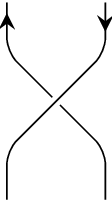}\\
$(b^{(-)}_i)^{-1}$\\
\includegraphics[width=0.5\linewidth]{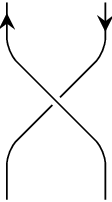}\\
$b^{(-)}_i$
\end{minipage}
\begin{minipage}{0.2\linewidth}
\center
\includegraphics[width=0.5\linewidth]{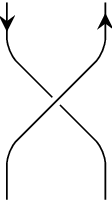}\\
$(b^{(-)}_i)^{-1}$\\
\includegraphics[width=0.5\linewidth]{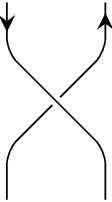}\\
$b^{(-)}_i$
\end{minipage}
\begin{minipage}{0.2\linewidth}
\center
\includegraphics[width=0.5\linewidth]{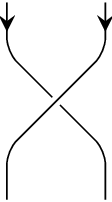}\\
$b^{(+)}_i$\\
\includegraphics[width=0.5\linewidth]{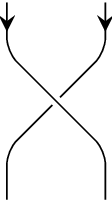}\\
$(b^{(+)}_i)^{-1}$
\end{minipage}
\caption{All possible braiding operators and their inverses.}\label{fig:Braidings}
\end{figure}

Given the isomorphism between the Hilbert space $\mathcal{H}$ and the space of conformal blocks, the former naturally has two types of basis, diagramatically represented in Figs.~\ref{fig:ConformalBlocks}. In the basis (1) the basis states $|\phi^{(1)}_{t,r_3r_4}(R_1,R_2,R_3,R_4)\rangle$ are labelled by the intermediate state $t$, where $t \in (R_1\otimes R_2)\cap(\bar{R}_3\otimes\bar{R}_4)$, as well as the multiplicity labels $r_3,r_4$ beside the vertices. If $R_1\otimes R_2$ contains $r$ copies of $t$ and $\bar{R}_r \otimes \bar{R}_4$ contains $r'$ copies of $t$, the label $r_4$ can take any integral value between $0$ and $r-1$, while $r_3$ any integral value between $0$ and $r'-1$. Similarly in the basis (2) the basis states $|\phi^{(2)}_{s,r_1r_2}(R_1,R_2,R_3,R_4)\rangle$ are labelled by the intermediate state $s$, where $s\in (R_2\otimes R_3) \cap (\bar{R}_1\otimes\bar{R}_4)$, as well as the multiplicity labels $r_1,r_2$.\footnote{The ordering of the multiplicity labels $r_4,r_3,r_2,r_1$ is to be consistent with the multiplicity labels on 6j--symbols presented in later sections.} Coincidentally the two types of basis states are also eigenstates of the so-called braiding operators (or half-monodromy operator) $b_a^{(\pm)}$. The braiding operator $b^{(\pm)}_a$ acts on the neighboring $a$-th and $(a+1)$-th strands, with the superscript $(+)$ if the two strands are parallel, or $(-)$ if the two strands are anti-parallel, by winding the two strands around each other once so that an over-crossing is produced, or an under-crossing is produced if the inverse operator is acted. Here an over-crossing and an under-crossing are conventionally defined as in Figs.~\ref{fig:Crossings}. In Figs.~\ref{fig:Braidings} all possibile braiding operators and their inverses are listed. 

Since in the basis (1) the first and the second primaries are fused together, so are the third and the fourth primaries, the basis states there are eigenstates of the braiding operators $b_1^{(\pm)}$ and $b_3^{(\pm)}$. Analogously the basis states in the basis (2) are eigenstates of the braiding operators $b_2^{(\pm)}$. The eigenvalues $\lambda^{(\pm)}_{R_i,R_j; R_k}$ depend on the two representations $R_i, R_j$ before the fusion, the intermediate representation $R_k$ and the multiplicity label $r_l$ after the fusion, as well as the label $(\pm)$ \cite{Moore:1988qv,Zodinmawia:2011oya}\footnote{Here and in eq.~\eqref{equ:TrivialCrossingMatrices} as well as in Table~\ref{fig:AllTopsA} and Table~\ref{fig:AllTopsB}, we treat the phases in a more systematically compared to ref.~\cite{Zodinmawia:2011oya} by relating them to the 3j--phases.}, i.e., 
\begin{equation}\label{equ:BraidingEigenvalues}
	\begin{aligned}
	b_1: \quad &\lambda^{(\pm)}_{R_i,R_j; R_k r_4} = \{R_i, R_j, \bar{R}_k, r_4\} q^{\pm (C_{R_i} + C_{R_j}-C_{R_k})/2}    \ ,\\
	b_2: \quad &\lambda^{(\pm)}_{R_i,R_j; R_k r_2} = \{R_i, R_j, \bar{R}_k, r_2\} q^{\pm (C_{R_i} + C_{R_j}-C_{R_k})/2}    \ ,\\
	b_3: \quad &\lambda^{(\pm)}_{R_i,R_j; R_k r_3} = \{R_i, R_j, \bar{R}_k, r_3\} q^{\pm (C_{R_i} + C_{R_j}-C_{R_k})/2}    \ .
	\end{aligned}
\end{equation}
The magnitudes $q^{\pm (C_{R_i} + C_{R_j}-C_{R_k})/2}$ of the eigenvalues are the square roots of the monodromies of the conformal blocks. The phases $\{R_i, R_j, \bar{R}_k,r_l\}=\pm 1$ are the 3j--phases \cite{Butler:1981} (the multiplicity label $r_l$ is sometimes omitted if it's trivial), the symmetry phases of the Clebsch--Gordon coefficients when the two coupling representations $R_i$ and $R_j$ are exchanged.
\begin{equation}
	\langle r_l R_k m_k| R_i m_i, R_j m_j \rangle = \{R_i, R_j, \bar{R}_k, r_l\} \langle r_l R_k m_k | R_j m_j , R_i m_i \rangle  \ ,
\end{equation}
where $m_i,m_j,m_k$ label some states in the respective representations. They are invariant under permutations and conjugations of representations
\[
		\{ R_1,R_2,R_3,r \} = \{\bar{R}_1,\bar{R}_2,\bar{R}_3,r\} = \{ R_1, R_3,R_2,r \} =\textrm{all possible permutations} \ .
\]
When $R_k$ is a singlet $0$, the 3j--phase is reduced to the so-called 2j--phase $\{R_i\}$
\[
	\{R_i,\bar{R}_i , 0, 0 \} = \{ R_i \} \ .
\]
There is some freedom in choosing the 3j--phases and 2j--phases. Their specification is given in Sec.~\ref{sec:3jPhases}. From the WZW model point of view the phases $\{R_i, R_j, \bar{R}_k, r_l\}=\pm 1$ are easy to understand as $R_k$ is in the tensor product of $R_i$ and $R_j$.\footnote{In principle fusion and tensor product are different in WZW models. They nonetheless coincide when $k$ is large; c.f., Sec.~\ref{sec:recoupling}.} The consistency of these phases is further illustrated in Sec.~\ref{sec:PhasesConsistency}. 

Note that among the eight braidings in Figs.~\ref{fig:Braidings}, all four right--handed braidings on the first row have the same type of eigenvalues, i.e., $\propto q^{(C_{R_i}+C_{R_j}-C_{R_k})/2}$, while all four left--handed braidings on the second row have the same type of eigenvalues, i.e., $\propto q^{-(C_{R_i}+C_{R_j}-C_{R_k})/2}$, regardless of the orientations of the strands.

%%%%%%%%%%%%%%%%%%%%%%%%%%%%%%%
\subsection{Basis transformation} \label{sec:BasisTransformation}
Another important concept in the WZW model --- or conformal field theories in general --- is the theory of crossing matrices (also called fusion matrices), which relate two types of conformal blocks. In the context of the Chern--Simons theory the crossing matrices $a^{t,r_3r_4}_{s,r_1r_2}\begin{bmatrix} R_1 & R_2 \\ R_3 & R_4 \end{bmatrix}$ (labelled by their representations) furnish a basis of transformation matrices
\begin{equation}
	|\phi^{(1)}_{t,r_3r_4}(R_1, R_2, R_3, R_4) \rangle = \sum_{s,r_1,r_2} a^{t,r_3r_4}_{s,r_1r_2}\begin{bmatrix} R_1 & R_2 \\ R_3 & R_4 \end{bmatrix} | \phi^{(2)}_{s,r_1r_2}(R_1, R_2, R_3, R_4) \rangle \ ,
\end{equation}
that map the side braiding eigenstates to the central braiding eigenstates. Due to unitarity of the crossing matrices
\begin{equation}
	\sum_{t,r_3,r_4} a^{t,r_3r_4}_{s,r_1r_2}\begin{bmatrix} R_1 & R_2 \\ R_3 & R_4 \end{bmatrix} a^{t,r_3r_4}_{s',r'_1r'_2}\begin{bmatrix} R_1 & R_2 \\ R_3 & R_4 \end{bmatrix}^* = \delta_{s,s'} \delta_{r_1,r'_1} \delta_{r_2,r'_2} \ ,
\end{equation}
we have the inverse relationship
\begin{equation}
	|\phi^{(2)}_{s,r_1r_2}(R_1, R_2, R_3, R_4) \rangle = \sum_{t,r_3,r_4} a^{t,r_3r_4}_{s,r_1r_2}\begin{bmatrix} R_1 & R_2 \\ R_3 & R_4 \end{bmatrix}^* | \phi^{(1)}_{t,r_3r_4}(R_1, R_2, R_3, R_4) \rangle \ .
\end{equation}
Note that for either $t$ or $s$ being a singlet the crossing matrices take the form \cite{Zodinmawia:2011oya}\footnote{Their values including the phases can also be deduced by relating them to the quantum 6j--symbols as described in Sec.~\ref{sec:toprec}.}
\begin{equation}\label{equ:TrivialCrossingMatrices}
\begin{aligned}
a^{0,00}_{s,r_1r_2}\begin{bmatrix} R_1 & \bar{R}_1 \\ R_2 & \bar{R}_2 \end{bmatrix} 
 &=  \{R_2\}\{R_1,\bar{R}_2,s,r_1\}\frac{\sqrt{\dim_q s}}{\sqrt{\dim_q R_1 \dim_q R_2}} \delta_{r_1,r_2}  \ , \\ a^{t,r_3r_4}_{0,00}\begin{bmatrix} R_1 & R_2 \\ \bar{R}_2 & \bar{R}_1 \end{bmatrix} 
 &=  \{R_2\}\{R_1,R_2,\bar{t},r_3\}\frac{\sqrt{\dim_q t}}{\sqrt{\dim_q R_1 \dim_q R_2}} \delta_{r_3,r_4}\ .
\end{aligned}
\end{equation}

%\begin{figure}[htbp]
%\center
%\subfloat[$\mathbf{4_1}$]{\includegraphics[width=0.2\linewidth]{Plat4s1}}
%\subfloat[$\mathbf{5_2}$]{\includegraphics[width=0.2\linewidth]{Plat5s2}}
%\subfloat[$\mathbf{6_1}$]{\includegraphics[width=0.2\linewidth]{Plat6s1}}
%\subfloat[$\mathbf{6_2}$]{\includegraphics[width=0.2\linewidth]{Plat6s2}} \\
%\subfloat[$\mathbf{6_3}$]{\includegraphics[width=0.2\linewidth]{Plat6s3}}
%\subfloat[$\mathbf{7_2}$]{\includegraphics[width=0.2\linewidth]{Plat7s2}}
%\subfloat[$\mathbf{7_3}$]{\includegraphics[width=0.2\linewidth]{Plat7s3}}
%\subfloat[$\mathbf{7_4}$]{\includegraphics[width=0.2\linewidth]{Plat7s4}} \\
%\subfloat[$\mathbf{7_5}$]{\includegraphics[width=0.2\linewidth]{Plat7s5}}
%\subfloat[$\mathbf{7_6}$]{\includegraphics[width=0.2\linewidth]{Plat7s6}}
%\subfloat[$\mathbf{7_7}$]{\includegraphics[width=0.2\linewidth]{Plat7s7}}
%\caption{Quasi--plat representations of hyperbolic knots with up to seven crossings.}\label{fig:4s1And6s1}
%\end{figure}

%%%%%%%%%%%%%%%%%%%%%%%%%
\subsection{Fixing the states}

The strategy to compute the $SU(N)$ knot quantum invariants is as follows. We first need to draw a knot in the plat or quasi--plat representation. A knot in a (quasi--)plat representation is a braid whose strands are closed off pairwise on either side of the braid. If in addition on both sides the strands are paired off in the following manner, $(1,2), (3,4), \ldots, (2m-1,2m)$, it is called a plat representation.  In Figs.~\ref{fig:4s1And6s1} we have the quasi--plat representations for the knots $\mathbf{4_1}$ and $\mathbf{6_1}$ \cite{Ramadevi:2000gq, Zodinmawia:2011oya}.\footnote{The (quasi--)plat representations of these knots represented here have been slightly changed compared to those in \cite{Zodinmawia:2011oya} in order to reduce the numbers of crossing matrices involved, so that the computation time for the colored HOMFLY invariants could be reduced.} A (quasi--)plat representation can only have even number of strands $2m$. A knot may have more than one  (quasi--)plat representation. The smallest $m$ in a plat representation is called the bridge number of the knot. In our computation, we need four strands in a (quasi--)plat representation. So the bridge number cannot be greater than two.

\begin{figure}[t]
\center
\includegraphics[width=0.4\linewidth]{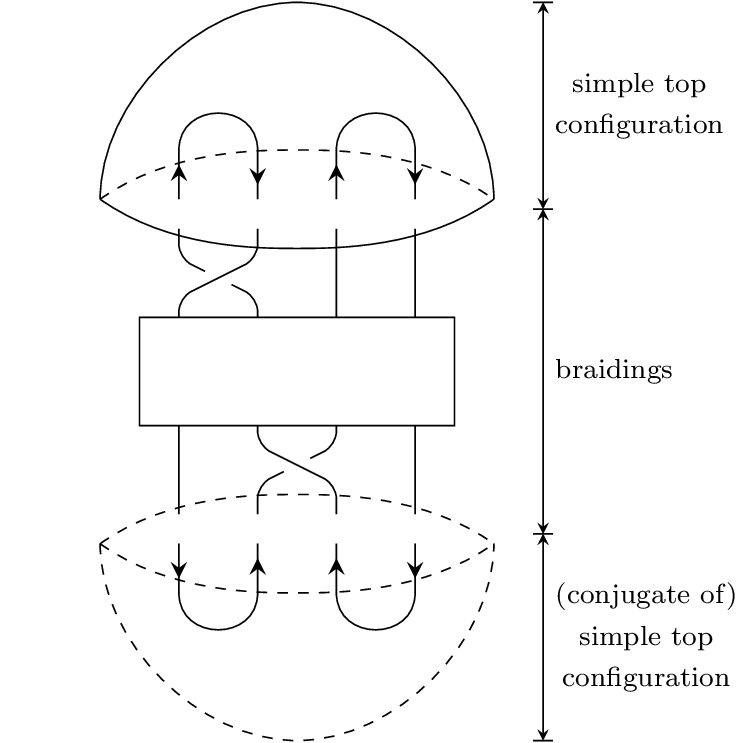}
\caption{A (quasi--)plat representation can be split into two simple top configurations and the braidings.}\label{fig:Sketch}
\end{figure}

Once a (quasi--)plat representation of a knot is drawn, we can break it apart into two simple top configurations and the braidings in the middle, as shown in Fig.~\ref{fig:Sketch}. We start with quantum state for the simple top configuration of a knot. Then we decompose it either in the basis~(1) or in the basis~(2) and apply the braiding operators in accordance with the (quasi--)plat representation. We use the crossing matrices if it is necessary to switch between central braidings and side braidings. In the last step we compute the inner product with the quantum state associated to the conjugate of a simple top configuration at the bottom of the knot.

In order to execute this procedure properly, we need to clarify two points. First, note that although the computation of the quantum knot invariant does not depend on which (quasi--)plat representation is used.  Different (quasi--)plat representations give rise to different framings, as the framing is minus the \emph{writhe} $w$ of the (quasi--)plat representation \cite{Zodinmawia:2011oya}.\footnote{One can understand this relationship by assigning the ``untwisted'' normal vector field to the (quasi--)plat representation in such a way that the normal vector at each point is parallel to the paper plane, and then by thickening the knot accordingly.} The writhe $w$ of a knot is defined by
\begin{equation}
		w =  \#(\textrm{over--crossing}) - \#(\textrm{under--crossing}) \ .
\end{equation}
So for a non--zero writhe of a (quasi--)plat representation the knot invariant is regularized by appropriate $U(1)$ factor as in eq.~\eqref{equ:U1Regularization}, so as to convert the colored HOMLFY invariant to zero framing with eq.~\eqref{equ:HOMFLYFramingTransform}. 

\begin{figure}[t]
\center
\includegraphics[width=0.255\textwidth]{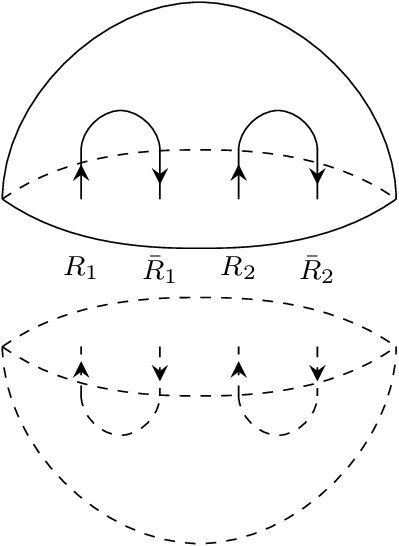}
\caption{The quantum state of the simple top configuration can be determined by gluing it to its conjugate configuration.}\label{fig:OneTop}
\end{figure}

Second, we still need to write down the quantum state for the simple top configurations. Given the top configuration in Fig.~\ref{fig:OneTop}, we observe that it is a trivial eigenstate of the braiding operator $b_1^{(-)}$ or $b_3^{(-)}$. So $|\phi^{(1)}_1\rangle$ is proportional to $|\phi^{(1)}_{0,00} (R_1,\bar{R}_1,R_2,\bar{R}_2) \rangle$. The proportionality constant is fixed by the requirement that the inner product of this state and its conjugate should yield the quantum knot invariant of two disconnected unknots
\begin{equation}
	\langle \Phi_1^{(1)} | \Phi_1^{(1)} \rangle  = \mathcal{W}^{SU(N)}_{R_1}( \bigcirc ) \mathcal{W}^{SU(N)}_{R_2}( \bigcirc ) = \dim_q R_1 \dim_q R_2	\ .
\end{equation}
With the natural normalization
\begin{equation}\label{equ:StateNormalization}
	\langle \phi^{(1)}_{0,00} (R_1,\bar{R}_1,R_2,\bar{R}_2)  |\phi^{(1)}_{0,00} (R_1,\bar{R}_1,R_2,\bar{R}_2) \rangle =1 \ , 
\end{equation}
 we find
\begin{equation}
	|\Phi^{(1)}_1\rangle = \sqrt{\dim_q R_1 \dim_q R_2} |\phi^{(1)}_{0,00} (R_1,\bar{R}_1,R_2,\bar{R}_2) \rangle \ .
\end{equation}
Note that in eq.\eqref{equ:StateNormalization} the conjugate configuration $R_i$ and $\bar{R}_i$ corresponds to strands coming out and going in, respectively.

Using eq.~\eqref{equ:TrivialCrossingMatrices} the quantum state $|\Phi^{(1)}_1\rangle$ can also be decomposed in the second basis with the help of the crossing matrices
\begin{align*}
 |\Phi^{(1)}_1\rangle =&  \sqrt{\dim_q R_1 \dim_q R_2} \, |\phi^{(1)}_{0,00} (R_1,\bar{R}_1,R_2,\bar{R}_2) \rangle \\
  =&\sum_{s,r_1,r_2} a^{0,00}_{s,r_1r_2}\begin{bmatrix}
 R_1 & \bar{R}_1 \\ R_2 & \bar{R}_2
  \end{bmatrix} \sqrt{\dim_q R_1 \dim_q R_2} \, |\phi^{(2)}_{s,r_1r_2}(R_1, \bar{R}_1, R_2, \bar{R}_2 ) \rangle \\
  =&\sum_{s,r_1,r_2} \{R_2\} \{ R_1,\bar{R}_2,s,r_1 \} \sqrt{\dim_q s} \; \delta_{r_1,r_2} \, |\phi^{(2)}_{s,r_1r_2}(R_1, \bar{R}_1, R_2, \bar{R}_2 ) \rangle \ .
\end{align*}
Therefore, the decomposition reads in the second basis
\begin{equation}
	|\Phi^{(1)}_1\rangle = \sum_{s,r_1,r_2} \{R_2\} \{ R_1,\bar{R}_2,s,r_1 \} \sqrt{\dim_q s} \; \delta_{r_1,r_2} \, |\phi^{(2)}_{s,r_1r_2}(R_1, \bar{R}_1, R_2, \bar{R}_2 ) \rangle \ .
\end{equation}
Note that this decomposition again fits well with the requirement that $\langle\Phi^{(1)}_1|\Phi^{(1)}_1\rangle$ is idential to $\mathcal{W}^{SU(N)}_{R_1}( \bigcirc ) \mathcal{W}^{SU(N)}_{R_2}( \bigcirc )$ because
\[
	\langle\Phi^{(1)}_1|\Phi^{(1)}_1\rangle = \sum_{s, r_1,r_2} \dim_q s \; \delta_{r_1,r_2} = \dim_q R_1 \dim_q R_2 \ .
\]

\begin{table}[t]
\center
\begin{tabular}{ c  l }
\parbox[c]{3.5cm}{\includegraphics[width=3.5cm]{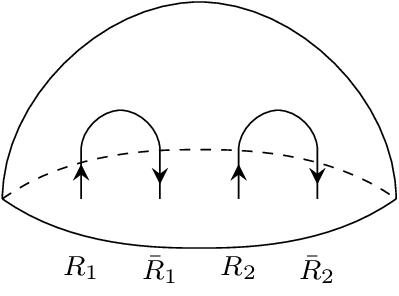}} & $\,\,\,\,\,\small\begin{aligned} &|\Phi^{(1)}_1\rangle =\sqrt{|R_1| |R_2|} |\phi^{(1)}_{0,00}(\ldots) \rangle \\ &=\{R_2\} \sum_{s,r_1,r_2} \{ \bar{R}_1,R_2,\bar{s},r_2\} \sqrt{ |s| } \; \delta_{r_1,r_2} |\phi^{(2)}_{s,r_1r_2}(\ldots) \rangle \end{aligned}$ \\
\parbox[c]{3.5cm}{\includegraphics[width=3.5cm]{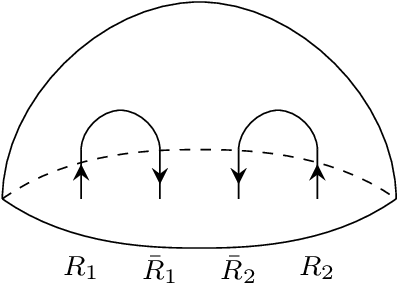}} & $\,\,\,\,\,\small\begin{aligned} &|\Phi^{(1)}_2\rangle =\{R_2\} \sqrt{|R_1| |R_2|} |\phi^{(1)}_{0,00}(\ldots) \rangle \\ &=\sum_{s,r_1,r_2} \{ \bar{R}_1,\bar{R}_2,\bar{s},r_2\} \sqrt{ |s| } \; \delta_{r_1,r_2} |\phi^{(2)}_{s,r_1r_2}(\ldots) \rangle \end{aligned}$\\
\parbox[c]{3.5cm}{\includegraphics[width=3.5cm]{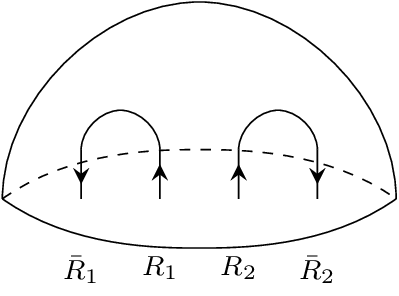}} & $\,\,\,\,\,\small\begin{aligned} &|\Phi^{(1)}_3\rangle =\{R_1\}\sqrt{|R_1| |R_2|} |\phi^{(1)}_{0,00}(\ldots) \rangle \\ &=\{R_1\}\{R_2\}\sum_{s,r_1,r_2} \{ R_1,R_2,\bar{s},r_2\} \sqrt{ |s| } \; \delta_{r_1,r_2} |\phi^{(2)}_{s,r_1r_2}(\ldots) \rangle \end{aligned}$\\
\parbox[c]{3.5cm}{\includegraphics[width=3.5cm]{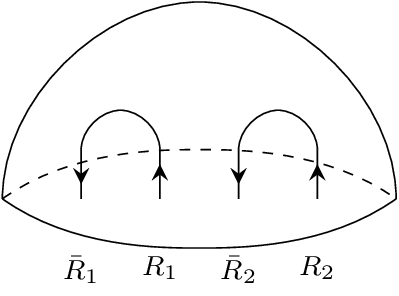}} & $\,\,\,\,\,\small\begin{aligned} &|\Phi^{(1)}_4\rangle =\{R_1\}\{R_2\}\sqrt{|R_1| |R_2|} |\phi^{(1)}_{0,00}(\ldots) \rangle \\ &=\{R_1\}\sum_{s,r_1,r_2} \{ R_1,\bar{R}_2,\bar{s},r_2\} \sqrt{ |s| } \; \delta_{r_1,r_2} |\phi^{(2)}_{s,r_1r_2}(\ldots) \rangle \end{aligned}$
\end{tabular}
\caption{Depicted are the first four simple top configurations together with their quantum states in terms of the quantum dimensions $|s|,|R_1|,|R_2|$. The ellipses in the parentheses are filled in with the appropriate WZW primaries.}\label{fig:AllTopsA}
\end{table}

\begin{table}[t]
\center
\begin{tabular}{ c  l }
\parbox[c]{3.5cm}{\includegraphics[width=3.5cm]{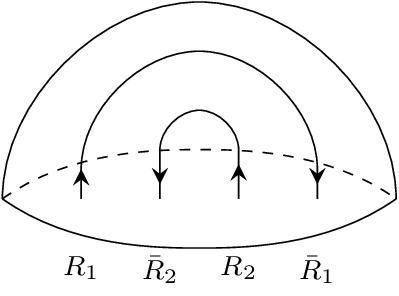}} & $\,\,\,\,\,\small\begin{aligned} &|\Phi^{(2)}_1\rangle =\{R_1\}\{R_2\}\sqrt{|R_1| |R_2|} |\phi^{(2)}_{0,00}(\ldots) \rangle \\ &=\{R_1\} \sum_{t,r_3,r_4} \{ R_1,\bar{R}_2,\bar{t},r_4\} \sqrt{ |t| } \; \delta_{r_3,r_4} |\phi^{(1)}_{t,r_3r_4}(\ldots) \rangle \end{aligned}$ \\
\parbox[c]{3.5cm}{\includegraphics[width=3.5cm]{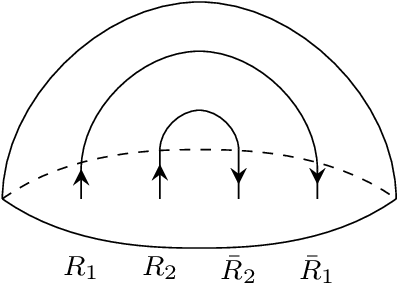}} & $\,\,\,\,\,\small\begin{aligned} &|\Phi^{(2)}_2\rangle =\{R_1\} \sqrt{|R_1| |R_2|} |\phi^{(2)}_{0,00}(\ldots) \rangle \\ &=\{R_1\}\{R_2\}\sum_{t,r_3,r_4} \{ R_1,R_2,\bar{t},r_4\} \sqrt{ |t| } \; \delta_{r_3,r_4} |\phi^{(1)}_{t,r_3r_4}(\ldots) \rangle \end{aligned}$\\
\parbox[c]{3.5cm}{\includegraphics[width=3.5cm]{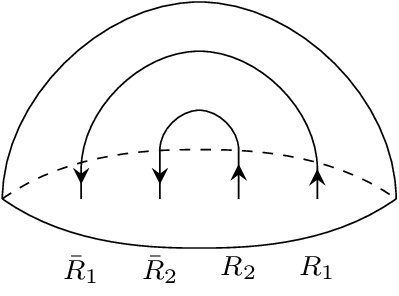}} & $\,\,\,\,\,\small\begin{aligned} &|\Phi^{(2)}_3\rangle =\{R_2\}\sqrt{|R_1| |R_2|} |\phi^{(2)}_{0,00}(\ldots) \rangle \\ &=\sum_{t,r_3,r_4} \{ \bar{R}_1,\bar{R}_2,\bar{t},r_4\} \sqrt{ |t| } \; \delta_{r_3,r_4} |\phi^{(1)}_{t,r_3r_4}(\ldots) \rangle \end{aligned}$\\
\parbox[c]{3.5cm}{\includegraphics[width=3.5cm]{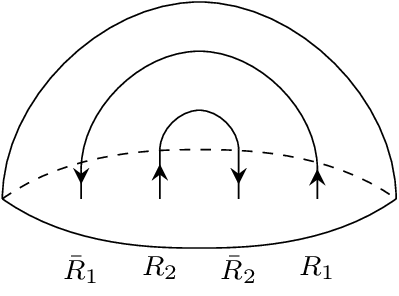}} & $\,\,\,\,\,\small\begin{aligned} &|\Phi^{(2)}_4\rangle =\sqrt{|R_1| |R_2|} |\phi^{(2)}_{0,00}(\ldots) \rangle \\ &=\{R_2\}\sum_{t,r_3,r_4} \{ \bar{R}_1,R_2,\bar{t},r_4\} \sqrt{ |t| } \; \delta_{r_3,r_4} |\phi^{(1)}_{t,r_3r_4}(\ldots) \rangle \end{aligned}$
\end{tabular}
\caption{The second four simple top configurations and the quantum states associated to them. $|s|,|R_1|,|R_2|$ denote the quantum dimensions. The ellipses in the parentheses are to be filled with the appropriate WZW primaries.}\label{fig:AllTopsB}
\end{table}

\begin{figure}[t]
\center
\subfloat[\label{fig:NormalizeTop1}]{\includegraphics[width=0.255\textwidth]{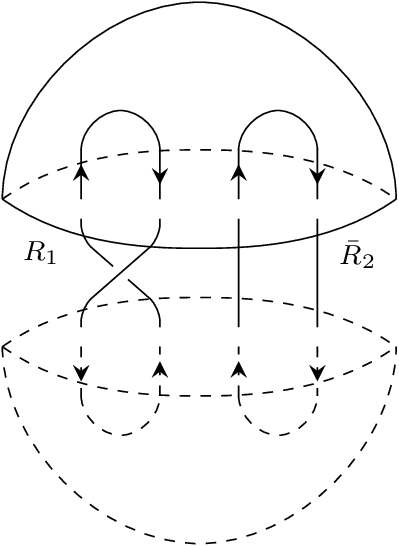}}\quad\quad\quad\quad 
\subfloat[\label{fig:NormalizeTop2}]{\includegraphics[width=0.255\textwidth]{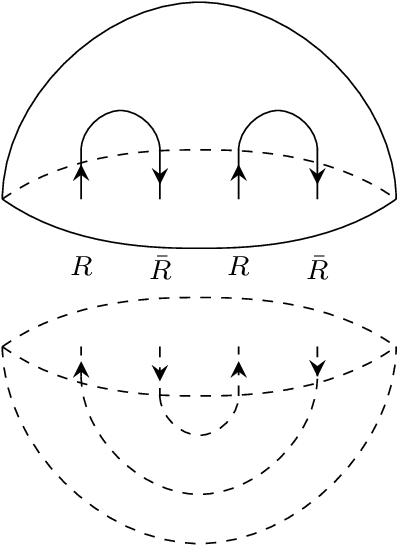}}
\caption{The relatives phases among configurations in Table~\ref{fig:AllTopsA} can be fixed by diagrams of type (a). The relative phases among configuraitons in Table~\ref{fig:AllTopsA} and configurations in Table~\ref{fig:AllTopsB} are determined by diagrams of type (b).}\label{fig:NormalizeTop}
\end{figure}

Similar arguments are used to write down the quantum states for all other simple top configurations in Table~\ref{fig:AllTopsA} and Table~\ref{fig:AllTopsB}. The relative phases among them are determined by gluing them together (if necessary with some twists) to build simple knots/links. For instance the relative phase between $|\Phi^{(1)}_1\rangle$  and $|\Phi^{(1)}_3\rangle$ can be derived by starting from the top configuration associated to $|\Phi^{(1)}_1\rangle$ in Table~\ref{fig:AllTopsA}, braiding the left two strands by $(b_1^{(-)})^{-1}$, and then closing it off with the conjugate of the top configuration associated to $|\Phi^{(1)}_3\rangle$ in Table~\ref{fig:AllTopsA}. The result as shown in Fig.~\ref{fig:NormalizeTop1} is two disconnected unknots with total framing +1, which is consistent with the computation
\begin{align*}
		\langle \Phi^{(1)}_3 | \Phi^{(1)}_1 \rangle &= \sqrt{ |R_1| |R_2| } \left( \{R_1,\bar{R}_1,0\} q^{-(2 C_{R_1}-0)}\right)^{-1}  \{R_1\} \sqrt{ |R_1| |R_2| } \\
		&=q^{2C_{R_1}}|R_1||R_2| = q^{2C_{R_1}} \mathcal{W}_{R_1}^{SU(N)}(\bigcirc) \mathcal{W}_{R_2}^{SU(N)}(\bigcirc) \ .
\end{align*}
Here we use $|R_1|, |R_2|$ as a shorthand notation for quantum dimensions. Analogously, we determine all the relative phases among the configurations in Table~\ref{fig:AllTopsA} and among the configurations in Table~\ref{fig:AllTopsB}. The relative phases between configurations in Table~\ref{fig:AllTopsA} and in Table~\ref{fig:AllTopsB} are fixed by computing the inner product of the first configuration in Table~\ref{fig:AllTopsA} and the first configuration in Table~\ref{fig:AllTopsB}. As depicted in Fig.~\ref{fig:NormalizeTop2}, it should yield the quantum invariant of a single unknot in framing zero. Indeed, we find
\begin{align*}
 \langle \Phi^{(2)}_1 | \Phi^{(1)}_1 \rangle &=\{R\}\sum_{s,r_1,r_2} \{\bar{R}, R, \bar{s}, r_1\} \sqrt{|s|} \delta_{r_1,r_2}  \{R\}^2|R| \; \langle \phi^{(2)}_{0,00}(\ldots) | \phi^{(2)}_{s,r_1r_2}(\ldots) \rangle\\
 & =|R|= \mathcal{W}^{SU(N)}_{R}(\bigcirc) \ .
\end{align*}

\subsection{Example}
As an illustrative example of the explained procedure, we briefly review the computation of the HOMFLY polynomial of the knot $\mathbf{6_1}$ colored with the fundamental representation. Fig.~\ref{fig:4s1And6s1} shows a convenient (quasi--)plat representation for the knot $\mathbf{6_1}$, which allows us to to compute $\mathcal{W}^{SU(N)}_{R}(\mathbf{6_1})$. We start with the first top configuration in Table~\ref{fig:AllTopsA}, decompose it in terms of the second basis, apply $b^{(-)}_2$ twice to produce the first two twists, switch to the first basis by crossing matrices, apply $(b^{(-)}_1)^{-1}$ four times to produce the remaining four under--crossings, and finally close it off with the conjugate of the first top configuration in Table~\ref{fig:AllTopsB}. Altogether we arrive at \cite{Ramadevi:2000gq}
\begin{align*}
\mathcal{W}^{SU(N)}_{R}(\mathbf{6_1},2) =&\{R \} \sum_{\substack{s^{(1)},r_1^{(1)},r_2^{(1)} \\ t^{(2)},r_3^{(2)},r_4^{(2)} }} \sqrt{|s^{(1)}|} \{R,\bar{R},s^{(1)},r_2^{(1)}\} \delta_{r^{(1)}_1,r^{(1)}_2}  \left(  \lambda^{(-)}_{R,\bar{R};s^{(1)},r^{(1)}_2}\right)^2 \\
a^{t^{(2)},r_3^{(2)}r_4^{(2)}}_{s^{(1)},r_1^{(1)}r_2^{(1)} }
&\begin{bmatrix}
 R & \bar{R} \\ R & \bar{R}
\end{bmatrix}^* 
\left( \lambda^{(-)}_{R,\bar{R};t^{(2)},r_4^{(2)}}\right)^{-4}  \sqrt{|t^{(2)}|} \{ R,\bar{R},t^{(2)},r_4^{(2)}\}  \delta_{r^{(2)}_3,r^{(2)}_4}\ ,
\end{align*}
where the argument indicates the framing $+2$ of the (quasi--)plat representation with writhe $w=-2$. Inserting the relevant crossing matrices $a^{t,r_3r_4}_{s,r_1r_2}\begin{bmatrix}
 \ydiagram{1} & \bar{\ydiagram{1}} \\ \ydiagram{1} & \bar{\ydiagram{1}}
\end{bmatrix}$ given in ref.~\cite{Ramadevi:2000gq} one arrives at
\begin{equation}
	\mathcal{W}^{SU(N)}_{\ydiagram{1}}(\mathbf{6_1},2) = - q^{-\frac{1}{N}}\frac{\lambda^{1/2}-\lambda^{-1/2}}{q^{1/2}-q^{-1/2}}\frac{1}{q \lambda}(-q+\lambda -  q \lambda + q^2 \lambda+\lambda^2-2 q \lambda^2 + q^2\lambda^2 -q \lambda^3) \ .
\end{equation}
Transforming to framing zero the normalized HOMFLY invariant for $\mathbf{6_1}$ becomes
\begin{equation}
	\bar{H}_{\ydiagram{1}}(\mathbf{6_1}) = \lambda  -(q+q^{-1}-2 )- (q+q^{-1}-1)\lambda^{-1}  +\lambda^{-2} \ ,
\end{equation}
in agreement with ref.~\cite{knotatlas} (with the identification $a = \lambda^{-1/2}$ and $z = q^{1/2} -q^{-1/2}$).

\bigskip

Analogously, one can determine the HOMFLY invariants colored by other representations for all two--bridge knots. The difficulty remains in explicitly computing the crossing matrices for the $SU(N)$ WZW models, as they are only known for simple representations $R_i$; for instance, for one-- or two--boxes representations $R_i$ the crossing matrices have been studied in refs.~\cite{RamaDevi:1992np, Zodinmawia:2011oya}. Explicit expressions for generic multiplicity--free crossing matrices have been conjectured in ref.~\cite{Nawata:2013ppa}, i.e., when $R_i$ are symmetric or anti--symmetric representations. However, cases with non--trivial multiplicities have not been treated before. Here we compute the crossing matrices when the representations $R_i$ are either $\ydiagram{2,1}$ or its conjugate, which are the simplest examples involving multiplicities. This allows us to derive the HOMFLY invariants for 2--bridge knots colored with $\ydiagram{2,1}$. 

%%%%%%%%%%%%%%%%%%%%%%%%%%%%%%
\section{Quantum and classical 6j--symbols for $SU(N)$} \label{sec:toprec}
%%%%%%%%%%%%%%%%%%%%%%%%%%%%%%

The crossing matrices in WZW models are directly related to the quantum 6j--symbols of quantum groups. It was first observed by physicists \cite{Moore:1988qv, Moore:1989vd,AlvarezGaume:1989aq} and later proved by mathematicians\cite{Kazhdan:1993,Kazhdan:1994a, Kazhdan:1994b, Finkelberg:1996} that the spectra of WZW models and the representations of quantum groups are closely related.  That is to say the WZW primaries of the $\widehat{su(N)}_k$ WZW model are in one-to-one correspondence with the finite dimensional representations of the quantum group $\mathcal{U}_q su(N)$ with $q=\exp(2\pi i/(k+N))$. Furthermore, the crossing matrices in the WZW~model are --- up to normalizations --- given in terms of the recoupling coefficients of the quantum group.

%%%%%%%%%%%%%%%%%%%%%%%%%%%%%%%%%%%%%%%%%%%%
\subsection{Quantum 6j--symbols and crossing matrices}\label{sec:recoupling}
So we need to determine the recoupling coefficints for the quantum group $\mathcal{U}_q su(N)$. First notice that when $q$ is generic or when $q$ is a root of unity and the order of $q$ is high, which is the case when either $k$ or $N$ is large, the finite dimensional representations of the quantum group $\mathcal{U}_q su(N)$ behave as their classical counterparts in the simple Lie algebra $su(N)$. So we will use the usual group representation notation with subscript $q$ here. Recoupling coefficients arise when three irreducible representations $\lambda_1,\lambda_2,\lambda_3$ couple to a fourth $\lambda$, or equivalently when four irreducible representations couple to singlets. From the former point of view, a resultant state in the representation $\lambda$ can be labelled by intermediate irreducible representations together with multiplicities, e.g.,
\begin{align*}
|\lambda_1 m_1\rangle_q &| \lambda_2 m_2\rangle_q |\lambda_3 m_3 \rangle_q \\
	=& \sum_{r_{12} \lambda_{12} m_{12}  } | (\lambda_1\lambda_2), r_{12} \lambda_{12} m_{12} \rangle_q |\lambda_3 m_3\rangle_q  \cdot \langle r_{12} \lambda_{12} m_{12} | \lambda_1m_1,\lambda_2m_2\rangle_q \\
	=& \sum_{\substack{r_{12}\lambda_{12}m_{12} \\ r\lambda m}} 
	| (\lambda_1\lambda_2)r_{12}\lambda_{12},\lambda_3,r\lambda m\rangle_q  \cdot
	\langle r_{12} \lambda_{12} m_{12} | \lambda_1 m_1, \lambda_2 m_2 \rangle_q \\ 
	&\qquad\qquad\times \langle r\lambda m | \lambda_{12} m_{12}, \lambda_3 m_3 \rangle_q \ .
\end{align*}
Here $m_i$ labels a state in the representation $\lambda_i$, and $\langle r_{12} \lambda_{12} m_{12} | \lambda_1m_1,\lambda_2m_2\rangle_q$ is a quantum Clebsch--Gordon coefficient, just like their classical counterpart in simple Lie algebra. One observes that in the end many states of the same type $|\lambda m\rangle_q$ appear, and they are distinguished by the intermediate representation $\lambda_{12}$ as well as the multiplicity labels $r_{12}, r$. On the other hand one can change the order of coupling and couple $|\lambda_2 m_2\rangle_q | \lambda_3m_3\rangle_q$ first instead
\begin{align*}
|\lambda_1 m_1\rangle_q | \lambda_2 m_2\rangle_q |\lambda_3 m_3 \rangle_q  
 =\sum_{\substack{\lambda_{23} \lambda_{23} m_{23} \\ r' \lambda m}} &
	| \lambda_1(\lambda_2\lambda_3)r_{23} \lambda_{23}, r'\lambda m\rangle_q \cdot \langle r'\lambda m| \lambda_1 m_1,\lambda_{23} m_{23} \rangle_q \\
 &\quad\times \langle r_{23} \lambda_{23}m_{23} | \lambda_2 m_2,\lambda_3 m_3\rangle_q \ .
\end{align*}
In this case the states $|\lambda m\rangle$ are distinguished by $\lambda_{23}r_{23}r'$. These two orthonormal basis are related by the quantum recoupling coefficients arising in
\begin{align*}
	| \lambda_1(\lambda_2\lambda_3)r_{23} \lambda_{23}, r'\lambda m\rangle_q =  \sum_{r_{12}\lambda_{12}r} &| (\lambda_1\lambda_2)r_{12}\lambda_{12},\lambda_3,r\lambda m\rangle_q  \\
	 &\ \times\langle (\lambda_1 \lambda_2 )r_{12}\lambda_{12} , \lambda_3; r\lambda | \lambda_1 (\lambda_2\lambda_3)r_{23}\lambda_{23} ; r' \lambda \rangle_q \ .
\end{align*}
Not surprisingly the quantum recoupling coefficients can be expressed in terms of the quantum Clebsch--Gordon coefficients, from which the symmetry properties of the recoupling coefficients are deduced
\begin{equation} \label{eq:recoup}
\begin{aligned}
	\langle (\lambda_1 \lambda_2 )r_{12}\lambda_{12} , \lambda_3; r\lambda | \lambda_1& (\lambda_2\lambda_3)r_{23}\lambda_{23} ; r' \lambda \rangle_q  \\ 
	 =\frac{1}{\dim_q \lambda}\sum_{\substack{m_1m_2m_3\\m,m_{12}m_{23}}}
	&\langle r_{12} \lambda_{12} m_{12} | \lambda_1 m_1, \lambda_2 m_2 \rangle_q
	\langle r \lambda m | \lambda_{12} m_{12}, \lambda_3 m_3 \rangle_q \\
	&\times\langle r' \lambda m | \lambda_1 m_1, \lambda_{23}m_{23} \rangle_q^*
	\langle r_{23} \lambda_{23} m_{23} | \lambda_2 m_2, \lambda_3 m_3 \rangle_q^* \ .
\end{aligned}
\end{equation}
The quantum recoupling coefficients are usually normalized to \emph{quantum 6j--symbols}
\begin{align}\label{equ:Recoupling6j}
	\prescript{}{q}{\begin{Bmatrix}
	\lambda_1 & \lambda_2 & \bar{\lambda}_{12} \nonumber\\
	\lambda_3 & \lambda & \lambda_{23} 
	\end{Bmatrix}_{r', r_{23},r, r_{12}} }
	\,=&\, \frac{\{\lambda_2 \} \{ \lambda_1\lambda_2 \bar{\lambda}_{12} r_{12} \} \{ \lambda_{12} \lambda_3 \bar{\lambda} r \} }{\sqrt{\dim_q \lambda_{12} \dim_q \lambda_{23} }}  \nonumber\\
	&\qquad \times\langle (\lambda_1 \lambda_2) r_{12} \lambda_{12}, \lambda_3 ; r \lambda | \lambda_1 (\lambda_2 \lambda_3) r_{23} \lambda_{23} ; r' \lambda \rangle_q \ .
\end{align}
Their symmetry properties can be presented in a more symmetric manner as discussed in detail in Sec.~\ref{sec:q6jSymmetry}. In general, from a quantum 6j--symbol
\[
	\prescript{}{q}{\begin{Bmatrix}
		\lambda_1 & \lambda_2 & \lambda_3 \\
		\mu_1 & \mu_2 & \mu_3
	\end{Bmatrix}_{r_1r_2r_3r_4}}
\]
one can easily read off all the four couplings involving the six representations $\lambda_i, \mu_j$. Diagrammatically they can be represented by
\vskip8pt
{\centering
\hfill\includegraphics[width=0.2\textwidth]{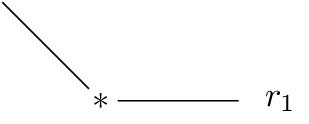} $\;$ \includegraphics[width=0.17\textwidth]{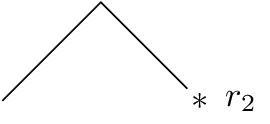} $\;$ \includegraphics[width=0.2\textwidth]{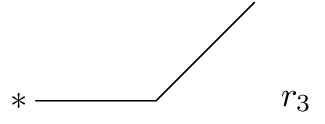} $\;$ \includegraphics[width=0.2\textwidth]{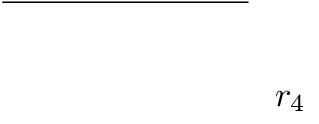}\hfill
}
\vskip8pt
\noindent where $*$ stands for conjugate representation. Here each path goes over three representations, which --- together with the multiplicity label, form a triad --- namely the tensor product of the three representations contains singlet. The multiplicity label are non--trivial if multiple copies of singlets occur. For instance, the first path represents the triad $(\lambda_1,\bar{\mu}_2,\mu_3;r_1)$, which is equivalent to saying that $\lambda_1$ and $\bar{\mu}_2$ couple to $\bar{\mu}_3$ with multiplicity label $r_1$. 

The values of the quantum 6j--symbols are known when any of the six representations is singlet (trivial quantum 6j--symbols), e.g.,
\begin{equation}\label{equ:trivial6jValue}
			\prescript{}{q}{\begin{Bmatrix}
			\lambda_1 & \lambda_2 & \lambda_3 \\
			\lambda_2^* & \lambda_1 & 0
			\end{Bmatrix}_{00rs}}
			= \frac{\{\lambda_1\lambda_2\lambda_3 r\} }{ \sqrt{\dim_q \lambda_1 \dim_q \lambda_2} }\delta_{rs}	\	.
\end{equation}

We can now state the relationship between quantum recoupling coefficients --- given in terms of quantum 6j--symbols --- and crossing matrices in WZW models
\begin{equation}\label{equ:Cross6j}
	a^{t,r_3r_4}_{s,r_1r_2}\begin{bmatrix} R_1 & R_2 \\ R_3 & R_4 \end{bmatrix} =  
	\{R_2 \} \{ R_1,R_2,\bar{t}, r_4 \} \{ \bar{R}_3, \bar{R}_4, \bar{t}, r_3 \}  \sqrt{\dim_q t \dim_q s }
	\prescript{}{q}{\begin{Bmatrix}
	R_1 & R_2 & \bar{t}\\
	R_3 & \bar{R}_4 & s 
	\end{Bmatrix}_{r_1r_2r_3r_4} } \ ,
\end{equation}
which maps the problem of determining crossing matrices to the computation of the quantum 6j--symbols. In this work, in order to compute the HOMFLY invariants colored with $R=\ydiagram{2,1}$, we need to determine the following two kinds of quantum 6j--symbols 
\begin{align}
		&\prescript{}{q}{\begin{Bmatrix}
			R & \bar{R} & \rho_1 \\
			R & R & \rho_2
			\end{Bmatrix}_{r_1r_2r_3r_4}} \hskip-40pt	
		&&\prescript{}{q}{\begin{Bmatrix}
			\bar{R} & R & \rho_3 \\
			R & R & \rho_4
			\end{Bmatrix}_{r'_1r'_2r'_3r'_4}} \label{equ:TwoKinds}\\
			&\quad\textrm{   first kind}  &&\quad\textrm{   second kind}\nonumber
\end{align}
To simplify the notation, we suppress from now on the subscript $q$. Whether we refer to classical or quantum 6j--symbols should become clear from the context.

\subsubsection{The symmetry properties of quantum 6j--symbols}\label{sec:q6jSymmetry}
We briefly review the symmetry properties of quantum 6j--symbols \cite{Lienert:1992}, as they are the pillar of our computational method: 
\begin{enumerate}
\item Cyclic permutation and exchange of columns
\begin{align}\label{equ:ColumnExchange}
	\begin{Bmatrix}
		\lambda_1 & \lambda_2 & \lambda_3 \\
		\mu_1 & \mu_2 & \mu_3
	\end{Bmatrix}_{r_1r_2r_3r_4} 
	=&\begin{Bmatrix}
		\lambda_2 & \lambda_3 & \lambda_1 \\
		\mu_2 & \mu_3 & \mu_1
	\end{Bmatrix}_{r_2r_3r_1r_4}   \nonumber\\
	=&\{\mu_1\} \{\mu_2\} \{\mu_3\} \{\lambda_1\bar{\mu}_2\mu_3r_1\}\{ \mu_1\lambda_2 \bar{\mu}_3r_2\} \nonumber\\
	&\times\{\bar{\mu}_1 \mu_2\lambda_3r_3 \}\{\lambda_1\lambda_2\lambda_3r_4\}
	\begin{Bmatrix}
	\lambda_2 & \lambda_1 & \lambda_3 \\
	\bar{\mu}_2 & \bar{\mu}_1 & \bar{\mu}_3
	\end{Bmatrix}_{r_2r_1r_3r_4} \ .
\end{align}
\item Rows exchange in two neighboring columns
\begin{align}
\begin{Bmatrix}
\lambda_1 & \lambda_2 & \lambda_3 \\
\mu_1 & \mu_2 & \mu_3
\end{Bmatrix}_{r_1r_2r_3r_4}
&=\begin{Bmatrix}
\bar{\lambda}_1 & \mu_2 & \bar{\mu}_3 \\
\bar{\mu}_1 & \lambda_2 & \bar{\lambda}_3
\end{Bmatrix}_{r_4r_3r_2r_1}
=\begin{Bmatrix}
\bar{\mu}_1 & \bar{\lambda}_2 & \mu_3 \\
\bar{\lambda}_1 & \bar{\mu}_2 & \lambda_3
\end{Bmatrix}_{r_3r_4r_1r_2} \nonumber\\
&=
\begin{Bmatrix}
\mu_1 & \bar{\mu}_2 & \bar{\lambda}_3 \\
\lambda_1 & \bar{\lambda}_2 & \bar{\mu}_3
\end{Bmatrix}_{r_2r_1r_4r_3}		\	.
\end{align}
\noindent These two symmetries together are also known as the \emph{tetrahedral symmetry}.
\item Complex conjugation\footnote{In ref.~\cite{Lienert:1992} $q$ is replaced by $1/q$ after complex conjugation. On the other hand it is shown in ref.~\cite{Pan:1993} the 6j--symbols are not changed by this replacement. Indeed a 6j--symbol consists of q--deformed numbers $[x]$ which are $q-1/q$ symmetric.}
\begin{equation}
			\begin{Bmatrix}
				\lambda_1 & \lambda_2 & \lambda_3 \\
				\mu_1 & \mu_2 & \mu_3 
			\end{Bmatrix}_{r_1r_2r_3r_4}^* 
			= \begin{Bmatrix}
				\bar{\lambda}_1 & \bar{\lambda}_2 & \bar{\lambda}_3	\\
				\bar{\mu}_1 & \bar{\mu}_2 & \bar{\mu}_3
			\end{Bmatrix}_{r_1r_2r_3r_4}		\	.
\end{equation}
\item Unitarity
\begin{equation}\label{equ:Unitarity}
	\sum_{\mu_3r_1r_2} |\lambda_3| |\mu_3| \begin{Bmatrix}
	\lambda_1 & \lambda_2 & \lambda_3 \\
	\mu_1 & \mu_2 & \mu_3
	\end{Bmatrix}_{r_1r_2r_3r_4}
	\begin{Bmatrix}
	\lambda_1 & \lambda_2 & \lambda_3' \\
	\mu_1 & \mu_2 & \mu_3
	\end{Bmatrix}_{r_1r_2r_3'r_4'}^*=\delta_{\lambda_3\lambda_3'}\delta_{r_3r_3'}\delta_{r_4r_4'}	\	.
\end{equation}
\item The generalized Racah backcoupling rule
\begin{align}\label{equ:Backcoupling}
	&q^{(C_{\lambda_1}+C_{\mu_1}+C_{\lambda_3}+C_{\mu_3})/2}	
	\begin{Bmatrix}
	\lambda_1 & \lambda_2 & \lambda_3\\
	\mu_1 & \mu_2 & \mu_3 
	\end{Bmatrix}_{r_1r_2r_3r_4} \nonumber\\
	=&\sum_{\nu r s} 
	q^{(C_{\nu}+C_{\lambda_2}+C_{\mu_2})/2}
	|\nu| \{\lambda_3\} \{ \lambda_1\bar{\mu}_2\mu_3r_1 \} \{ \mu_1\lambda_2\bar{\mu}_3r_2\} \nonumber\\
	&\quad\quad\quad\quad\cdot \{ \bar{\lambda}_1\mu_1\nu r\}
	\begin{Bmatrix}
	\mu_3 & \nu & \lambda_3 \\
	\mu_1 & \mu_2 & \lambda_1
	\end{Bmatrix}_{r_1 r r_3 s}
	\begin{Bmatrix}
	\lambda_1 & \lambda_2 & \lambda_3\\
	\bar{\mu}_3 & \nu & \bar{\mu}_1
	\end{Bmatrix}_{r r_2 s r_4}	\ .
\end{align}
\noindent This symmetry reflects the fact the conformal blocks of \textit{t}--, \textit{s}-- and \textit{u}--channels are related to each other via a triangular commutative diagram.
\item The Biedenharn--Elliott sum rule
\begin{align}\label{equ:Pentagon}
\begin{Bmatrix}
	\lambda_1 & \lambda_2 & \lambda_3 \\
	\mu_1 & \mu_2 & \mu_3 
\end{Bmatrix}_{r_1r_2r_3r_4}
&= \sum_{\substack{\lambda \nu_3 \\t_1t_2t_3s_1s_2}} |\lambda_3| |\nu_3| |\lambda| \{\lambda_1\} \{\nu_1\} \{\lambda_1\bar{\mu}_2\mu_3r_1\} \nonumber\\
&\cdot \{ \mu_1\lambda_2\bar{\mu}_3r_2\} \{\bar{\mu}_1\mu_2\lambda_3r_3\} \{\lambda\bar{\mu}_1\nu_1t_1\}\{ \lambda\bar{\mu}_2\nu_2t_2\}\{ \lambda\bar{\mu}_3\nu_3t_3\} \nonumber\\
&\cdot \begin{Bmatrix}
\nu_2 & \bar{\mu}_2 & \lambda \\
\mu_3 & \nu_3 & \bar{\lambda}_1
\end{Bmatrix}_{s_1r_1t_3t_2}
\begin{Bmatrix}
\nu_3 & \bar{\mu}_3 & \lambda\\
\mu_1 & \nu_1 & \bar{\lambda}_2
\end{Bmatrix}_{s_2r_2t_1t_3}\nonumber\\
&\cdot \begin{Bmatrix}
\nu_1 & \bar{\mu}_1 & \lambda \\
\mu_2 & \nu_2 & \bar{\lambda}_3
\end{Bmatrix}_{s_3r_3t_2t_1}
\begin{Bmatrix}
\lambda_1 & \lambda_2 & \lambda_3 \\
\nu_1 & \nu_2 & \nu_3
\end{Bmatrix}_{s_1s_2s_3r_4}	\	.
\end{align}
On the right hand side the representations $\nu_1$ and $\nu_2$ are chosen arbitrarily, only subject to the constraint that the right hand side does not vanish identically, i.e., the triads involving $\nu_1$ or $\nu_2$ should all be valid in the sense that the tensor product of the three representations in each of these triads contains a singlet. This symmetry property is also known as the \emph{pentagon relation}.
\end{enumerate}
As shorthand notation we have introduced here the notation $|\lambda|$ for the quantum dimension $\dim_q \lambda$.

\subsubsection{Consistency of the braiding eigenvalue phases}\label{sec:PhasesConsistency}
\begin{figure}
\center
\includegraphics[width=0.3\textwidth]{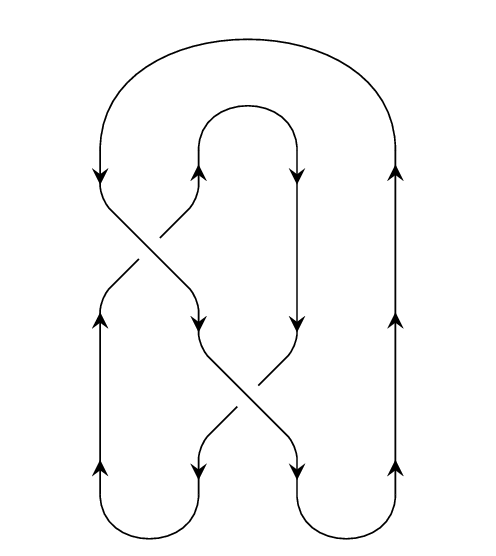}
\caption{Computing the knot invariant of this plat representation can demonstrate the consistency of the braiding phase.}\label{fig:BraidingPhaseConsistency}
\end{figure}

Let us first test the powers of the symmetry properties of the quantum 6j--symbols by using them to demonstrate the consistency of the braiding phases~\eqref{equ:BraidingEigenvalues}. We will use the procedure explained in Sec.~\ref{sec:curves} to compute the knot invariant of two disconnected unknots with the (quasi--)plat representation given in Fig.~\ref{fig:BraidingPhaseConsistency}, where braidings appear twice and their phases do not trivially cancel. In this way we can confirm the proper and consistent assignment of the previously stated braiding phases.

The (quasi--)plat representation has writhe 0, i.e., framing 0. So we expect for the quantum invariant
\begin{equation} \label{eq:Wlink}
	\mathcal{W}^{SU(N)}_{R_1}(\bigcirc) \mathcal{W}^{SU(N)}_{R_2}(\bigcirc) =\dim_q R_1 \dim_q R_2 \ .
\end{equation}
Following the procedure introduced in Sec.~\ref{sec:curves}, the formula to compute the quantum invariant of this untangled link $\mathcal{L}$ is 
\begin{align*}
	\mathcal{W}^{SU(N)}_{R_1,R_2}(\mathcal{L}) =&\sum_{\substack{t,s\\r_1,r_2,r_3,r_4}} \{R_2\} \{\bar{R}_1,R_2,\bar{t},r_4\} \sqrt{\dim_q t}\;\delta_{r_3,r_4}
	\lambda^{(-)}_{\bar{R}_1R_2;t,r_4}
	a^{t,r_4r_4}_{s,r_2r_2}\begin{bmatrix}
	R_2 & \bar{R}_1 \\
	\bar{R}_2 & R_1
	\end{bmatrix}\\
	&\times\left( \lambda^{(+)}_{\bar{R}_1\bar{R}_1;s,r_2} \right)^{-1}
	\{\bar{R}_1,\bar{R}_2,\bar{s},r_2\}\sqrt{\dim_q s} \; \delta_{r_1,r_2} \\
	=&\sum_{\substack{t,s\\r_2,r_4}}\{R_1\}\{R_2\}q^{-(C_{R_1}+C_{R_2})+(C_s+C_t)/2}  \dim_q t\dim_q s 
	\begin{Bmatrix}
		R_2 & \bar{R}_1 & \bar{t}\\
		\bar{R}_2 & \bar{R}_1 & s
	\end{Bmatrix}_{r_2r_2r_4r_4} \ .
\end{align*}
Using the cyclic permutation symmetry and the generalized Racah backcoupling rule,  we find
\begin{align*}
	\begin{Bmatrix}
		R_1 & \bar{R}_1 & 0 \\
		\bar{R}_2 & \bar{R}_2 & s
	\end{Bmatrix}_{r_1r_200} = &
	\begin{Bmatrix}
		\bar{R}_1 & 0 & R_1\\
		\bar{R}_2 & s & \bar{R}_2
	\end{Bmatrix}_{r_20r_10}\\ = &
	\sum_{t,r_3,r_4} q^{-(C_{R_1}+C_{R_2})+(C_s+C_t)/2} \dim_q t \{R_1\} \{\bar{R}_1,\bar{s},\bar{R}_2,r_2\} \{R_2\}\\
	 &\{\bar{R}_1,\bar{t},R_2,r_3\}
	\begin{Bmatrix}
		\bar{R}_2 & t & R_1 \\
		\bar{R}_2 & s & \bar{R}_1
	\end{Bmatrix}_{r_2r_3r_1r_4} 
	\begin{Bmatrix}
		\bar{R}_1 & 0 & R_1\\
		R_2 & t & R_2
	\end{Bmatrix}_{r_30r_40} \ .
\end{align*}
Plug in the values of trivial quantum 6j--symbols, we finally arrive at
\[
	\{R_1\}\{R_2\}\sum_{t,r_3,r_4} q^{-(C_{R_1}+C_{R_2})+(C_s+C_t)/2}\dim_q t \; \delta_{r_3,r_4} 
	\begin{Bmatrix}
		R_1 & \bar{R}_2 & t\\
		\bar{R}_1 & \bar{R}_2 & s
	\end{Bmatrix}_{r_1r_2r_3r_4} = \delta_{r_1,r_2} \ .
\]
Finally, with $\dim_q t =\dim_q \bar{t}$ we infer 
\begin{align*}
	\mathcal{W}^{SU(N)}_{R_1,R_2}(\mathcal{L}) = \sum_s \dim_q s = \dim_q R_1 \dim_q R_2 \ ,
\end{align*}
which is in agreement with eq.~\eqref{eq:Wlink}.

\subsubsection{Phase specification}\label{sec:3jPhases}
In this section we specify our choice of 2j--phases and 3j--phases. When $k$ or $N$ is large --- as is the case in this note --- the representations of $\mathcal{U}_qsu(N)$ behave as those of $su(N)$. Therefore, we follow the same choice made for 2j--phases and 3j--phases for the classical Lie algebra $su(N)$ \cite{Butler:1981, Haase:1983}.

For 2j--phases: It's easy to see that $\{\lambda\} = \{\bar{\lambda}\}$. Furthermore $su(N)$ is a so-called \emph{quasi-ambivalent} group, which means consistent choices for 2j--phases can be made so that
\begin{equation}
		\{\lambda_1\} \{\lambda_2\} \{\lambda_3\} = 1 \ ,
\end{equation}
whenever $\lambda_1\lambda_2\lambda_3$ is a valid triad (with trivial multiplicity label). Repetitively using this rule all the 2j--phases can be reduced to either 1, if $\lambda$ has even number of boxes, or $\{ \ydiagram{1}\}$, if $\lambda$ has odd number of boxes. We find in our computation that  it is self--consistent and convenient to choose $\{\ydiagram{1}\}=1$ since we work with generic rank $N$. 

As for 3j--phases: A 3j--phase $\{\lambda_1\lambda_2\lambda_3r\}$ is fixed if $\lambda_1=\lambda_2$, being $+1$ if $\bar{\lambda}_3$ appears in the symmetric tensor product $\textrm{Sym}^2\lambda_1$, or $-1$ if $\bar{\lambda}_3$ appears in the antisymmetric tensor product $\wedge^2\lambda_1$. In the generic case $\lambda_1 \neq \lambda_2 \neq\lambda_3$, the 3j--phase is a priori undetermined. However, a consistent choice can be made for 3j--phases so that the column exchange symmetry for 6j--symbols is simplified, i.e.,
\begin{equation}\label{equ:3jPhases}
		\{\mu_1\} \{\mu_2\} \{\mu_3\} \{\lambda_1\bar{\mu}_2\mu_3r_1\}\{ \mu_1\lambda_2 \bar{\mu}_3r_2\} \{\bar{\mu}_1 \mu_2\lambda_3r_3 \} \{ \lambda_1\lambda_2\lambda_3r_4\} = (-1)^{r_1 + r_2 + r_3 + r_4} \ .
\end{equation}
Then eq.~\eqref{equ:ColumnExchange} becomes
\begin{align}\label{equ:ColumnExchangeSimp}
	\begin{Bmatrix}
		\lambda_1 & \lambda_2 & \lambda_3 \\
		\mu_1 & \mu_2 & \mu_3
	\end{Bmatrix}_{r_1r_2r_3r_4}
	=(-1)^{r_1+r_2+r_3+r_4}
	\begin{Bmatrix}
	\lambda_2 & \lambda_1 & \lambda_3 \\
	\bar{\mu}_2 & \bar{\mu}_1 & \bar{\mu}_3
	\end{Bmatrix}_{r_2r_1r_3r_4} \ .
\end{align}
Furthermore, we specify that 
\begin{equation}
	\{\lambda_1\lambda_2\lambda_3r\} = (-1)^r \{\lambda_1\lambda_2\lambda_30\} \equiv (-1)^r \{\lambda_1\lambda_2\lambda_3\} \ .
\end{equation}
By using these rules one can reduce a generic 3j--phase to those containing a fundamental representation $\{ \ydiagram{1} \lambda_2\lambda_3 \}$, and further reduce them to one 3j--phase $\{\ydiagram{1} \lambda' \lambda \}$ for each non--fundamental representation $\lambda$, where $\lambda'$ is some lower representation than $\lambda$. The values $(\pm 1)$ of these 3j--phases are free to choose. The first few 3j--phases are tabulated in the appendix of ref.\cite{Haase:1983}.

\subsubsection{Composite labelling of representations}\label{sec:compos}
In our computation we use the composite labelling of the partition associated to an irreducible representation of $\mathcal{U}_qsu(N)$ or $su(N)$. The partition label $\lambda  = \lambda_1 \geqslant \lambda_2 \geqslant \cdots \geqslant \lambda_{N-1} \geqslant \lambda_N = 0$ can be recast in a composite manner,
\begin{align*}
	(\lambda) &= (\lambda_1, \lambda_2, \cdots, \lambda_N) \\
	&=( \mu_1, \mu_2 , \cdots, \mu_p, 0, \cdots, 0, -\nu_q, -\nu_{q-1}, \cdots, -\nu_1  ) \\
	&=( \mu; \nu )		\	,
\end{align*}
where $p+q \leqslant N$. Here the second line is obtained by subtracting the same integer from each $\lambda_i$.

On the one hand, the composite labelling seems to suffer from an integral shift ambiguity: $\mu_i \mapsto \mu_i + n, \nu_j \mapsto \nu_j + n, \forall n\in \mathbb{Z}$, all corresponding to the same $\lambda$. In fact the composite labellings are one-to-one correspondent to the irreducible representations of $U(N)$, and the shift freedom arises from discarding the $U(1)$ factor. More precisely, the relation between a composite representation of $U(N)$ and an irreducible representation of $SU(N)$ is
\begin{align*}
			U(N)  &= U(1) \times SU(N)		\\
		(\lambda) &=  (f) \times (\lambda')
\end{align*}
where $f = \sum_{i=1}^N \lambda_i$  and $\lambda'_i = \lambda_i -\lambda_N$ for $i=1,\cdots, N$. On the other hand, when the rank $N$ is generic, for a low dimensional or low power representation the composite labelling without explicit $N$ is unique. Here we define the power $p(\lambda)$ of a representation $\lambda$ to be the minimum number of fundamentall or anti-fundamentals so that $\lambda$ is included in the tensor product
\[
		\lambda \in \underbrace{\ydiagram{1} \otimes \cdots\otimes \ydiagram{1} \otimes \bar{\ydiagram{1}}\otimes \cdots\otimes\bar{\ydiagram{1}}}_{p(\lambda)} \ .
\]
It is easy to show $p(\lambda)$ is the same as the number of boxes in the \emph{unique} composite labelling of $\lambda = (\mu;\nu)$.

The advantage of using the composite labelling is that the conjugate representations can be easily expressed. The conjugate  of $(\mu;\nu)$ is nothing else but $(\nu;\mu)$. Since the rank of the group $N$ does not appear explicitly, the composite labelling is particularly useful in computation with generic rank $N$.

The tensor product of two composite representations is given by \cite{Butler:1974}
\begin{equation} \label{eq:tprod}
		(\mu;\nu) \otimes (\rho;\sigma) = \sum_{\zeta,\eta} \{ (\mu/\zeta)\cdot (\rho/\eta) ; (\nu/\eta) \cdot (\sigma/\zeta) \}	\	,
\end{equation}
where both the division and multiplication are given by the Littlewood--Richardson coefficients $c_{i,j}^k$,
\[
	\lambda_i \cdot \lambda_j = \sum_k c_{ij}^k \lambda_k  \ ,\qquad
	\lambda_k / \lambda_i = \sum_j c_{ij}^k \lambda_j \ .
\]
The quantum dimension of a composite representation is
\[
			\dim_q (\mu;\nu) = N_n(\mu;\nu)/H(\mu) H(\nu)	\	,
\]
where $H(\mu)$ and $H(\nu)$ are the hook length products of the respective Young diagrams with each factor $x$ promoted to the q--ormed number
\[
		[x] = \frac{q^{x/2} - q^{-x/2}}{q^{1/2} -q^{-1/2}} \ ,
\]
and the numerator is
\[
			N_n(\mu;\nu) = \prod_{i,j;k,l} [N-i-j + \mu_i + \nu_j +1 ] [N+k+l -\tilde{\mu}_k - \tilde{\nu}_l -1 ] \ .
\]
Here $\mu_i, \nu_j$ are the number of boxes on the $i$-th row of $\mu$ and $j$-th row of $\nu$ respectively. $\tilde{\mu}, \tilde{\nu}$ are the transposed Young diagrams. And the product is multiplied over all the cells with the row and column labels $(i,j)$ in $\mu$ and all the cells with the row and column labels $(k,l)$ in $\nu$. When $j$ exceeds the height of $\nu$ or $k$ exceeds the height of $\tilde{\mu}$, the corresponding $\nu_j$ and $\tilde{\mu}_k$ vanish.

%%%%%%%%%%%%%%%%%%%%%%%%%%%%%%
\subsection{Bootstrap: Build up quantum 6j--symbols}
%%%%%%%%%%%%%%%%%%%%%%%%%%%%%%
\subsubsection{From non-primitives to cores}
The bootstrap method was developed in refs.~\cite{Butler:1981, Haase:1983, Searle:1988} for the computation of classical 6j--symbols in classical Lie algebras. Later it was pointed out in ref.~\cite{Lienert:1992} that the same method can be generalized to quantum groups, and examples were given for $\mathcal{U}_qsl(2)$. Here we directly apply the bootstrap method to compute quanum 6j--symbols for the quantum group $\mathcal{U}_qsu(N)$ with general $N$. 

The basic idea of bootstrap is that the symmetry properties of quantum 6j--symbols are so powerful that the values of the quantum 6j--symbols can be computed with the help of these symmetry properties from basic input data like quantum dimensions and fusion rules alone, at least for 6j--symbols with relatively small representations. In order to do so, we first classify quantum 6j--symbols to three types: trivial, primitive, and non-primitive. Trivial 6j--symbols contain at least one singlet, and their values are known, as given in \eqref{equ:trivial6jValue}.  A primitive 6j--symbol contains a fundamental or anti-fundamental representation. All rest 6j--symbols are non--primitive. 

Non-primitive 6j--symbols can always be converted to primitive 6j--symbols \cite{Butler:1981}. Given an arbitrary non-primitive 6j--symbol
\[
\begin{Bmatrix}
	\lambda_1 & \lambda_2 & \lambda_3 \\
	\mu_1 & \mu_2 & \mu_3
\end{Bmatrix}_{r_1r_2r_3r_4} \ ,
\]
we can always make $\lambda_3$ the smallest representation (with the smallest number of boxes) by using the tetrahedral symmetry. Then apply the pentagon relation, where $\nu_1$ is chosen to be either the fundamental or anti-fundamental, and $\nu_2$ one box fewer than $\lambda_3$ so that $(\bar{\nu}_1\nu_2\lambda_3)$ is a legitimate triad. Then the right hand side of the pentagon relation does not vanish identically. We have,
\[
\begin{Bmatrix}
	\lambda_1 & \lambda_2 & \lambda_3 \\
	\mu_1 & \mu_2 & \mu_3
\end{Bmatrix}_{r_1r_2r_3r_4} = \sum_{\substack{\lambda \nu_3\\r'_2r'_3r'_1}} c \;
\begin{Bmatrix}
	\lambda_1 & \nu_3 & \bar{\nu}_2 \\
	\lambda & \mu_2 & \mu_3
\end{Bmatrix}_{r_1r'_3r'_2r'_1} \ ,
\]
where in each summand $c$ is the product of some primitive 6j--symbols, while in the last 6j--symbol, the smallest representation cannot have more boxes than $\bar{\nu}_2$, which in turn has one box fewer than $\lambda_3$ in the original 6j--symbol. Repeat this step and one will end up with primitive 6j--symbols only.

Furthermore it was shown in \cite{Searle:1988} that every primitive 6j--symbol can be converted by using the pentagon relation and the generalized Racah backcoupling rule to one of the following two types:
\begin{itemize}
\item Type II (Core 6j--symbols)
\[
\begin{Bmatrix}
\lambda_1 & \lambda_2 & \lambda_3 \\
\mu_1 & \epsilon & \mu_3
\end{Bmatrix}_{0\,r_20\,r_4}	\	.
\]
Here $\epsilon$ is either fundamental or anti--fundamental. The triad $(\lambda_1\lambda_2\lambda_3r_4)$ is (partially) ordered, meaning that $p(\lambda_1)\geqslant p(\lambda_2) \geqslant p(\lambda_3)$; furthermore it satisfies $p(\lambda_3)>p(\mu_1)$ or $p(\lambda_1)>p(\mu_3)$. It implies that among the four triads in the 6j--symbol, $(\lambda_1\lambda_2\lambda_3r_4)$ is the greatest one. Two (partially) ordered triads are ordered by the following rules: $(\lambda_1\lambda_2\lambda_3)>(\mu_1\mu_2\mu_3)$ if $p(\lambda_3)>p(\mu_3)$ or when $p(\lambda_3)=p(\mu_3)$ if $p(\lambda_2)>p(\mu_2)$ or when $p(\lambda_3)=p(\mu_3)$ and $p(\lambda_2)=p(\mu_2)$ if $p(\lambda_1)>p(\mu_1)$.
\item Type IV
\[
\begin{Bmatrix}
\lambda_1 & \lambda_2 & \epsilon_1 \\
\mu_1 & \mu_2 & \epsilon_2
\end{Bmatrix}_{0000}	\	,
\]
where $\epsilon_1$ and $\epsilon_2$ are either fundamental or anti-fundamental.
\end{itemize}

The basic idea is first to count the number of primitive triads, those triads with a fundamental or anti-fundamental, in a primitive 6j--symbol. If all four triads are primitive, it is a type~IV, otherwise use the symmetry properties to arrange the six representations so that the top row triad $(\lambda_1\lambda_2\lambda_3r_4)$ in
\[
\begin{Bmatrix}
\lambda_1 & \lambda_2 & \lambda_3 \\
\mu_1 & \mu_2 & \mu_3
\end{Bmatrix}_{r_1r_2r_3r_4}	\	,
\]
is (partially) ordered and the greatest one. Then the fundamental or anti-fundamental representation $(\epsilon)$ in the primitive 6j--symbol must appear on the bottom row. If $\epsilon$ appears as $\mu_1$ and $p(\lambda_1)>p(\lambda_2)$ (type I), permute the three columns of the 6j--symbol by the cycle (132) and apply the generalized backcoupling rule. If $\epsilon$ appears as $\mu_3$ and $p(\lambda_2)>p(\lambda_3)$ (type III), apply the pentagon relation and chose $\nu_2$ to be fundamental or anti-fundamental representation. All the rest cases can be converted by simple permutation to a type~II. This recursive algorithm will terminate and in end yields only type IV and type II 6j--symbols.

%A useful trick when applying the generalized backcoupling rule is to use the $q-1/q$ symmetry of quantum 6j--symbols. Apply the backcoupling rule once and rescale everything by $q^{-(c_1(\nu')+c_1(\lambda_2)+c_1(\mu_2))/2}$, where $\nu'$ has the greatest Casimir among all possible $\nu$'s. Apply the backcoupling rule again with $q$ replaced by $1/q$, and rescale everything by $q^{(c_1(\nu')+c_1(\lambda_2)+c_1(\mu_2))/2}$. In the difference of these two formulae, not only the terms involving the most complicated $\nu$ are cancelled, all the powers of $q$ can be converted to q--deformed numbers. The computation can proceed without the explicit appearance of $q$.

\subsubsection{Crushing the cores}\label{sec:Cores}
A type IV 6j--symbol is readily soluble. Let us assume that $\epsilon_1$ and $\epsilon_2$ are conjugate to each other (If not we can permute the first two columns to make them so). If $\lambda_1$ and $\mu_1$ are identical, we apply the backcoupling rule and get
\begin{align*}
q^{C_{\lambda_1}+C_{\epsilon_1}}\begin{Bmatrix}
		\lambda_1 & \lambda_2 & \epsilon_1 \\
		\lambda_1 & \mu_2 & \epsilon_2
\end{Bmatrix}_{0000} &= \\
\sum_{\nu,r}
&q^{(C_{\nu}+C_{\lambda_2}+C_{\nu_2})/2}|\nu| \{\ldots\}
\begin{Bmatrix}
		\epsilon_2 & \nu & \epsilon_1 \\
		\lambda_1 & \mu_2 & \lambda_1
\end{Bmatrix}_{0r00}
\begin{Bmatrix}
		\lambda_1 & \lambda_2 & \epsilon_1 \\
		\bar{\epsilon}_2 & \nu & \bar{\lambda}_1
\end{Bmatrix}_{r000} \ .
\end{align*}
We indicate a bunch of phases factors by $\{\ldots\}$. Here $\nu$ can be either the singlet $[0;0]$ or the adjoint $[1;1]$. We can rescale the identity to normalize the coefficients in front of the combinations with $\nu$ being the adjoint to 1. Then we can get a second identity by replacing $q$ with $q^{-1}$, which does not change the values of the 6j--symbols and only changes coefficients in the identity. Subtracting the second identity from the first identity the combinations with $\nu$ being the adjoint drop out and we have effectively expressed the original 6j--symbol in terms of trivial 6j--symbols only. If $\lambda_1$ and $\mu_1$ are not identical, it can be shown that fixing the other five representations, the representation at position (2,1) (where $\mu_1$ is occupying) is either unique or can only be either the original $\mu_1$ or the same as $\lambda_1$. In the former case, the absolute value is solved directly as unitarity indicates
\[
	\dim_q \lambda_1 \dim_q \mu_1 \left| \begin{Bmatrix}
		\lambda_1 & \lambda_2 & \epsilon_1 \\
		\mu_1 & \mu_2 & \epsilon_2
	\end{Bmatrix} \right|^2 = 1 \ .
\]
In the latter case unitarity relates the absolute value of the 6j--symbol to the one with $\mu_1$ being replaced by $\lambda_1$, which has already been solved,
\[
	\dim_q \lambda_1 \dim_q \mu_1 \left| \begin{Bmatrix}
		\lambda_1 & \lambda_2 & \epsilon_1 \\
		\mu_1 & \mu_2 & \epsilon_2
	\end{Bmatrix} \right|^2 +
	(\dim_q \lambda_1)^2 \left| \begin{Bmatrix}
		\lambda_1 & \lambda_2 & \epsilon_1 \\
		\lambda_1 & \mu_2 & \epsilon_2
	\end{Bmatrix} \right|^2 =1 \ .
\]
As the last step you can freely choose the phase of this type IV 6j--symbol.

Core 6j--symbols are trickier. The relative easy cases are when the other five representations are fixed, the representation at position (1,3) (where $\lambda_3$ is occupying) can only be the $\lambda_3$ or some smaller representations. A simple example is
\[
\begin{Bmatrix}
	\lambda_1 & \bar{\lambda}_1 & 1;1\\
	1;0 & 1;0 & \mu_3
\end{Bmatrix}_{000r_4} \ ,
\]
where the only other alternative at position (1,3) is singlet. These core 6j--symbols are called \emph{descendable} at position (1,3). They (absolute values) can always be related by the unitarity to simpler 6j--symbols, hence are recursively soluble. Among all the descendable-at-position-(1,3) 6j--symbols which only differ by $\mu_3$, their phases can be calibrated by using the orthogonality relation, 
\begin{equation}\label{equ:PhaseCalibration}
\sum_{\lambda_3 r_4}|\lambda_3||\mu_3|
\begin{Bmatrix}
	\lambda_1 & \lambda_2 & \lambda_3\\
	\mu_1 & \epsilon & \mu_3
\end{Bmatrix}_{0 r_2 0 r_4}
\begin{Bmatrix}
	\lambda_1 & \lambda_2 & \lambda_3\\
	\mu_1 & \epsilon & \underline{\mu_3}
\end{Bmatrix}_{0 r_2 0 r_4}^*  = 0 \ ,
\end{equation}
where $\underline{\mu_3}$ is the smallest possible representation at this position. Then among all the 6j--symbols with different $\mu_3$'s, only the phase of the 6j--symbol with $\underline{\mu_3}$ is free to choose. Similar discussions can be made on the core 6j--symbols descendable at position (1,1).

However, unitarity relation or any other symmetry properties cannot separate 6j--symbols which only differ by a non--trivial multiplicity label. In order to fix their values, we need to choose a \emph{Multiplicity Separation Scheme}. First to be sure of being consistent with the 3j--phase convention specified by eqs.~\eqref{equ:3jPhases} and \eqref{equ:ColumnExchangeSimp}, we require that 6j--symbols with odd total multiplicity $\sum_{i=1}^4 r_i$ be imaginary. Then, suppose that there are $m$ copies of $\bar{\lambda}_3$ in $\lambda_1\otimes\lambda_2$ so that in the 6j--symbol
\[
\begin{Bmatrix}
\lambda_1 & \lambda_2 & \lambda_3\\
\mu_1 & \epsilon & \mu_3
\end{Bmatrix}_{0r_20r_4} \ ,
\]	
$r_4$ can take the value from $0$ to $m-1$. Also let us suppose that 
\[
\sum_{r_4=0}^{m-1}\Big| \begin{Bmatrix}
\lambda_1 & \lambda_2 & \lambda_3\\
\mu_1 & \epsilon & \mu_3
\end{Bmatrix}_{0r_20r_4} \Big|^2
\]
for various $\mu_3$, as well as
\[
\sum_{r_4=0}^{m-1}  \begin{Bmatrix}
\lambda_1 & \lambda_2 & \lambda_3\\
\mu_1 & \epsilon & \mu_3
\end{Bmatrix}_{0r_20r_4} 
\begin{Bmatrix}
\lambda_1 & \lambda_2 & \lambda_3\\
\mu_1 & \epsilon & \mu'_3
\end{Bmatrix}_{0r_20r_4}^*
\]
with $\mu_3$ being different from $\mu'_3$ have been computed. Let us call them unitarity sums and orthogonality sums respectively. We notice the fact that the values of the 6j--symbols
\[
\begin{Bmatrix}
\lambda_1 & \lambda_2 & \lambda_3\\
\mu_1 & \epsilon & \mu_3
\end{Bmatrix}_{0r_20r_4} \ , \quad r_4 = 0,\ldots, m-1
\]
are not fixed. There is a $SU(m)$ freedom which can rotate their values to a set of new values,
\begin{equation}
\prescript{}{\textrm{new}}{\begin{Bmatrix}
\lambda_1 & \lambda_2 & \lambda_3\\
\mu_1 & \epsilon & \mu_3
\end{Bmatrix}_{0r_20r'_4}}
=\sum_{r_4} M_{r'_4, r_4}
\prescript{}{\textrm{old}}{\begin{Bmatrix}
\lambda_1 & \lambda_2 & \lambda_3\\
\mu_1 & \epsilon & \mu_3
\end{Bmatrix}_{0r_20r_4}} , \quad M_{r'_4,r_4} \in SU(m) \ .
\end{equation}
Therefore one needs to specify a multiplicity separation scheme in order to fix the values of these 6j--symbols. Our choice of the multiplicity separation scheme is as follows. Choose one $\mu_3$, and set the values of the 6j--symbols with $r_4$ being from 1 to $m-1$ to 0. The absolute value of the 6j--symbol with $r_4=0$ is fixed by the unitarity sum, and we assign an arbitrary phase to this 6j--symbol. Next choose a different $\mu_3$. The value of the first 6j--symbol is now fixed by the orthogonality sum. We set the 6j--symbols with $r_4$ being from 2 to $m-1$ to 0, and the absolute value of the 6j--symbol with $r_4=1$ is then fixed by the unitarity sum. We again assign an arbitrary phase to this 6j--symbol. Repeat this procedure the values of all the 6j--symbols can be fixed, as summarized in the following table:
\begin{center}
\begin{tabular}{c|c:c:c:c}
$r_4\backslash \mu_3$ & 1 & 2 & 3 & $\cdots$\\\hline
0 & $*$ & $-$ & $-$ & \\
1 & 0 & $*$ & $-$ & \\
2 & 0 & 0 & $*$ & \\
3 & 0 & 0 & 0 & \\
$\vdots$ & $\vdots$ & $\vdots$ & $\vdots$ &\\
$m-1$ & 0 & 0 & 0 &
\end{tabular}
\quad
\begin{tabular}{c | l}
$*$ & absolute value fixed by unitarity sums\\
$-$ & value fixed by orthogonality sums
\end{tabular}
\end{center}
Note that the values of the 6j--symbols computed in this way depend on the order of $\mu_3$ in the table above. In this process, we have used $2(m-1)+1$ degrees of freedom in the first column, $2(m-2)+1$ degrees of freedom in the second column, so and so forth. At most we need to use $m^2-1$ degrees of freedom, which are precisely the degrees of freedom in the symmetry group of $SU(m)$.

Non--descendable core 6j--symbols are much more difficult to solve. It is in fact not known whether all core 6j--symbols can be solved by using symmetry properties of quantum 6j--symbols, although in practice there does not seem to be an exception, at least for core 6j--symbols with relative low representations. For some 6j--symbols, the representation at one position is unique, as in the first situation in case 2 of Type~IV core 6j--symbols, then the absolute value is solved by unitarity. For some other 6j--symbols, the representation at some position has two distinct possibilities, yet the corresponding two 6j--symbols can be shown to be linearly related. For instance, given a core 6j--symbol of the type
\[
\begin{Bmatrix}
\lambda_1 & \lambda_2 & 2;0 \\
1;0 & 0;1 & \nu
\end{Bmatrix}_{0000}  \quad \textrm{ or }\quad
\begin{Bmatrix}
\lambda_1 & \lambda_2 & 1^2;0 \\
1;0 & 0;1 & \nu
\end{Bmatrix}_{0000}\ ,
\]
it can be shown that fixing the other five representations either $\nu$ is unique, or it has only two possibilities $\nu_1,\nu_2$. In the latter case, one can prove that the generalized backcoupling rule gives a linear relation between the two 6j--symbols with $\nu_1$ and $\nu_2$. Plugging the linear relation into the unitarity sum one can solve either of the two 6j--symbols. Another trick is if $p(\lambda_2)=p(\lambda_3)$, the following two types of core 6j--symbols,
\[
\begin{Bmatrix}
\lambda_1 & \lambda_2 & \lambda_3 \\
\mu_1 & \epsilon & \mu_3
\end{Bmatrix}_{0r_20r_4}  \quad \textrm{ and }\quad
\begin{Bmatrix}
\lambda_1 & \lambda_3 & \lambda_2 \\
\mu'_1 & \epsilon'  & \mu'_3
\end{Bmatrix}_{0r'_20r'_4}\ ,
\]
can be related to each other by permuting the last two columns of either 6j--symbol and applying the pentagon relation with $\nu_2$ being fundamental or anti-fundamental.

For many non-descendable core 6j--symbols, by using the methods explained above, only their absolute values can be determined. Nonetheless not all the phases of these core 6j--symbols can be freely chosen, as they may be related via orthogonality relation as in eq.~\eqref{equ:PhaseCalibration}.  All possible such relations must be sought out in order to consistently choose phases for non-descendable core 6j--symbols. On the other hand, since our goals are the two kinds of 6j--symbols in eq.~\eqref{equ:TwoKinds}, when they are expanded in terms of core 6j--symbols, some core 6j--symbols only appear in even powers. Then the phases of these 6j--symbols need not be chosen. 

By using these tricks, we can compute all the core 6j--symbols, or their absolute values, relevant for the computation of the two kinds of 6j--symbols in eq.~\eqref{equ:TwoKinds} when $R=\ydiagram{2,1}$.

%%%%%%%%%%%%%%%%%%%%%%%%%%%%%%
\subsection{Eigenvector method to relate two kinds of 6j--symbols}
%%%%%%%%%%%%%%%%%%%%%%%%%%%%%%

\begin{figure}
\center
\subfloat[s--channel\label{fig:sChannel}]{\includegraphics[width=0.25\linewidth]{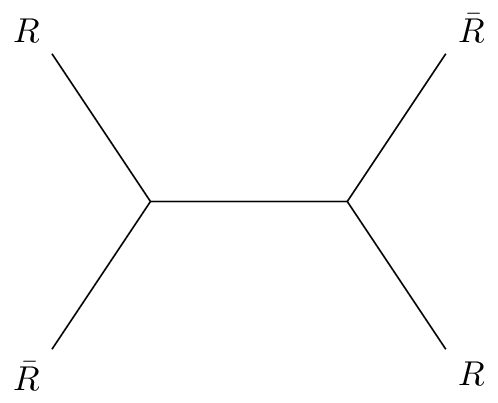}}\quad\quad
\subfloat[t--channel\label{fig:tChannel}]{\includegraphics[width=0.17\linewidth]{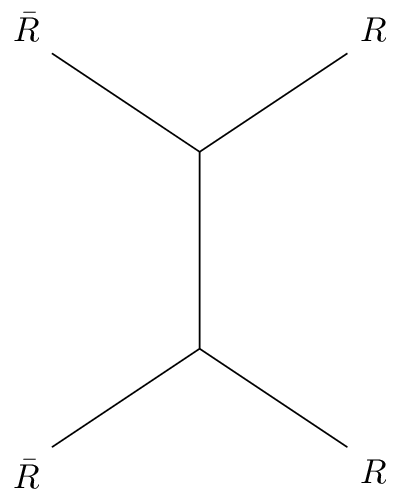}}\quad\quad
\subfloat[u--channel\label{fig:uChannel}]{\includegraphics[width=0.25\linewidth]{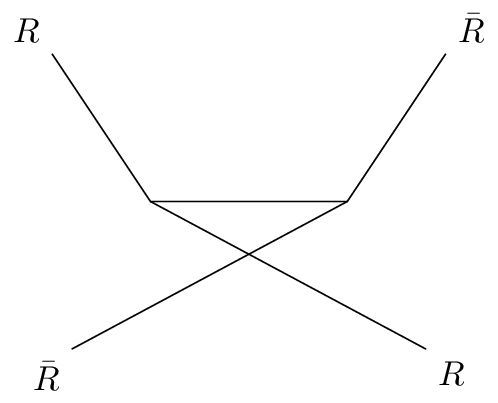}}
\caption{The three types of conformal blocks of four point correlation functions.}\label{fig:3Channels}
\end{figure}

There are three types of conformal blocks of four point correlation functions, corresponding to the s--, t--, and u--channels, respectively. When the four fields are the WZW primaries $\bar{R}, R, \bar{R}$, and $R$ as shown in Figs.~\ref{fig:3Channels}, the s-channel conformal blocks are related to the t--channel conformal blocks by the quantum 6j--symbols of the first kind as in eq.~\eqref{equ:TwoKinds}, while both the s--channel and t--channel are related to the u--channel by the quantum 6j--symbols of the second kind. Use the matrix notation $T^{\rho_1, r_3r_4}_{\rho_2,r_1r_2}$ and $U^{\rho_3,r'_3r'_4}_{\rho_4,r'_1r'_2}$ for the families of quantum 6j--symbols of the first and second kind, respectively, where the superscripts label rows and the subscripts label columns. Then the recoupling relations translate into the matrix equation
\[
		T \cdot U = U \ ,
\]
which gives rise to eigenvector equations for the column vectors of the matrix $U$. More explicitly --- by using the columns--permutation symmetry and the generalized backcoupling rule --- we find
\begin{align*}
	\begin{Bmatrix}
		\bar{R} & R & \rho_3\\
		R & R & \rho_4
	\end{Bmatrix}_{r'_1r'_2r'_3r'_4}
	=&\sum_{\nu,r,r'}(-1)^{r+r'+r'_3+r'_4}\{R\} \{R,R,\bar{\rho}_4,r'_2 \} \{R,\bar{R},\rho_3, r'_3\} \{R,\bar{R},\nu,r\} \\
	&q^{-2C_R +(C_{\rho_3}+C_{\rho_4}+C_{\nu})/2 } \dim_q \nu
	\begin{Bmatrix}
		R & \bar{R} & \rho_3\\
		R & R & \bar{\nu}
	\end{Bmatrix}_{rr'r'_3r'_4}
	\begin{Bmatrix}
		\bar{R} & R & \nu \\
		R & R & \rho_4
	\end{Bmatrix}_{r'_1r'_2rr'} \ .
\end{align*}
Now define 
\begin{align}
	T^{\rho_3,r'_3r'_4}_{\nu,rr'} = &(-1)^{r+r'+r'_3+r'_4}\{R\}  \{R,\bar{R},\rho_3, r'_3\} \{R,\bar{R},\nu,r\} q^{-2C_R +(C_{\rho_3}+C_{\rho_4}+C_{\nu})/2 } \nonumber\\
	&\sqrt{\dim_q \rho_3 \dim_q \nu} \begin{Bmatrix}
		R & \bar{R} & \rho_3\\
		R & R & \bar{\nu}
	\end{Bmatrix}_{rr'r'_3r'_4} \ , \\
	U^{\nu,rr'}_{\rho_4,r'_1r'_2} =& \sqrt{\dim_q \nu \dim_q \rho_4}
	\begin{Bmatrix}
		\bar{R} & R & \nu \\
		R & R & \rho_4
	\end{Bmatrix}_{r'_1r'_2rr'} \ ,
\end{align}
we have the eigenvector matrix equation
\begin{equation}
	U^{\rho_3,r'_3r'_4}_{\rho_4,r'_1r'_2} = \{R,R,\bar{\rho}_4,r'_2\}\; T^{\rho_3,r'_3r'_4}_{\nu,rr'} U^{\nu,rr'}_{\rho_4,r'_1r'_2} \ .
\end{equation}
Recall that the value of the 3j--phase $\{R,R,\bar{\rho}_4,r'_2\}=\pm 1$ depends on whether $\rho_4$ is in the symmetric or anti-symmetric double tensor product of $R$. So the matrix $T$ has degenerate eigenvalues $\pm 1$, and its eigenvectors are the column vectors of the matrix~$U$. 

Thus, one can use other methods to solve for the quantum 6j--symbols of the first kind, which are more symmetric and easier to solve.\footnote{Since both $\rho_3$ and $\nu$ are in the tensor product $R\otimes \bar{R}$ the $T$ matrix is more or less a symmetric matrix.} Then one uses the eigenvector equation to find relations among the quantum 6j--symbols of the second kind. In this way the number of 6j--symbols that explicitly need to be determined can be reduced by at least a factor of two. Alternatively, one can first solve for the quantum 6j--symbols of the second kind, and use the eigenvector equation to build the eigenvalue matrix $T$, so as to determine all 6j--symbols of the first kind at once.

In practice, one can use the classical version of the eigenvector equation in the limit $q\rightarrow 1$, in order to determine the classical 6j--symbols first. Then one  tries to promote the resulting classical 6j--symbols to quantum 6j--symbols. The ambiguities that arise from recovering the quantum 6j--symbols from their classical counterparts can eventually be eliminated by demanding that they satisfy unitary relation or some other symmetry properties combined with the explicit knowledge of some other quantum 6j--symbols.

%%%%%%%%%%%%%%%%%%%%%%%%%%%%%%
\subsection{Projector method for classical 6j--symbols} \label{sec:proj}
%%%%%%%%%%%%%%%%%%%%%%%%%%%%%%
As a non--trivial check for our computations of quantum 6j--symbols, we develop here another independent method to calculate classical 6j--symbols explicitly. This allows us to compare classical 6j--symbols to quantum 6j--symbols in the limit $q\to 1$. The knowledge of classical 6j--symbols provides for a useful guideline for the methods presented in the previous subsections. The technique discussed here is inspired by the projector method for $U(N)$ representations developed in refs.~\cite{MR2418111,Elvang:2003ue}.

%%%%%%%%%%%%%%%%%%%%%%%%%%%%%%%%%%%
\subsubsection{Projectors for representations of Lie groups}
Let us consider a representation $\rho$ of a Lie group $G$ on a finite dimensional vector space~$V$
\begin{equation}
     \rho:\ G \to \operatorname{GL}(V), \ g \mapsto \rho(g) \ .
\end{equation}     
Furthermore, let $\lambda$ be a subrepresentation of $\rho$ on the vector space $V_\lambda\subset V$. Then we can define a projector $P(\lambda): V \to V$ for $\lambda$ with
\begin{equation}
  P(\lambda)^2 \,=\, c_\lambda P(\lambda)  \ , \quad V_\lambda\,=\,\operatorname{Im} P(\lambda) \ ,
\end{equation}
with $c_\lambda\ne0$ such that the representation $\lambda$ is given by
\begin{equation}
  \lambda:\ G \to \operatorname{GL}(V_\lambda), \ g \mapsto \lambda(g)= \rho(g) \circ P(\lambda) \ .
\end{equation}
For non-zero projectors $P(\lambda)$ the constant $c_\lambda$ is determined by $c_\lambda=\frac{\operatorname{tr}_V P(\lambda)^2}{\operatorname{tr}_V P(\lambda)}$. In the following we mainly use the normalized projectors
\begin{equation} \label{eq:pnorm}
    \mathcal{P}(\lambda)\,=\,\frac1{c_\lambda} P(\lambda) \ , \quad \mathcal{P}(\lambda)^2= \mathcal{P}(\lambda) \ , \quad
    \operatorname{dim} \lambda \,=\, \operatorname{tr}_V\mathcal{P}(\lambda) \ .
\end{equation}
It is convenient to introduce the normalization operator $\mathcal{N}$, which maps $\mathcal{N}[\mathbf{0}]=\mathbf{0}$ and non-zero projectors to
\begin{equation}
 \mathcal{P}(\lambda) \,=\,\mathcal{N}[P(\lambda)]\,=\, \frac{\operatorname{dim} \lambda}{\operatorname{tr}_VP(\lambda)} P(\lambda)\,=\,
  \frac{\operatorname{tr}_V P(\lambda)}{\operatorname{tr}_V P(\lambda)^2} P(\lambda) \ .
\end{equation}

Projectors $\mathcal{P}: V \to V$ decompose the vector space $V$ into the direct sum $V= V_{\mathcal{P}} \oplus V^{\perp}_{\mathcal{P}}$, where $V_{\mathcal{P}}$ and $V^{\perp}_{\mathcal{P}}$ are the eigenspaces to the eigenvalues one and zero. Two projectors $\mathcal{P}_1$ and $\mathcal{P}_2$ are orthogonal, i.e., $\mathcal{P}_1 \perp \mathcal{P}_2$, if $V_1 \subset V_2^\perp$ and $V_2 \subset V_1^\perp$. It is straightforward to check that orthogonal projectors obey
\begin{equation}
  \mathcal{P}_1 \perp \mathcal{P}_2 \ \Longleftrightarrow\ \mathcal{P}_1 \mathcal{P}_2 \,=\, \mathcal{P}_2 \mathcal{P}_1 \,=\, 0 \ .
\end{equation}

We call two subrepresentations $\lambda_1$ and $\lambda_2$ orthogonal if the associated projectors $P(\lambda_1)$ and $P(\lambda_2)$ are orthogonal. Schur's lemma ensures that projectors $\mathcal{P}(\lambda_1)$ and $\mathcal{P}(\lambda_2)$ for subrepresentation $\lambda_1$ and $\lambda_2$ are automatically orthogonal if the representations $\lambda_1$ and $\lambda_2$ are irreducible and inequivalent. In general, however, two representations $\lambda_1$ and $\lambda_2$ need not be orthogonal. Nevertheless, it is straight forward to construct from a pair of projectors $\mathcal{P}(\lambda_1)$ and $\mathcal{P}(\lambda_2)$ a pair of orthogonal projectors $\mathcal{P}(\tilde\lambda_1)$ and $\mathcal{P}(\tilde\lambda_2)$ --- and hence a pair of orthogonal representations $\tilde\lambda_1$ and $\tilde\lambda_2$ --- according to
\begin{equation}
     \mathcal{P}(\tilde\lambda_1) \,=\, \mathcal{P}(\lambda_1) \ , \qquad \mathcal{P}(\tilde\lambda_2) \,=\,\mathcal{N}\left[
      \left(\mathbf{1} - \mathcal{P}(\lambda_1)\right) \mathcal{P}(\lambda_2)  \left(\mathbf{1} - \mathcal{P}(\lambda_1)\right) \right] \ ,
\end{equation}
with $V_1 \cup V_2 = \widetilde V_1 \cup \widetilde V_2$. Note that the zero projector $\mathbf{0}$ is orthogonal to any projector~$\mathcal{P}$, i.e., $\mathcal{P}\perp \mathbf{0}$. Analogously to the Gram--Schmidt process, which constructing an orthogonal basis with respect to a scalar product, this construction generalizes to more than two projectors. Namely, given a set projectors $\{ \mathcal{P}(\lambda_i) \}$ we can construct an orthogonal set of projectors $\{ \mathcal{P}(\tilde\lambda_i) \}$ such that $\bigcup_i V_i = \bigcup_i \widetilde V_i$. 

With the help of projectors, we can also describe a decomposition of the representation $\rho$ into irreducible representations 
\begin{equation}
  \rho \,=\, \bigoplus_{i,r_i} \rho_{i,r_i} \ ,
\end{equation}
where the label $i$ labels inequivalent irreducible representations, while the label $r_i$ captures their multiplicities. Again, due to Schur's lemma the representations $\rho_{i,r_i}$ and $\rho_{j,r_j}$ are automatically orthogonal for distinct $i\ne j$. Using the associated projectors $\mathcal{P}(\rho_{i,r_i})$ and the above described algorithm, we can always orthogonalize the equivalent irreducible components that appear with multiplicities, such that all projectors $\mathcal{P}(\rho_{i,r_i})$ are mutually orthogonal. The normalized projectors of such an orthogonal decomposition obey
\begin{equation}
    \sum_{i,r_i} \mathcal{P}(\rho_{i,r_i}) \,=\, \mathbf{1} \, \in \, \operatorname{GL}(V) \ .
\end{equation}

Similarly, from an orthogonal irreducible decomposition $\rho \,=\, \bigoplus_{i,r_i} \rho_{i,r_i}$, we can algorithmically construct an orthogonal decomposition of any (reducible) subrepresentation $\lambda$ of $\rho$. Firstly, we compute the projectors
$$
  \mathcal{P}(\hat\lambda_{r,r_i}) \,=\, \mathcal{N}\left[\mathcal{P}(\rho_{i,r_i}) \mathcal{P}(\lambda) \mathcal{P}(\rho_{i,r_i}) \right] \ ,
$$  
in terms of the irreducible projectors $\mathcal{P}(\rho_{i,r_i})$ of the representations $\rho_{i,r_i}$. Secondly, from the set of projectors $\{ \mathcal{P}(\hat\lambda_{i,r_i}) \}$ we compute a set of orthogonal projectors (with all zero projectors $\mathbf{0}$ removed). The obtained set of non-zero and mutual orthogonal projectors $\{ \mathcal{P}(\lambda_{j,s_j}) \}$ obeys this time
$$
  \mathcal{P}(\lambda) \,=\, \sum_{j,s_j} \mathcal{P}(\lambda_{j,s_j}) \ ,
$$  
and describes an orthogonal decomposition $\lambda \,=\, \bigoplus_{j,s_j} \lambda_{j,s_j}$.

Finally, we note that projectors are a powerful means to study tensor products. Namely, for two projectors $\mathcal{P}(\lambda_1)$ and $\mathcal{P}(\lambda_2)$ of two (not necessarily irreducible) subrepresentations $\lambda_1$ and $\lambda_2$ of $\rho_1$ and $\rho_2$, respectively, the tensor product $\lambda_1\otimes\lambda_2$ arises as a subrepresentation of $\rho_1\otimes\rho_2$ and its projector is simply given by
\begin{equation}
  \mathcal{P}(\lambda_1\otimes\lambda_2) \,=\, \mathcal{P}(\lambda_1) \otimes \mathcal{P}(\lambda_2) \ .
\end{equation}  
Using the above described algorithm, it is straight forward to construct an orthogonal irreducible decomposition for the tensor product $\lambda_1\otimes\lambda_2=\bigoplus_{j,s_j}(\lambda_1\otimes\lambda_2)_{j,s_j}$.

So far we have only reformulated various concepts in representation theory of Lie groups --- such as decompositions in irreducible representations or tensor products of representations --- in terms of projectors onto subrepresentations of reducible representations. In practice this is beneficial, if the discussed projectors furnish a convenient and applicable realization for the representations of interest. In this note, we will see that projectors provide for a powerful tool to study (for general $N$) composite representations of $SU(N)$ as subrepresentations of $\rho=\ydiagram{1}^{\otimes k} \otimes \bar{\ydiagram{1}}^{\otimes \ell}$. 

%%%%%%%%%%%%%%%%%%%%%%%%%%%%%%%%%%%
\subsubsection{Classical 6j--symbols from projectors}
We have assembled all the necessary ingredients to compute classical 6j--symbols from projectors. We recall that the 6j--symbols can be interpreted as the normalized recoupling coefficients arising from recoupling of representations in the trilinear tensor product $\lambda_1 \otimes \lambda_2 \otimes \lambda_3$ of the irreducible representations $\lambda_\ell$, $\ell=1,2,3$. To spell out the classical 6j--symbols from projectors, we first need to construct the list of projectors described below:
\begin{enumerate}
\item Choose three convenient representations $\rho_\ell$, $\ell=1,2,3$, on the vector spaces $V^{(\ell)}$, which contain $\lambda_\ell$ as subrepresentation of $\rho_\ell$. Then we describe the representations $\lambda_\ell$ with the projectors $\mathcal{P}(\lambda_\ell): V^{(\ell)} \to V^{(\ell)}$. 
\item  Determine orthogonal decompositions $\lambda_{12}=\bigoplus_{i,r_i} (\lambda_{12})_{i,r_i}$ and $\lambda_{23}=\bigoplus_{j,s_j} (\lambda_{23})_{j,s_j}$ of the tensor products $\lambda_{12} = \lambda_1 \otimes \lambda_2$ and $\lambda_{23} = \lambda_2 \otimes \lambda_3$ in terms of their projectors $\mathcal{P}((\lambda_{12})_{i,r_i}): V^{(1)} \otimes V^{(2)} \to V^{(1)} \otimes V^{(2)} $  and $\mathcal{P}((\lambda_{23})_{j,s_j}):V^{(2)} \otimes V^{(3)} \to V^{(2)} \otimes V^{(3)}$.
\item Determine orthogonal decompositions $\lambda(i,r_i) = \bigoplus_{\alpha,t_\alpha} \lambda(i,r_i)_{\alpha,t_\alpha}$ and $\lambda(j,s_j) = \bigoplus_{\beta,u_\beta}\lambda(j,s_j)_{\beta,u_\beta}$ of the tensor products $\lambda(i,r_i)=  (\lambda_{12})_{i,r_i} \otimes \lambda_3$ and $\lambda(j,s_j)= \lambda_3 \otimes (\lambda_{23})_{j,s_j}$ given in terms of the projectors $\mathcal{P}(\lambda(i,r_i)_{\alpha,t_\alpha})$ and $\mathcal{P}(\lambda(j,s_j)_{\beta,u_\beta})$. 
\end{enumerate}
Now we are ready to spell out the classical 6j--symbols in terms of projectors. First note that in the state notation of quantum mechanics a projector $\mathcal{P}(\lambda)$ of a representation $\lambda$ takes the simple form
\begin{equation}
  \mathcal{P}(\lambda) \,=\, \sum_{m=1}^{\dim\lambda} \left| \lambda,m \middle\rangle \middle\langle \lambda,m \right| \ ,
\end{equation}  
where $m$ labels the states within the representation $\lambda$. Then we can write the square of the matrix element~\eqref{eq:recoup} --- identifying the labels $(i,r_i,t_\alpha)$ with $(\lambda_{12},r_{12},r)$ and the labels $(j,s_j,u_\beta)$ with $(\lambda_{23},r_{23},r')$ --- in terms of projectors
\begin{align}
    &\left| \left\langle \lambda(i,r_i)_{\alpha,t_\alpha} \middle| \lambda(j,s_j)_{\beta,u_\beta} \right\rangle \right|^2  \nonumber \\
    &\quad\,=\, \frac{1}{\dim \lambda_\alpha}
    \sum_{\psi,m} \left\langle \psi \middle| \lambda(i,r_i)_{\alpha,t_\alpha},m \middle\rangle \middle\langle  \lambda(i,r_i)_{\alpha,t_\alpha},m  \middle |
    \lambda(j,s_j)_{\beta,u_\beta},m \middle\rangle \middle\langle  \lambda(j,s_j)_{\beta,u_\beta},m  \middle |
    \psi\right\rangle \nonumber \\
    &\quad\,=\,\frac{1}{\dim \lambda_\alpha}\operatorname{tr}_{V^{(1)}\otimes V^{(2)}\otimes V^{(3)}}\mathcal{P}(\lambda(i,r_i)_{\alpha,t_\alpha})\mathcal{P}(\lambda(j,s_j)_{\beta,u_\beta}) \ . \label{eq:square}
\end{align}   
Here $\{ \left| \psi \right\rangle \}$ is complete set of states, i.e., $\sum \left| \psi \middle\rangle \middle\langle \psi \right|$ is the identity on $V^{(1)}\otimes V^{(2)}\otimes V^{(3)}$. Again with the above identification of labels, we insert this matrix element into the definition~\eqref{equ:Recoupling6j} to arrive at the classical 6j--symbol
\begin{equation} \label{eq:cl6j}
\begin{aligned}
   \begin{Bmatrix}
		\lambda_1 & \lambda_2 & (\bar\lambda_{12})_i \\
		\lambda_3 & \lambda_\alpha & (\lambda_{23})_j
   \end{Bmatrix}_{u_\alpha,s_j,t_\alpha,r_i}
   \,&=\, e^{i \varphi}\sqrt{\frac{\operatorname{tr}\mathcal{P}(\lambda(i,r_i)_{\alpha,t_\alpha})\mathcal{P}(\lambda(j,s_j)_{\alpha,u_\alpha})}
   {\dim(\lambda_{12})_i \cdot \dim (\lambda_{23})_j \cdot \dim \lambda_\alpha}}\ . 
\end{aligned}   
\end{equation}   
There is some freedom in choosing the phases $e^{i\varphi}$, which arise here from taking the square root of eq.~\eqref{eq:square}. For a consistent choice of phases the matrix elements $\left\langle \lambda(i,r_i)_{\alpha,t_\alpha} \middle| \lambda(j,s_j)_{\beta,u_\beta} \right\rangle$ must form a unitary matrix; c.f., the detailed discussion in Sec.~\ref{sec:3jPhases}. 

Note that the projector expression for 6j--symbol given in refs.~\cite{MR2418111,Elvang:2003ue} in the context of the Lie Group $U(N)$ is in agreement with our findings. The expression derived here, however, is more general as it is also applicable for sectors with non--trivial multiplicities. However, we should stress that --- compared to the choice of bases of states made in Sect.~\ref{sec:recoupling} --- the constructed orthogonal decompositions of tensor products generically give rise to a different multiplicity separation scheme in sectors with non--trivial multiplicities.

%%%%%%%%%%%%%%%%%%%%%%%%%%%%%%%%%%%%%%%%%%%
\subsubsection{Projectors for finite irreducible representations of $SU(N)$}
Now we specialize to the construction of projectors for finite irreducible representations of $SU(N)$ (for general $N$). Our method builds upon refs.~\cite{MR2418111,Elvang:2003ue}, where projectors are constructed for representations of $U(N)$. We generalize this construction to representations of $SU(N)$ by constructing projectors for composite representations discussed in Sec.~\ref{sec:compos}.

The basic building blocks for the projector method are the fundamental and anti--fundamental representations ${\ydiagram{1}}$ and ${\overline{\ydiagram{1}}}$ of $SU(N)$, which canonically induce the (reducible) tensor product representations
\begin{equation}
\begin{aligned}
  &{\ydiagram{1}^{\otimes k} \,\otimes\, \overline{\ydiagram{1}}^{\otimes \ell}} : \ SU(N) \times \mathbb{C}^{(k+\ell)N}  \to   \mathbb{C}^{(k+\ell)N}  \ , \\
  &\ \left( M, v_1 \otimes \ldots \otimes v_k \otimes \overline{v}_1 \otimes \ldots \otimes \overline{v}_\ell \right)
  \mapsto  M v_1 \otimes \ldots \otimes M v_k \otimes M^\dagger \overline{v}_1 \otimes \ldots \otimes M^\dagger \overline{v}_\ell \ .
\end{aligned}  
\end{equation}
It decomposes into irreducible representations, which we label by composite representations. These irreducible representations  can be worked out explicitly and algorithmically by applying the Littlewood--Richardson rule in two steps. Firstly, we compute the decomposition 
\begin{equation} \label{eq:decomp1}
   \ydiagram{1}^{\otimes k} \,=\, \bigoplus_r (\rho_r;0) \ , \qquad  \overline{\ydiagram{1}}^{\otimes \ell} \,=\, \bigoplus_s (0;\sigma_s) \ .
\end{equation}
Secondly, using the formula~ \eqref{eq:tprod} for the tensor product of two composite representations, we arrive at
\begin{equation}
   (\rho_r;0) \otimes (0;\sigma_s)  \,=\, \bigoplus_{\zeta_{rs}} \left( \rho_r/\zeta_{rs} ; \sigma_s/\zeta_{rs} \right) \ ,
\end{equation}
and altogether we obtain the decomposition into irreducible representations
\begin{equation} \label{eq:decomp}
  {\ydiagram{1}^{\otimes k} \,\otimes\, \overline{\ydiagram{1}}^{\otimes \ell}} \,=\, \bigoplus_i \lambda_i \,=\,\bigoplus_{r,s}\bigoplus_{\zeta_{rs}} \left( \rho_r/\zeta_{rs} ; \sigma_s/\zeta_{rs} \right) \ .
\end{equation}
Note that this decomposition implies a decomposition of the vector space $\mathbb{C}^{(k+\ell)N}$ into subvector spaces $V_i$
$$
  \mathbb{C}^{(k+\ell)N} \,=\, \bigoplus_i V_i \ ,
$$
on which the representation $\lambda_i$ acts irreducibly with $\dim \lambda_i = \dim V_i$. It is the projectors $\mathcal{P}(\lambda_i)$ on these subspaces $V_i$, which we wish to determine.

To determine projectors onto subrepresentations, we represent the vector space associated to the representation ${\ydiagram{1}^{\otimes k} \,\otimes\, \overline{\ydiagram{1}}^{\otimes \ell}}$ in terms of the tensor $R^{i_1\ldots i_k}_{\bar\imath_1\ldots\bar\imath_\ell}$, where the superscript indices and the subscript indices transform in the fundamental and anti--fundamental representation. Then we define symmetrization and anti--symmetrization operators $\mathcal{S}^{(i_{s_1} \ldots i_{s_n})}$ and $\mathcal{A}^{[i_{s_1} \ldots i_{s_n}]}$ for fundamental indices, i.e.,
\begin{equation}
\begin{aligned}
   \mathcal{S}^{(i_{s_1} \ldots i_{s_n})}( R^{i_1\ldots i_k}_{\bar\imath_1\ldots\bar\imath_\ell} ) \,&=\, 
     \frac1{n!} \sum_{\tau \in S_{(i_{s_1} \ldots i_{s_n})}} R^{i_{\tau(1)}\ldots i_{\tau(k)}}_{\bar\imath_1\ldots\bar\imath_\ell} \ ,\\
   \mathcal{A}^{[i_{s_1} \ldots i_{s_n}]}( R^{i_1\ldots i_k}_{\bar\imath_1\ldots\bar\imath_\ell} ) \,&=\, 
     \frac1{n!} \sum_{\tau \in S_{(i_{s_1} \ldots i_{s_n})}} (-1)^{\operatorname{sign}\tau}R^{i_{\tau(1)},\ldots,i_{\tau(k)}}_{\bar\imath_1\ldots\bar\imath_\ell} \ ,
\end{aligned}
\end{equation}   
where $S_{(i_{s_1} \ldots i_{s_n})}$ is the permutation group of $n!$ elements that permutes only the indices $i_{s_1},\ldots,i_{s_n}$ and $\operatorname{sign}\tau$ is signum of the permutation $\tau$. Analogously, we can also define the symmetrization and anti--symmetrization operators $S_{(\bar\imath_{s_1} \ldots \bar\imath_{s_n})}$ and $A_{[\bar\imath_{s_1} \ldots \bar\imath_{s_n}]}$ for anti-fundamental indices. Finally, we define the trace operator $\mathcal{T}^{i_{s}}_{\bar\imath_t}$ 
\begin{equation}
   \mathcal{T}^{i_{s}}_{\bar\imath_t}( R^{i_1 \ldots i_k}_{\bar\imath_1\ldots \bar\imath_\ell} ) \,=\, \delta^{i_s}_{\bar\imath_t} \sum_{j=1}^N 
   R^{i_1 \ldots i_{s-1} j i_{s+1} \ldots i_k}_{\bar\imath_1 \ldots \bar\imath_{t-1} j \bar\imath_{t+1} \ldots \bar\imath_k} \ .
\end{equation}   

The operators $\mathcal{S}^{(i_{s_1} \ldots i_{s_n})}$, $\mathcal{A}^{[i_{s_1} \ldots i_{s_n}]}$, $\mathcal{S}_{(\bar\imath_{s_1} \ldots \bar\imath_{s_n})}$, $\mathcal{A}_{[\bar\imath_{s_1} \ldots \bar\imath_{s_n}]}$, and $\mathcal{T}^{i_{s}}_{\bar\imath_t}$ commute with the group action of $SU(N)$, and as we sketch below, they are suitable to construct all irreducible subrepresentations of ${\ydiagram{1}^{\otimes k} \,\otimes\, \overline{\ydiagram{1}}^{\otimes \ell}}$ in an algorithmic way. First, we realize the decompositions \eqref{eq:decomp1}. The Littlewood--Richardson rule decomposes the tensor product  $\ydiagram{1}^{\otimes k}$ into Young tableaus with the boxes labelled by integers $1$ through $k$, e.g.,
\begin{equation}
  \ydiagram{1}^{\otimes 3} \,=\, 
    \ystdnum
      \begin{ytableau} 1&2&3 \end{ytableau} \, \oplus \,
      \begin{ytableau} 1&2\\3 \end{ytableau} \, \oplus \,
      \begin{ytableau} 1&3\\2 \end{ytableau} \, \oplus \,
      \begin{ytableau} 1\\2\\3 \end{ytableau}
    \ystd \ .
\end{equation}
We can then construct a projector for each Young tableau by first symmetrizing with respect to the indices assigned to the individual rows and then anti--symmetrizing with respect to all indices in the columns \cite{MR2418111,Elvang:2003ue}. For instance, for the above example we get the projectors 
\begin{equation} \label{eq:dec}
\begin{aligned}
  \ystdnum
  &\mathcal{P}\big( \, \begin{ytableau} 1&2&3 \end{ytableau}  \, \big) ( R^{i_1i_2i_3} ) \,=\, \mathcal{N}[\mathcal{S}^{(i_1i_2i_3)}] ( R^{i_1i_2i_3}) \ , \\
  &\mathcal{P}\big( \, \begin{ytableau} 1&2\\3  \end{ytableau} \, \big) ( R^{i_1i_2i_3} ) \,=\, \mathcal{N}[\mathcal{A}^{[i_1i_3]} \circ \mathcal{S}^{(i_1i_2)}] ( R^{i_1i_2i_3}) \ , \\
  &\mathcal{P}\big( \, \begin{ytableau} 1&3\\2 \end{ytableau} \, \big) ( R^{i_1i_2i_3} ) \,=\, \mathcal{N}[\mathcal{A}^{[i_1i_2]} \circ \mathcal{S}^{(i_1i_3)}] ( R^{i_1i_2i_3}) \ , \\
  &\mathcal{P}\big( \, \begin{ytableau} 1\\2\\3 \end{ytableau} \, \big) ( R^{i_1i_2i_3} ) \,=\, \mathcal{N}[\mathcal{A}^{[i_1i_2i_3]}] ( R^{i_1i_2i_3}) \ .
  \ystd
\end{aligned}
\end{equation}
In this basic example, we get already two distinct projectors for the representation~$\ydiagram{2,1}$, reflecting the multiplicity two for this representation in the discussed tensor product. We should point out that the presented construction of projectors does generically not yield orthogonal projectors. Thus, in order to arrive at an orthogonal decomposition, we must employ the algorithm discussed in the previous subsection. Recall that such an orthogonal decomposition is not unique. For instance, employing the orthogonalization algorithm of the previous subsection, the decomposition depends on the chosen order of projectors. Different choices give rise to distinct multiplicity separation schemes. 

In the same fashion we can construct projectors for Young tableaus with respect to anti--fundamental indices. Simultaneously applying projections for fundamental and anti--fundamental indices, we obtain projectors for the (reducible) subrepresentations $(\rho_r; 0) \otimes (0; \sigma_r)$ of ${\ydiagram{1}^{\otimes k} \,\otimes\, \overline{\ydiagram{1}}^{\otimes \ell}}$. In order to further decompose them into irreducible representations, we observe that taking traces with respect to pairs of fundamental and anti--fundamental indices project onto yet smaller representations. As a consequence, we can arrive at irreducible representations and ultimately at an irreducible decomposition, if we consecutively remove traces from the tensors associated to the representation $(\rho_r; 0) \otimes (0; \sigma_r)$. The systematic removal of traces is again governed by the Littlewood--Richardson rule applied to the quotient of Young tableaus appearing in eq.~\eqref{eq:decomp}. For instance, the decomposition
$$
    \ystdnum
      \left(\,\begin{ytableau} 1&2\\3 \end{ytableau}\,;0\right) 
      \otimes \left(  0 ; \begin{ytableau} \bar 1 \end{ytableau}\, \right) \,=\,    
      \left(\,\begin{ytableau} 1&2 \end{ytableau}\,;0\right) \, \oplus \,
       \left(\,\begin{ytableau} 1\\3 \end{ytableau}\,;0\right) \, \oplus \,
        \left(\,\begin{ytableau} 1&2\\3 \end{ytableau}\,;\begin{ytableau} \bar 1 \end{ytableau}\,\right)
     \ystd \ .
$$
is given by the projectors
\begin{equation} \label{eq:dectr}
\begin{aligned}
  \ystdnum
  &\mathcal{P}\big( \,\begin{ytableau} 1&2 \end{ytableau}\,;0 \, \big) ( R^{i_1i_2i_3}_{\bar\imath_1} ) 
    \,=\, \mathcal{N}[ \mathcal{T}^{i_3}_{\bar\imath_1}\circ \mathcal{A}^{[i_1i_3]} \circ S^{(i_1i_2)}] ( R^{i_1i_2i_3}_{\bar\imath_1} ) \ , \\
  &\mathcal{P}\big(\,\begin{ytableau} 1\\3 \end{ytableau}\,;0 \big) ( R^{i_1i_2i_3}_{\bar\imath_1} ) 
    \,=\, \mathcal{N}[\mathcal{T}^{i_2}_{\bar\imath_1}\circ \mathcal{A}^{[i_1i_3]} \circ \mathcal{S}^{(i_1i_2)}] ( R^{i_1i_2i_3}_{\bar\imath_1}) \ , \\
  &\mathcal{P}\big(\,\begin{ytableau} 1&2\\3 \end{ytableau}\,;\begin{ytableau} \bar 1 \end{ytableau}\,\big) ( R^{i_1i_2i_3}_{\bar\imath_1} ) 
    \,=\, \mathcal{N}[\mathcal{A}^{[i_1i_3]} \circ \mathcal{S}^{(i_1i_2)}] ( R^{i_1i_2i_3}_{\bar\imath_1} ) \\
   &\qquad\qquad\qquad\qquad\qquad\qquad
    - \mathcal{P}\big( \,\begin{ytableau} 1&2 \end{ytableau}\,;0 \, \big) ( R^{i_1i_2i_3}_{\bar\imath_1} ) 
    - \mathcal{P}\big(\,\begin{ytableau} 1\\3 \end{ytableau}\,;0 \big) ( R^{i_1i_2i_3}_{\bar\imath_1} )
  \ystd \ .
\end{aligned}
\end{equation}
This basic example demonstrates the algorithm to arrive in general at the irreducible decomposition \eqref{eq:decomp} in terms of projectors. As before, to arrive at an orthogonal decomposition, it is necessary to orthogonalize projectors in non--trivial multiplicity sectors.

\begin{figure}[t]
\center
\includegraphics[width=0.3\linewidth]{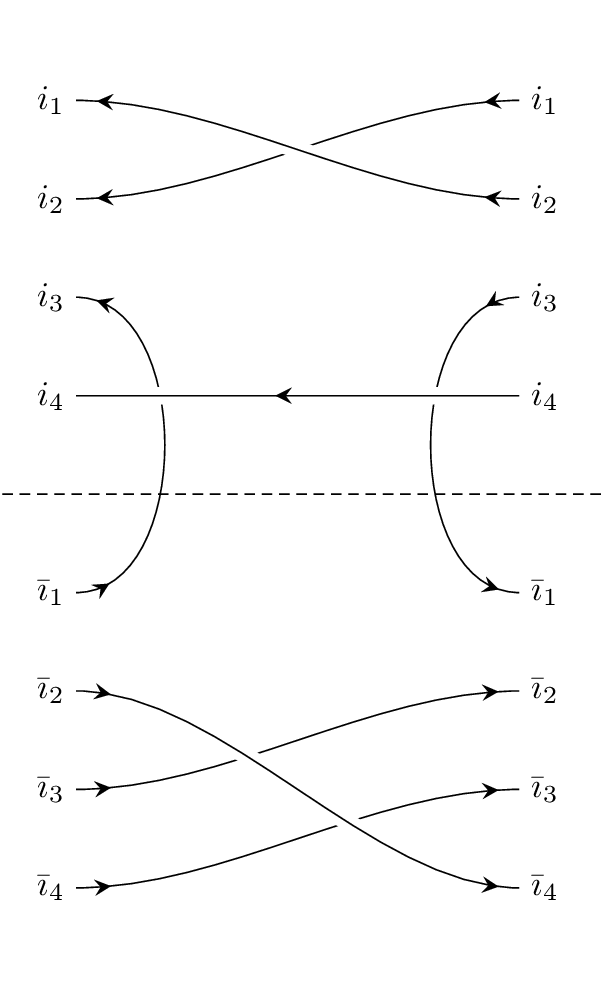}
\caption{Shown is the oriented string graph $\mathcal{G}_\tau$ assigned to permutation element $\tau \in S_{(i_1\ldots i_4 \bar\imath_1\ldots \bar\imath_4)}$, which is given in terms of the cycle representation $\tau=\{i_1i_2\}\{i_3\bar\imath_1\}\{\bar\imath_2\bar\imath_4\bar\imath_3\}$.}\label{fig:Projector1}
\end{figure}

To calculate the projectors in practice --- that is to say to realize the products of the various operators appearing in eqs.~\eqref{eq:dec} and \eqref{eq:dectr} --- it is convenient to adopt a uniform graphical representation of the trace operators and of the summands appearing the (anti--)symmetrization operators, which we call \emph{oriented string graphs} $\mathcal{G}$. All oriented string graphs are in one--to--one correspondence with permutation elements of the symmetric group $S_{(i_1\ldots i_k \bar\imath_1 \ldots \bar\imath_\ell)}$. For each permutation element $\tau \in S_{(i_1\ldots i_k \bar\imath_1 \ldots \bar\imath_\ell)}$ an oriented string graph $\mathcal{G}_\tau$ is drawn by connecting two columns of fundamental and anti--fundamental indices by oriented open paths in the following way: Each fundamental index in the right column and each anti--fundamental index in the left column is a source for an oriented open path, while each fundamental in the left column and each anti--fundamental index in the right column is a drain for an oriented open path. Sources and drains are connected according to the permutation group element $\tau$. An illustrative oriented string graphs is depicted in Fig.~\ref{fig:Projector1}. 

In this way we can uniformly represent the operators $\mathcal{S}$ and $\mathcal{A}$ as a formal sum of oriented string graphs
$$
\begin{aligned}
    \mathcal{S}^{(i_{s_1} \ldots i_{s_n})} \,&=\, \frac1{n!} \sum_{\tau \in S_{(i_{s_1} \ldots i_{s_n})}} \mathcal{G}_\tau \ , 
    &\mathcal{S}_{(\bar\imath_{s_1} \ldots \bar\imath_{s_n})} \,&=\, \frac1{n!} \sum_{\tau \in S_{(\bar\imath_{s_1} \ldots \bar\imath_{s_n})}} \mathcal{G}_\tau \ , \\
    \mathcal{A}^{[i_{s_1} \ldots i_{s_n}]} \,&=\, \frac1{n!} \sum_{\tau \in S_{(i_{s_1} \ldots i_{s_n})}} (-1)^{\operatorname{sign}\tau}\mathcal{G}_\tau \ , 
    &\mathcal{A}_{[\bar\imath_{s_1} \ldots \bar\imath_{s_n}]} \,&=\, \frac1{n!} \sum_{\tau \in S_{(\bar\imath_{s_1} \ldots \bar\imath_{s_n})}}(-1)^{\operatorname{sign}\tau} \mathcal{G}_\tau \ ,
\end{aligned}
$$
and $\mathcal{T}$ as
$$
   \mathcal{T}^{i_s}_{\bar\imath_t} \,=\, \mathcal{G}_{\{i_s \bar\imath_t \}} \ .
$$
Here $\{i_s \bar\imath_t \}$ is the permutation element that permutes the indices $i_s$ and $\bar\imath_t$ given in the cycle representation.

\begin{figure}[t]
\center
\includegraphics[width=0.9\linewidth]{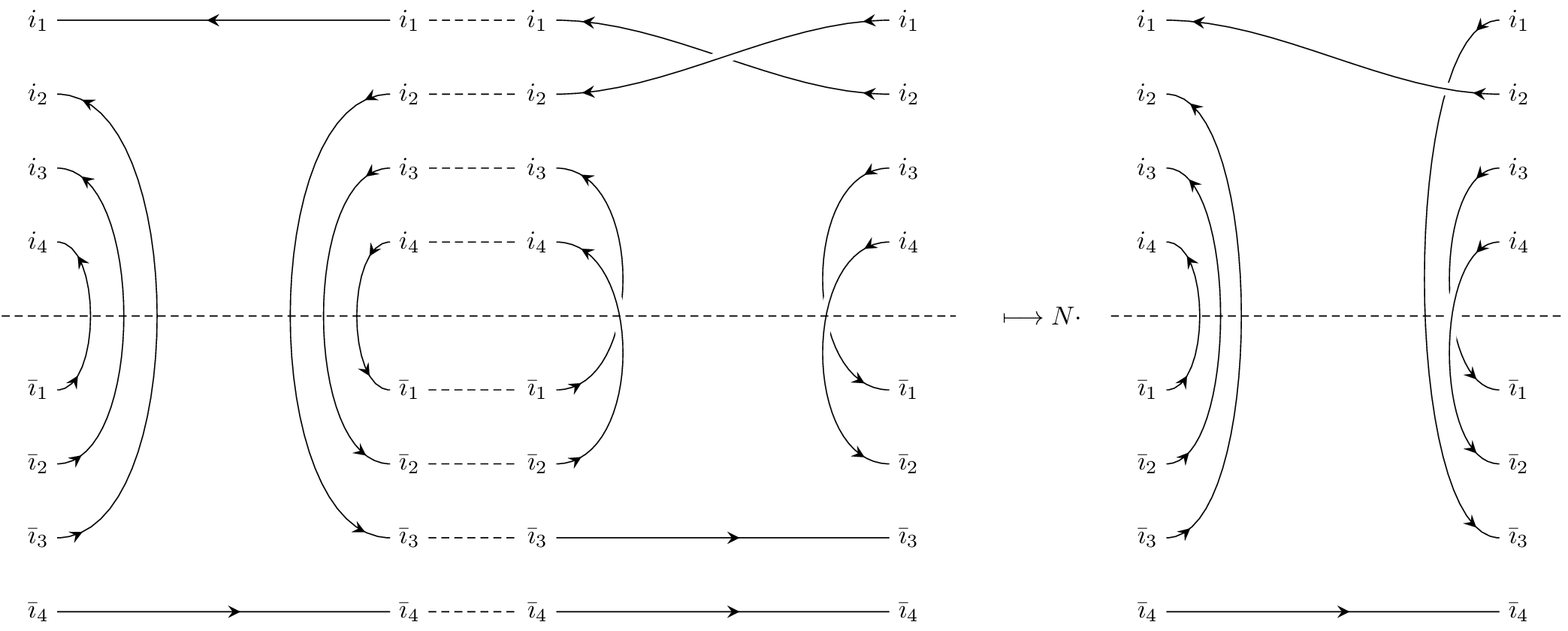}
\caption{Shown is the product $\mathcal{G}_{\tau_1} \circ \mathcal{G}_{\tau_2} = N \,\mathcal{G}_{\tau_3}$ of the oriented string graphs $\mathcal{G}_{\tau_1}$ and $\mathcal{G}_{\tau_2}$ of  $\tau_1,\tau_2 \in S_{(i_1\ldots i_4 \bar\imath_1\ldots \bar\imath_4)}$. The permutation elements read in the cycle representation $\tau_1=\{i_2\bar\imath_3\}\{i_3\bar\imath_2\}\{i_4\bar\imath_1\}$ and $\tau_2=\{i_1i_2\}\{i_3\bar\imath_1\}\{i_4\bar\imath_2\}$, whereas the resulting permutation element becomes $\tau_3=\{i_1\bar\imath_3i_2\}\{i_3\bar\imath_1i_4\bar\imath_2\}$.}\label{fig:Projector2a3}
\end{figure}

In order to construct projectors as in eqs.~\eqref{eq:dec} and \eqref{eq:dectr} we need to define the product of two oriented string graphs
\begin{equation}
  \mathcal{G}_{\tau_1} \circ \mathcal{G}_{\tau_2} \mapsto N^{\#(\textrm{loops})} \mathcal{G}_{\tau_3} \ .
\end{equation}
As illustrated in Fig.~\ref{fig:Projector2a3} we connect horizontally the graphs $\mathcal{G}_{\tau_1}$ and $\mathcal{G}_{\tau_2}$ and obtain the result of the product by first enumerating the number $\#(\textrm{loops})$ of closed oriented paths, which yields the numerical prefactor $N^{\#(\textrm{loops})}$. The permutation element $\tau_3$ is obtained from the oriented open paths that connect sources and drains of the resulting graph. Note that for permutation group elements $\tau_1$ and $\tau_2$ that permute only the fundamental and anti--fundamental indices among themselves, there appear never any closed loops and the resulting permutation element $\tau_3$ arises simply from the permutation product $\tau_3 = \tau_1 \circ \tau_2$. 

Finally, we define the trace of oriented string graphs $\operatorname{tr} \mathcal{G}_{\tau}$, which for instance is required for computing dimensions of representations according to eq.~\eqref{eq:pnorm} and which appear in the formula~\eqref{eq:cl6j} for the classical 6j--symbol. The trace can graphically be evaluated by connecting the sources and drains of a oriented string graph in accordance with their indices. Then the number of closed paths raised to the power $N$ yields the trace. It is not difficult to see that alternatively we arrive at the trace by the simple formula 
\begin{equation}
     \operatorname{tr} \mathcal{G}_\tau \,=\, N^{\#\text{cycles}(\tau)} \ ,
\end{equation}     
where $\#\text{cycles}(\tau)$ enumerates the number of cycles --- including the trivial one--cycles --- in the cycle representation of the permutation element $\tau$.  The defined trace is now extended to a formal sum of oriented string graphs in the obvious way, namely
$$
  \operatorname{tr} \left( \sum_\tau c_\tau \mathcal{G}_\tau \right) \,=\, \sum_\tau c_\tau \operatorname{tr} \mathcal{G}_\tau \ .
$$  

Now we have assembled all the ingredients to successfully employ the outlined projector method to compute with eq.~\eqref{eq:cl6j} classical 6j--symbols of $SU(N)$. This provides for us a non--trivial check on the $\mathcal{U}_qsu(N)$ quantum 6j--symbols in the limit $q\to 1$. We hope that the presented projector method techniques prove useful in other contexts as well, for instance, in determining color factors for $SU(N)$ Yang Mills amplitudes.

%%%%%%%%%%%%%%%%%%%%%%%%%%%%%%
\section{Results for two--bridge hyperbolic knots} \label{sec:knotinv}
%%%%%%%%%%%%%%%%%%%%%%%%%%%%%%
\subsection{The quantum 6j--symbols}
\definecolor{light-gray}{gray}{0.8}

Using the first three symmetry properties of quantum 6j--symbols: permutation of columns, exchange of rows, and complex conjugation, one can show that many quantum 6j--symbols of the first kind are identical. Thus, the number of independent 6j--symbols that we need to compute is actually greatly reduced. For instance, for the quantum 6j--symbols of the first kind
\[
	\hat{T}^{\rho_i,r_3r_4}_{\rho_j,r_1r_2} = 
	\begin{Bmatrix}
		R & \bar{R} & \rho_i\\
		R & R & \rho_j
	\end{Bmatrix}_{r_1r_2r_3r_4}	\ ,
\]
the matrix $\hat{T}$ exhibits the following symmetries
\begin{align}\label{equ:SymmetricTHat}
	\hat{T}^{\rho_i,r_3r_4}_{\rho_j,r_1r_2} = \hat{T}^{\bar{\rho}_i,r_4r_3}_{\bar{\rho}_j,r_2r_1} = \hat{T}^{\bar{\rho}_i,r_3r_4}_{\rho_j,r_2r_1} = \left(\hat{T}^{\bar{\rho}_j,r_1r_2}_{\bar{\rho}_i,r_3r_4} \right)^*  \ .
\end{align}
For our case of interest, i.e., $R=\ydiagram{2,1}$, we need to consider 
\begin{align*}
	\rho_i,\rho_j \in &(21;0)\otimes (0;21) \\
	&= (0;0)\oplus 2(1;1)\oplus (2;2) \oplus (2;1^2) \oplus (1^2;2) \oplus (1^2;1^2) \oplus (21;21) \ .
\end{align*}
which implies that $\hat{T}$ is a $10\times 10$ matrix. The symmetries reduce the number of independent entries from 100 to 37 to be listed below. The stated values for the quantum 6j--symbols with non--trivial multiplicity labels depend on the used multiplicity separation scheme.  Note, however, that physical quantities --- such as the colored HOMFLY invariants --- do \emph{not} depend on a choice of multiplicity separation scheme. In fact one can choose the most convenient multiplicity separation schemes so that the quantum 6j--symbols with non--trivial multiplicity labels are as simple as possible.

\bigskip
Divided into five blocks, we explicitly spell out the results for the quantum 6j--symbols of the first kind for $R = \ydiagram{2,1}$ :

\vskip4ex
\noindent\underline{Trivial quantum 6j--symbols:}
\vskip1ex

\noindent\resizebox{\linewidth}{!}{
\begin{tabular}{c|ccccc}
\raisebox{-0.5ex}[1ex][1.8ex]{$(\rho_j)_{r_1r_2}$}$\backslash$ \raisebox{0.5ex}[3ex][1ex]{$(\rho_i)_{r_3r_4}$} & $ (1;1)_{00} $ & $ (1;1)_{01} $ & $ (1;1)_{10} $ & $(1;1)_{11}$ & $ (0;0) $ \\ \hline
\raisebox{0ex}[3ex][2ex]{$(0;0)$}  & $\frac{[3]}{[N-1] [N] [N+1]}$ & 0 & \cellcolor{light-gray} 0 & $-\frac{[3]}{[N-1] [N] [N+1]}$ & $\frac{[3]}{[N-1] [N] [N+1]}$  \\ \hline\hline
\raisebox{-0.5ex}[1ex][1.8ex]{$(\rho_j)_{r_1r_2}$}$\backslash$ \raisebox{0.5ex}[3ex][1ex]{$(\rho_i)_{r_3r_4}$}  & $ (2;2) $ & $ (2;1^2) $ & $ (1^2;2) $ & $(1^2;1^2)$ & $(21;21)$\\ \hline
\raisebox{0ex}[3ex][2ex]{$(0;0)$} & $\frac{[3]}{[N-1] [N] [N+1]}$ & $-\frac{[3]}{[N-1] [N] [N+1]}$ & \cellcolor{light-gray}$-\frac{[3]}{[N-1] [N] [N+1]}$ & $\frac{[3]}{[N-1] [N] [N+1]}$ & $\frac{[3]}{[N-1] [N] [N+1]}$
\end{tabular}
}
\vskip2ex\noindent
In the table above, the multiplicity labels are omitted if they are trivial. Further more, the cells containing the quantum 6j--symbols which are related to other 6j--symbols via the first three symmetry properties are colored in light gray. Only the 6j--symbols in the white cells are independent. The trivial quantum 6j--symbols with $\rho_i=(0;0)$ can be obtained by using the last equality in eq.~\eqref{equ:SymmetricTHat}.

\vskip4ex
\noindent\underline{$\rho_i = \rho_j = (1;1)$:}
\vskip1ex
\noindent\resizebox{\linewidth}{!}{
\begin{tabular}{c|cccc}
\raisebox{-0.5ex}[1ex][1.8ex]{$(\rho_j)_{r_1r_2}$}$\backslash$ \raisebox{0.5ex}[3ex][1ex]{$(\rho_i)_{r_3r_4}$}  & $(1;1)_{00}$ & $(1;1)_{01}$ & $(1;1)_{10}$ & $(1;1)_{11}$ \\ \hline
\raisebox{0ex}[3ex][2ex]{$(1;1)_{00}$} & long expression & $ \frac{i [2 N]}{[N-1] [N]^2 [N+1] \sqrt{[N-2] [N+2]}} $ & \cellcolor{light-gray} $ \frac{i [2 N]}{[N-1] [N]^2 [N+1]  \sqrt{[N-2] [N+2]}} $ & $ -\frac{[4]}{[2] [N-1] [N] [N+1]}$ \\
\raisebox{0ex}[3ex][2ex]{$(1;1)_{01}$} & \cellcolor{light-gray} $-\frac{i [2 N]}{[N-1] [N]^2 [N+1] \sqrt{[N-2] [N+2]}} $ & $ -\frac{1}{[N-1] [N] [N+1]} $ & \cellcolor{light-gray} $ -\frac{1}{[N-1] [N] [N+1]} $ & $ 0$ \\
\raisebox{0ex}[3ex][2ex]{$(1;1)_{10}$} & \cellcolor{light-gray} $-\frac{i [2 N]}{[N-1] [N]^2 [N+1] \sqrt{[N-2] [N+2]}} $ & \cellcolor{light-gray} $ -\frac{1}{[N-1] [N] [N+1]} $ & \cellcolor{light-gray}$ -\frac{1}{[N-1] [N] [N+1]} $ & \cellcolor{light-gray} $0$ \\
\raisebox{0ex}[3ex][2ex]{$(1;1)_{11}$} & \cellcolor{light-gray}$-\frac{[4]}{[2] [N-1] [N] [N+1]} $ & \cellcolor{light-gray} $0$ & \cellcolor{light-gray} $0$ & $ -\frac{1}{[N-1] [N] [N+1]}$ 
\end{tabular}
}
\vskip2ex\noindent
The quantum 6j--symbol with the entry `long expression' explicitly reads
\begin{align*}
	&\begin{Bmatrix}
		21;0 & 0;21 & 1;1\\
		21;0 & 21;0 & 1;1
	\end{Bmatrix}_{0000} \\
	&= \tiny\frac{[3]^2 ([N-3] [N+2]+[N-2] [N+3])}{[N-2] [N-1] [N] [N+1] [N+2] ([N-2] [N+1]+[N-1] [N+2])}\\
	&+\frac{[2] [3]^2 [2 N]^2}{[N-2] [N-1]^3 [N]^3 [N+1]^3 [N+2] ([N-2] [N+1]+[N-1] [N+2])}\\
	&+\frac{[2][N-2] [N+2]([N-2][N+1]^2+[N-1]^2[N+2]) - 2 [3] ([N-1]^3+[N+1]^3)  }{[2] [N-1]^3 [N]^2 [N+1]^3} \ .
\end{align*}
Furthermore, we note that suitable multiplicity separation schemes have been chosen so that many zero entries appear in the above table. Let us view this table as a $4\times 4$ complex matrix $M$. Altering multiplicity separation schemes amounts to transforming the matrix $M$ with the adjoint action of a group element of $U(4)$, i.e., 
\[			M \mapsto U \, M \, U^{\dagger}, \quad\quad U\in U(4)	.	\]
This has been explicitly checked by observing that the four Casimirs of $U(4)$ --- i.e., the coefficients of the characteristic polynomial $P(t)=\det( t - M)$ --- do not depend on any choices of the multiplicity separation scheme.

\vskip4ex
\noindent\underline{$\rho_i = (2;2),(2;1^2), (1^2;2), (1^2;1^2); \rho_j = (1;1)$:}
\vskip1ex
\noindent\resizebox{\linewidth}{!}{
\begin{tabular}{c|cccc}
\raisebox{-0.5ex}[1ex][1.8ex]{$(\rho_j)_{r_1r_2}$}$\backslash$ \raisebox{0.5ex}[3ex][1ex]{$(\rho_i)_{r_3r_4}$}  & $ (2;2) $ & $ (2;1^2) $ & $ (1^2;2) $ & $(1^2;1^2)$ \\ \hline
\raisebox{0ex}[3ex][2ex]{$(1;1)_{00}$} & long expression & long expression & \cellcolor{light-gray} long expression  & long expression \\
\raisebox{0ex}[3ex][2ex]{$(1;1)_{01}$} & $ \frac{i \sqrt{[N-2]}}{[N-1] [N]^2 \sqrt{[N+2]}} $ & $ -\frac{i}{[N] [N+1] \sqrt{[N-2] [N+2]}} $ & $ \frac{i}{[N-1] [N] \sqrt{[N-2] [N+2]}} $ & $ -\frac{i \sqrt{[N+2]}}{[N]^2 [N+1] \sqrt{[N-2]}}$ \\
\raisebox{0ex}[3ex][2ex]{$(1;1)_{10}$} & \cellcolor{light-gray} $\frac{i \sqrt{[N-2]}}{[N-1] [N]^2 \sqrt{[N+2]}}$ & \cellcolor{light-gray} $\frac{i}{[N-1] [N] \sqrt{[N-2] [N+2]}}$ & \cellcolor{light-gray} $-\frac{i}{[N] [N+1] \sqrt{[N-2] [N+2]}}$ & \cellcolor{light-gray} $-\frac{i \sqrt{[N+2]}}{[N]^2 [N+1] \sqrt{[N-2]}}$ \\
\raisebox{0ex}[3ex][2ex]{$(1;1)_{11}$} & $ -\frac{[N-2]}{[N-1] [N]^2 [N+1]} $ & $ -\frac{1}{[N-1] [N] [N+1]} $ & \cellcolor{light-gray} $-\frac{1}{[N-1] [N] [N+1]}$ & $ -\frac{[N+2]}{[N-1] [N]^2 [N+1]}$ \\
\end{tabular}
}
\vskip2ex\noindent
The values of the quantm 6j--symbols on the first row are given below. The values of the quantum 6j--symbol with $\rho_i=(1;1)$ and $\rho_j$ being one of the four representations $(2;2),(2;1^2),(1^2;2), (1^2;1^2)$ can all be obtained via the last equality in the symmetry relation~\eqref{equ:SymmetricTHat}.

\noindent\begin{align*}
	&\begin{Bmatrix}
		21;0 & 0;21 & 2;2\\
		21;0 & 21;0 & 1;1
	\end{Bmatrix}_{0000} \\
	&= -\left(-[2] [N-2]^3[N+1] [N+2]^2 +2 [3] [N-2] [N+2]  ([3][N-1]^2+ [N+1]^2 ) \right.\\
	&\left. +[2]^2 [3]^2  -[3]^2 [N-1] [N] ([N-3][N+2]+  [N-2] [N+3])  \right)\\
	& / ([2] [N-2] [N-1]^3 [N]^2 [N+1]^2 [N+2]) \ , \\ \\
	&\begin{Bmatrix}
		21;0 & 0;21 & 2;1^2\\
		21;0 & 21;0 & 1;1
	\end{Bmatrix}_{0000} =
	\begin{Bmatrix}
		21;0 & 0;21 & 1^2;2\\
		21;0 & 21;0 & 1;1
	\end{Bmatrix}_{0000}\\
	=& -\left([3]^2 [N] ([N-3]  [N+2] +[N-2] [N+3]) - [3] [N-2] [N+2] ([N-3]+ [N+3]) \right.\\
	&\left. -[2][3]^2 [N-2] [N] [N+2] +[2] [N-2]^2 [N] [N+2]^2 \right)\\
	&/([2] [N-2] [N-1]^2 [N]^2 [N+1]^2 [N+2]) \ , \\ \\
	&\begin{Bmatrix}
		21;0 & 0;21 & 1^2;1^2\\
		21;0 & 21;0 & 1;1
	\end{Bmatrix}_{0000} \\
	&=-\left(-[2] [N-2]^2 [N-1] [N+2]^3+2 [3] [N-2] [N+2] ( [N-1]^2+[3][N+1]^2) \right.\\
	&\left.+[2]^2 [3]^2-[3]^2  [N] [N+1] ( [N-3][N+2]+ [N-2] [N+3]) \right)\\
	&/([2] [N-2] [N-1]^2 [N]^2 [N+1]^3 [N+2])
\end{align*}

\vskip4ex
\noindent\underline{$\rho_i, \rho_j = (2;2),(2;1^2), (1^2;2), (1^2;1^2)$:}
\vskip1ex

\noindent\resizebox{\linewidth}{!}{
\begin{tabular}{c|cccc}
\raisebox{-0.5ex}[1ex][1.8ex]{$(\rho_j)_{r_1r_2}$}$\backslash$ \raisebox{0.5ex}[3ex][1ex]{$(\rho_i)_{r_3r_4}$}  & $ (2;2) $ & $ (2;1^2) $ & $ (1^2;2) $ & $(1^2;1^2)$ \\ \hline
\raisebox{0ex}[3ex][2ex]{$(2;2)$} & $\frac{[N-1] [2]^2+[N+1] [2]^2-[N-4] [N] [N+3]}{[N-2] [N-1] [N]^3 [N+1] [N+2] [N+3]} $ & $ \frac{[3]}{[N-1] [N]^2 [N+1] [N+2]} $ & \cellcolor{light-gray} $\frac{[3]}{[N-1] [N]^2 [N+1] [N+2]} $ & $ -\frac{[3]^2}{[N-2] [N-1] [N] [N+1] [N+2]}$ \\
\raisebox{0ex}[3ex][2ex]{$(2;1^2)$} & \cellcolor{light-gray} $\frac{[3]}{[N-1] [N]^2 [N+1] [N+2]} $ & $ \frac{[3]}{[N-2] [N-1] [N] [N+1] [N+2]} $ & \cellcolor{light-gray} $\frac{[3]}{[N-2] [N-1] [N] [N+1] [N+2]} $ & $ \frac{[3]}{[N-2] [N-1] [N]^2 [N+1]}$ \\
\raisebox{0ex}[3ex][2ex]{$(1^2;2)$} & \cellcolor{light-gray} $\frac{[3]}{[N-1] [N]^2 [N+1] [N+2]} $ & \cellcolor{light-gray} $\frac{[3]}{[N-2] [N-1] [N] [N+1] [N+2]} $ & \cellcolor{light-gray} $\frac{[3]}{[N-2] [N-1] [N] [N+1] [N+2]} $ & \cellcolor{light-gray} $\frac{[3]}{[N-2] [N-1] [N]^2 [N+1]}$ \\
\raisebox{0ex}[3ex][2ex]{$(1^2;1^2)$} & \cellcolor{light-gray} $-\frac{[3]^2}{[N-2] [N-1] [N] [N+1] [N+2]} $ & \cellcolor{light-gray} $\frac{[3]}{[N-2] [N-1] [N]^2 [N+1]} $ & \cellcolor{light-gray} $\frac{[3]}{[N-2] [N-1] [N]^2 [N+1]} $ & $\frac{[N-1] [2]^2+[N+1] [2]^2-[N-3] [N] [N+4]}{[N-3] [N-2] [N-1] [N]^3 [N+1] [N+2]}$
\end{tabular}
}

\vskip4ex
\noindent\underline{$\rho_j = (21;21)$:}
\vskip1ex

\noindent\resizebox{\linewidth}{!}{
\begin{tabular}{c|ccccc}
\raisebox{-0.5ex}[1ex][1.8ex]{$(\rho_j)_{r_1r_2}$}$\backslash$ \raisebox{0.5ex}[3ex][1ex]{$(\rho_i)_{r_3r_4}$}  & $ (1;1)_{00} $ & $ (1;1)_{01} $ & $ (1;1)_{10} $ & $(1;1)_{11}$ & \\ \hline
\raisebox{0ex}[3ex][2ex]{$(21;21)$} & $-\frac{[3]^2}{[N-2] [N-1] [N] [N+1] [N+2]}$ & 0 & \cellcolor{light-gray} 0 & 0  &    \\ \hline\hline
\raisebox{-0.5ex}[1ex][1.8ex]{$(\rho_j)_{r_1r_2}$}$\backslash$ \raisebox{0.5ex}[3ex][1ex]{$(\rho_i)_{r_3r_4}$}  & $ (2;2) $ & $ (2;1^2) $ & $ (1^2;2) $ & $(1^2;1^2)$ & $(21;21)$ \\ \hline
\raisebox{0ex}[3ex][2ex]{$(21;21)$} & $ \frac{[2] [3]^2}{[N-2] [N-1] [N]^2 [N+1] [N+2] [N+3]}$ & 0 & \cellcolor{light-gray} 0 & $\frac{[2] [3]^2}{[N-3] [N-2] [N-1] [N]^2 [N+1] [N+2]}$ & $-\frac{[3]^3}{[N-3] [N-2] [N-1]^2 [N] [N+1]^2 [N+2] [N+3]}$
\end{tabular}
}
\vskip2ex\noindent
Again the quantum 6j--symbols with $\rho_i=(21;21)$ can be obtained from the 6j--symbols in the table above via the symmetry properties in eq.~\eqref{equ:SymmetricTHat}.

\bigskip
The quantum 6j--symbols of the second kind enjoy less symmetry. Using the matrix notation
\[
	\hat{U}^{\rho_i,r_3r_4}_{\rho_j,r_1r_2} = 
	\begin{Bmatrix}
		\bar{R} & R & \rho_i\\
		R & R & \rho_j
	\end{Bmatrix}_{r_1r_2r_3r_4}	\ ,
\]
the matrix $\hat{U}$ satisfies the following symmetry properties
\begin{equation}\label{equ:SymmetryUHat}
	\hat{U}^{\rho_i,r_3r_4}_{\rho_j,r_1r_2} = \hat{U}^{\bar{\rho}_i,r_3r_4}_{\rho_j,r_2r_1} = \left(\hat{U}^{\rho_i,r_4r_3}_{\rho_j,r_2r_1} \right)^* \ .
\end{equation}
For $R=\ydiagram{2,1}$ the relevant representations $\rho^i$ and $\rho^j$ are
\begin{align*}
	\rho^i\in & (21;0)\otimes (0;21)\\
	 &= (0;0)\oplus 2(1;1)\oplus (2;2) \oplus (2;1^2) \oplus (1^2;2) \oplus (1^2;1^2) \oplus (21;21) \ , \\
	\rho^j\in &(21;0)\otimes (21;0) \\
	&= (42;0)\oplus (2^3;0) \oplus (31^3;0) \oplus 2(321;0)\oplus (41^2;0) \oplus (3^2;0)\oplus (2^21^1;0) \ ,
\end{align*}
in terms of the the $10\times 10$ matrix $\hat U$. But its symmetry properties only reduce the number of independent entries to 66. On the other hand, for the quantum 6j--symbols with $\rho_j=(321;0)$ and $(r_1,r_2) = (1,0)$ or $(0,1)$, the symmetry properties together with our convention that 6j--symbols with $\sum_i r_i = 1 \mod 2$ are imaginary constrain almost all of them to be zero except for
\begin{align}
	&\begin{Bmatrix}
		0;21 & 21;0 & 1;1\\
		21;0 & 21;0 & 321;0
	\end{Bmatrix}_{1001}
	=\begin{Bmatrix}
		0;21 & 21;0 & 1;1\\
		21;0 & 21;0 & 321;0
	\end{Bmatrix}_{1010}\nonumber\\
	=&\begin{Bmatrix}
		0;21 & 21;0 & 1;1\\
		21;0 & 21;0 & 321;0
	\end{Bmatrix}_{0101}
	=\begin{Bmatrix}
		0;21 & 21;0 & 1;1\\
		21;0 & 21;0 & 321;0
	\end{Bmatrix}_{0110} \nonumber\\
	=& \frac{[3] \sqrt{[5]}}{2 [N-1] [N] [N+1] \sqrt{[N-2] [N+2]}} \ , \label{equ:321B}
\end{align}
\begin{align}
	&\begin{Bmatrix}
		0;21 & 21;0 & 2;1^2\\
		21;0 & 21;0 & 321;0
	\end{Bmatrix}_{1000}
	=-\begin{Bmatrix}
		0;21 & 21;0 & 1^2;2\\
		21;0 & 21;0 & 321;0
	\end{Bmatrix}_{1000}\nonumber\\
	=&-\begin{Bmatrix}
		0;21 & 21;0 & 2;1^2\\
		21;0 & 21;0 & 321;0
	\end{Bmatrix}_{0100}
	=\begin{Bmatrix}
		0;21 & 21;0 & 1^2;2\\
		21;0 & 21;0 & 321;0
	\end{Bmatrix}_{0100} \nonumber \\
	=& -\frac{i [2] [3] \sqrt{[5]}}{2 [N-2] [N-1] [N] [N+1] [N+2]} \ . \label{equ:321A}
\end{align}

In the following we list the values of the remaining quantum 6j--symbols of the second kind with $R=\ydiagram{2,1}$. This time, we divide them into four blocks:

\vskip4ex
\noindent\underline{Trivial 6j--symbols:}
\vskip1ex

{\center
\hfill\resizebox{0.6\linewidth}{!}{
\begin{centering}
\begin{tabular}{c|c||c|c}
\raisebox{-0.5ex}[1ex][1.8ex]{$(\rho_j)_{r_1r_2}$}$\backslash$ \raisebox{0.5ex}[3ex][1ex]{$(\rho_i)_{r_3r_4}$}  & $ (0;0) $ &  & $ (0;0) $ \\ \hline
\raisebox{0ex}[3.5ex][2ex]{$(42;0)$} & $\frac{[3]}{[N-1] [N] [N+1]}$ & \raisebox{0ex}[3.5ex][2ex]{$(41^2;0)$} & $-\frac{[3]}{[N-1] [N] [N+1]}$ \\
\raisebox{0ex}[3.5ex][2ex]{$(2^3;0)$} & $\frac{[3]}{[N-1] [N] [N+1]}$ & \raisebox{0ex}[3.5ex][2ex]{$(3^2;0)$} & $-\frac{[3]}{[N-1] [N] [N+1]}$ \\
\raisebox{0ex}[3.5ex][2ex]{$(31^3;0)$} & $\frac{[3]}{[N-1] [N] [N+1]}$ & \raisebox{0ex}[3.5ex][2ex]{$(2^21^2;0)$} & $-\frac{[3]}{[N-1] [N] [N+1]}$ \\
\raisebox{0ex}[3.5ex][2ex]{$(321;0)_{00}$} & $\frac{[3]}{[N-1] [N] [N+1]}$ & \raisebox{0ex}[3.5ex][2ex]{$(321;0)_{11}$} & $-\frac{[3]}{[N-1] [N] [N+1]}$
\end{tabular}
\end{centering}}
\hfill}

\vskip4ex
\noindent\underline{$\rho_i = (1;1)$ and $\rho_j \neq (321;0)$:}
\vskip1ex

\noindent\resizebox{\linewidth}{!}{
\begin{tabular}{c|cccc}
\raisebox{-0.5ex}[1ex][1.8ex]{$(\rho_j)_{r_1r_2}$}$\backslash$ \raisebox{0.5ex}[3ex][1ex]{$(\rho_i)_{r_3r_4}$}  & $ (1;1)_{00} $ & $ (1;1)_{01} $ & $ (1;1)_{10} $ & $(1;1)_{11}$ \\ \hline
\raisebox{0ex}[3.5ex][2ex]{$(42;0)$} & $\frac{[2] [3] [N-1]+[4] [N+1]}{[2] [N-1] [N]^2 [N+1] [N+2]}$ & $-\frac{i \sqrt{[N-2]}}{[N-1] [N]^2 [N+1] \sqrt{[N+2]}}$ & \cellcolor{light-gray} & $\frac{[2]}{[N-1] [N]^2 [N+1]}$ \\
\raisebox{0ex}[3.5ex][2ex]{$(2^3;0)$} & $-\frac{[3]}{[N-2] [N]^2 [N+1]}$ & $\frac{i [3] \sqrt{[N+2]}}{[N-1] [N]^2 [N+1] \sqrt{[N-2]}}$ & \cellcolor{light-gray} & $\frac{[2] [3]}{[N-1] [N]^2 [N+1]}$ \\
\raisebox{0ex}[3.5ex][2ex]{$(31^3;0)$} & $\frac{[2] [3] [N-2] [N-1]-[6] [N-1] [N+2]-[3] [4] [N+1] [N+2]}{[4] [N-2] [N-1] [N]^2 [N+1] [N+2]}$ & $-\frac{i [3]}{[N-1] [N] [N+1] \sqrt{[N-2] (N+2]}}$ & \cellcolor{light-gray} & $0$ \\
\raisebox{0ex}[3.5ex][2ex]{$(41^2;0)$} & $-\frac{[3] [4] [N-2] [N-1]+[6] [N-2] [N+1]-[2] [3] [N+1] [N+2]}{[4] [N-2] [N-1] [N]^2 [N+1] [N+2]}$ & $\frac{i [3]}{[N-1] [N] [N+1] \sqrt{[N-2] [N+2]}}$ & \cellcolor{light-gray} & $0$ \\
\raisebox{0ex}[3.5ex][2ex]{$(3^2;0)$} & $-\frac{[3]}{[N-1] [N]^2 [N+2]}$ & $-\frac{i [3] \sqrt{[N-2]}}{[N-1] [N]^2 [N+1] \sqrt{[N+2]}}$ & \cellcolor{light-gray} & $\frac{[2] [3]}{[N-1] [N]^2 [N+1]}$ \\
\raisebox{0ex}[3.5ex][2ex]{$(2^21^2;0)$} & $\frac{[4] [N-1]+[2] [3] [N+1]}{[2] [N-2] [N-1] [N]^2 [N+1]}$ & $\frac{i \sqrt{[N+2]}}{[N-1] [N]^2 [N+1] \sqrt{[N-2]}}$ & \cellcolor{light-gray} & $\frac{[2]}{[N-1] [N]^2 [N+1]}$
\end{tabular}
}
\vskip2ex\noindent
The quantum 6j--symbols in the gray cells (omitted) are constrained by symmetry properties to be equal to minus the 6j--symbols immediately to their left.

\vskip4ex
\noindent\underline{$\rho_i=(2;2),(2;1^2),\ldots$ and $\rho_j \neq (321;0)$:}
\vskip1ex

\noindent\resizebox{\linewidth}{!}{
\begin{tabular}{c|ccccc}
\raisebox{-0.5ex}[1ex][1.8ex]{$(\rho_j)_{r_1r_2}$}$\backslash$ \raisebox{0.5ex}[3ex][1ex]{$(\rho_i)_{r_3r_4}$}  & $(2;2)$ & $(2;1^2)$ & $(1^2;2)$ & $(1^2;1^2)$ & $(21;21)$\\ \hline
\raisebox{0ex}[3.5ex][2ex]{$(42;0)$} & $\frac{[2] [3] [N-1]+[4] [N+1]}{[2] [N-1] [N]^2 [N+1] [N+2] [N+3]}$ & $\frac{[3]}{[N-1] [N]^2 [N+1] [N+2]}$ & \cellcolor{light-gray} & $\frac{[3]^2}{[N-1] [N]^2 [N+1] [N+2]}$ & $\frac{[3]^2}{[N-1] [N]^2 [N+1] [N+2] [N+3]}$ \\
\raisebox{0ex}[3.5ex][2ex]{$(2^3;0)$} & $\frac{[3]^2}{[N-2] [N-1] [N]^2 [N+1]}$ & $-\frac{[3]^2}{[N-2] [N-1] [N]^2 [N+1]}$ & \cellcolor{light-gray} & $-\frac{[3]}{[N-2] [N-1] [N]^2 [N+1]}$ & $\frac{[3]^2}{[N-2] [N-1]^2 [N]^2 [N+1]}$ \\
\raisebox{0ex}[3.5ex][2ex]{$(31^3;0)$} & $-\frac{[3]^2}{[N-2] [N-1] [N] [N+1] [N+2]}$ & $-\frac{[3]}{[N-2] [N-1] [N] [N+1] [N+2]}$ & \cellcolor{light-gray} & $-\frac{[2] [3] [N-2] [N-1]-[6] [N+2] [N-1]-[3] [4] [N+1] [N+2]}{[4] [N-3] [N-2] [N-1] [N]^2 [N+1] [N+2]}$ & $\frac{[3]^2}{[N-3] [N-2] [N-1] [N] [N+1] [N+2]}$ \\
\raisebox{0ex}[3.5ex][2ex]{$(41^2;0)$} & $-\frac{[3] [4] [N-2] [N-1]+[6] [N-2] [N+1]-[2] [3] [N+1] [N+2]}{[4] [N-2] [N-1] [N]^2 [N+1] [N+2] [N+3]}$ & $\frac{[3]}{[N-2] [N-1] [N] [N+1] [N+2]}$ & \cellcolor{light-gray} & $\frac{[3]^2}{[N-2] [N-1] [N] [N+1] [N+2]}$ & $\frac{[3]^2}{[N-2] [N-1] [N] [N+1] [N+2] [N+3]}$ \\
\raisebox{0ex}[3.5ex][2ex]{$(3^2;0)$} & $\frac{[3]}{[N-1] [N]^2 [N+1] [N+2]}$ & $\frac{[3]^2}{[N-1] [N]^2 [N+1] [N+2]}$ & \cellcolor{light-gray} & $-\frac{[3]^2}{[N-1] [N]^2 [N+1] [N+2]}$ & $\frac{[3]^2}{[N-1] [N]^2 [N+1]^2 [N+2]}$ \\
\raisebox{0ex}[3.5ex][2ex]{$(2^21^2;0)$} & $-\frac{[3]^2}{[N-2] [N-1] [N]^2 [N+1]}$ & $-\frac{[3]}{[N-2] [N-1] [N]^2 [N+1]}$ & \cellcolor{light-gray} & $-\frac{[4] [N-1]+[2] [3] [N+1]}{[2] [N-3] [N-2] [N-1] [N]^2 [N+1]}$ & $\frac{[3]^2}{[N-3] [N-2] [N-1] [N]^2 [N+1]}$
\end{tabular}
}
\vskip2ex\noindent
The quantum 6j--symbols in the gray cells (omitted) are constrained by symmetry properties to be the same as the 6j--symbols immediately to their left.

\vskip4ex
\noindent\underline{$\rho_j = (321;0):$}
\vskip1ex

\noindent\resizebox{\linewidth}{!}{
\begin{tabular}{c|cccc}
\raisebox{-0.5ex}[1ex][1.8ex]{$(\rho_j)_{r_1r_2}$}$\backslash$ \raisebox{0.5ex}[3ex][1ex]{$(\rho_i)_{r_3r_4}$}  & $ (1;1)_{00} $ & $ (1;1)_{01} $ & $ (1;1)_{10} $ & $(1;1)_{11}$ \\ \hline
\raisebox{0ex}[3.5ex][2ex]{$(321;0)_{00}$} & $-\frac{[3] ([2 N]+1)}{[N-2] [N-1] [N]^2 [N+1] [N+2]}$ & $\frac{i ([3] [N-2]+[6] [N]-[3] [N+2])}{2 [2] [N-1] [N]^2 [N+1] \sqrt{[N-2] [N+2]}}$ & \cellcolor{light-gray} & $-\frac{[3]}{[N-1] [N]^2 [N+1]}$ \\
\raisebox{0ex}[3.5ex][2ex]{$(321;0)_{11}$} & $\frac{[3] ([2 N]-1)}{[N-2] [N-1] [N]^2 [N+1] [N+2]}$ & $\frac{i ([3] [N-2]-[6] [N]-[3] [N+2])}{2 [2] [N-1] [N]^2 [N+1] \sqrt{[N-2] [N+2]}}$ & \cellcolor{light-gray} & $-\frac{[3]}{[N-1] [N]^2 [N+1]}$
\end{tabular}
}
\vskip2ex
\noindent\resizebox{\linewidth}{!}{
\begin{tabular}{c|ccccc}
\raisebox{-0.5ex}[1ex][1.8ex]{$(\rho_j)_{r_1r_2}$}$\backslash$ \raisebox{0.5ex}[3ex][1ex]{$(\rho_i)_{r_3r_4}$}  & $(2;2)$ & $(2;1^2)$ & $(1^2;2)$ & $(1^2;1^2)$ & $(21;21)$\\ \hline
\raisebox{0ex}[3.5ex][2ex]{$(321;0)_{00}$} & $-\frac{[3] ([N-2]-[N+1])}{[N-2] [N-1] [N]^2 [N+1] [N+2]}$ & $-\frac{[3] ([3] [N-2]-[4] [N]-[3] [N+2])}{2 [2] [N-2] [N-1] [N]^2 [N+1] [N+2]}$ & \cellcolor{light-gray} & $-\frac{[3] ([N-1]+[N+2])}{[N-2] [N-1] [N]^2 [N+1] [N+2]}$ & $\frac{[3]^2}{[N-2] [N-1] [N]^2 [N+1] [N+2]}$ \\
\raisebox{0ex}[3.5ex][2ex]{$(321;0)_{11}$} & $\frac{[3] ([N-2]+[N+1])}{[N-2] [N-1] [N]^2 [N+1] [N+2]}$ & $-\frac{[3] ([3] [N-2]+[4] [N]-[3] [N+2])}{2 [2] [N-2] [N-1] [N]^2 [N+1] [N+2]}$ & \cellcolor{light-gray} & $-\frac{[3] ([N-1]-[N+2])}{[N-2] [N-1] [N]^2 [N+1] [N+2]}$ & $\frac{[3]^2}{[N-2] [N-1] [N]^2 [N+1] [N+2]}$
\end{tabular}
}
\vskip2ex\noindent
Due to the symmetry properties, the quantum 6j--symbols in the gray cells (omitted) in the first table are equated with minus the 6j--symbols immediately to their left, while those in the gray cells (omitted) in the second table are identical to the 6j--symbols immediately to their left. Finally, as stated before for quantum 6j--symbols of the second kind with $\rho_j=(321;0)$ and $(r_1,r_2) = (1,0)$ or $(0,1)$, only eight of them --- given in eqs.~\eqref{equ:321B} and \eqref{equ:321A} --- are non--zero (falling into two independent classes). The remaining once are constrained by symmetry.

Note that in the limit $q\to 1$ we can compare the quantum 6j--symbols to their classical counterparts. Using the projector method of Sec.~\ref{sec:proj}, we have independently computed the relevant classical 6j--symbols and find agreement. This serves as a highly non--trivial consistency check both on the general presented methods for computing 6j--symbols and the explicitly stated results in this subsection.

\subsection{The colored HOMFLY invariants}
Using (quasi--)plat representations of the two-bridge hyperbolic knots with up to eight crossings in Figs.~\ref{fig:4s1And6s1} and Figs.~\ref{fig:8sx}, one can write down the formulae to compute their HOMFLY invariants, which are explicitly spelled out in Appendix~\ref{sec:HOMFLYFormulae}. Plugging in the quantum 6j--symbols of the first and second kinds with $R=\ydiagram{2,1}$, applying the proper framing transformations and normalization, the \emph{normalized} HOMFLY invariants colored by $\ydiagram{2,1}$ with framing 0 are explicitly computed. The final results are listed below:

\begin{align}
 \bar{H}&_{\ydiagram{2,1}}(\mathbf{4_1}) = \frac{1}{\lambda ^3 q^5}
 	\left( q^5 \lambda ^6   + (-q^8-q^6+q^5-q^4-q^2) \lambda ^5 \right.\nonumber\\
 	&+ (q^{10}-q^9+3 q^8-3 q^7+5 q^6-4 q^5+5 q^4-3 q^3+3 q^2-q+1 ) \lambda ^4  \nonumber\\
 	&+ (-2 q^{10}+2 q^9-5 q^8+6 q^7-8 q^6+7 q^5-8 q^4+6 q^3-5 q^2+2 q-2 ) \lambda ^3\nonumber\\
 	&+ (q^{10}-q^9+3 q^8-3 q^7+5 q^6-4 q^5+5 q^4-3 q^3+3 q^2-q+1 ) \lambda ^2 \nonumber\\
 	&\left. +  (-q^8-q^6+q^5-q^4-q^2 ) \lambda + q^5 \right)
\end{align}

\begin{align}
 \bar{H}&_{\ydiagram{2,1}}(\mathbf{5_2}) = \frac{1}{\lambda ^9 q^7}
 \left( (q^{12}-2 q^{11}+3 q^{10}-4 q^9+5 q^8-5 q^7+5 q^6 -4 q^5+3 q^4-2 q^3+q^2 ) \lambda ^6 \nonumber \right. \\
 &+ (q^{13}-2 q^{12}+5 q^{11}-7 q^{10}+9 q^9-11 q^8+13 q^7 -11 q^6+9 q^5-7 q^4+5 q^3-2 q^2+q ) \lambda ^5\nonumber\\ 
 & + (q^{14}-2 q^{13}+4 q^{12}-6 q^{11}+9 q^{10} -12 q^9+13 q^8-14 q^7+13 q^6-12 q^5+9 q^4-6 q^3 \nonumber\\ 
 &+4 q^2-2 q+1 ) \lambda ^4 +(-q^{14}+q^{13}-3 q^{12}+3 q^{11}-5 q^{10}+7 q^9-8 q^8+7 q^7-8 q^6+7 q^5 \nonumber\\ 
 &\left. -5 q^4 +3 q^3  -3 q^2+q-1 ) \lambda ^3+(q^{10}-2 q^7+q^4 ) \lambda ^2 +  (q^{10}+q^8-q^7+q^6+q^4 ) \lambda -q^7  \right)
\end{align}

\begin{align}
 \bar{H}&_{\ydiagram{2,1}}(\mathbf{6_1}) =\frac{1}{\lambda ^6 q^7}
 \left( q^7 \lambda ^9 + (-q^{10}+2 q^7-q^4 ) \lambda ^8 + (q^{12}-2 q^{11}+q^{10}-4 q^9+4 q^8-3 q^7  \nonumber \right.\\
 &+4 q^6-4 q^5+q^4-2 q^3+q^2 )\lambda ^7 + (q^{13}-2 q^{12}+4 q^{11}-6 q^{10}+8 q^9-9 q^8+11 q^7-9 q^6 \nonumber\\
 &+8 q^5-6 q^4+4 q^3 -2 q^2+q )\lambda ^6 + (q^{14}-2 q^{13}+4 q^{12}-7 q^{11}+11 q^{10}-12 q^9+15 q^8 \nonumber\\
 &-17 q^7+15 q^6 -12 q^5+11 q^4-7 q^3+4 q^2-2 q+1 ) \lambda ^5+ (-2 q^{14}+3 q^{13}-7 q^{12}+9 q^{11} \nonumber\\
 &-13 q^{10} +17 q^9-19 q^8+18 q^7-19 q^6+17 q^5-13 q^4+9 q^3-7 q^2+3 q-2 ) \lambda ^4 \nonumber\\
 & + (q^{14}-2 q^{13}+4 q^{12}-5 q^{11}+8 q^{10}-10 q^9+11 q^8-12 q^7+11 q^6-10 q^5+8 q^4-5 q^3 \nonumber\\
 & +4 q^2-2 q+1 ) \lambda ^3 + (q^{11}-q^{10}+q^9-q^8+3 q^7-q^6+q^5-q^4+q^3 )\lambda ^2 \nonumber\\
 &\left. +  (-q^{10}-q^8+q^7-q^6-q^4 )\lambda +q^7 \right)
\end{align}

\begin{align}
 \bar{H}&_{\ydiagram{2,1}}(\mathbf{6_2}) = \frac{1}{\lambda ^6 q^{10}} 
 \left(  (q^{15}+2 q^{13}-q^{12}+2 q^{11}+2 q^9-q^8+2 q^7+q^5 ) \lambda ^6+ (-q^{18}-3 q^{16} \nonumber \right. \\
 &+2 q^{15}-5 q^{14}+3 q^{13}-8 q^{12}+4 q^{11}-8 q^{10}+4 q^9-8 q^8+3 q^7-5 q^6+2 q^5-3 q^4-q^2 ) \lambda ^5 \nonumber\\
 & + (q^{20}-q^{19}+5 q^{18}-7 q^{17}+13 q^{16}-14 q^{15}+24 q^{14}-22 q^{13}+29 q^{12}-26 q^{11}+32 q^{10} \nonumber\\
 &-26 q^9+29 q^8-22 q^7+24 q^6-14 q^5+13 q^4-7 q^3 +5 q^2-q+1 ) \lambda ^4 + (-2 q^{20}+3 q^{19} \nonumber\\
 & -8 q^{18}+12 q^{17}-22 q^{16}+24 q^{15}-33 q^{14}+35 q^{13}-42 q^{12}+39 q^{11}-44 q^{10}+39 q^9-42 q^8\nonumber\\
 &+35 q^7-33 q^6+24 q^5-22 q^4+12 q^3-8 q^2+3 q-2 ) \lambda ^3 + (q^{20}-2 q^{19}+5 q^{18}-6 q^{17} \nonumber\\
 &+12 q^{16}-14 q^{15}+17 q^{14}-16 q^{13}+21 q^{12}-18 q^{11}+18 q^{10}-18 q^9+21 q^8-16 q^7+17 q^6 \nonumber\\
 &-14 q^5+12 q^4-6 q^3+5 q^2-2 q+1 ) \lambda ^2 +  (-q^{18}+q^{17}-q^{16}-q^{14}-2 q^{13}+5 q^{12}-6 q^{11} \nonumber\\
 &+4 q^{10}-6 q^9+5 q^8-2 q^7-q^6-q^4+q^3-q^2 ) \lambda +q^{15}-2 q^{14}+3 q^{13}-4 q^{12}+5 q^{11} \nonumber\\
 &\left. -5 q^{10}+5 q^9-4 q^8+3 q^7-2 q^6+q^5 \right)
\end{align}

\begin{align}
 \bar{H}&_{\ydiagram{2,1}}(\mathbf{6_3}) = \frac{1}{\lambda ^3 q^{10}} 
 \left(  (-q^{15}+2 q^{14}-3 q^{13}+4 q^{12}-5 q^{11}+5 q^{10}-5 q^9+4 q^8-3 q^7+2 q^6 \nonumber \right. \\
 & -q^5 ) \lambda ^6+ (q^{18}-q^{17}+2 q^{16}-2 q^{15}+4 q^{14}-4 q^{13}+6 q^{12}-5 q^{11}+7 q^{10}-5 q^9+6 q^8 \nonumber\\
 &-4 q^7+4 q^6-2 q^5+2 q^4-q^3+q^2 ) \lambda ^5 + (-q^{20}+2 q^{19}-6 q^{18}+9 q^{17}-17 q^{16}+23 q^{15} \nonumber\\
 &-36 q^{14}+41 q^{13}-55 q^{12}+56 q^{11}-62 q^{10}+56 q^9-55 q^8+41 q^7-36 q^6+23 q^5-17 q^4 \nonumber\\
 &+9 q^3-6 q^2+2 q-1 ) \lambda ^4 + (2 q^{20}-4 q^{19}+10 q^{18}-16 q^{17}+31 q^{16}-40 q^{15}+60 q^{14} \nonumber\\
 &-71 q^{13}+90 q^{12}-92 q^{11}+105 q^{10}-92 q^9+90 q^8-71 q^7+60 q^6-40 q^5+31 q^4-16 q^3 \nonumber\\
 & +10 q^2-4 q+2 ) \lambda ^3+ (-q^{20}+2 q^{19}-6 q^{18}+9 q^{17}-17 q^{16}+23 q^{15}-36 q^{14}+41 q^{13} \nonumber\\
 &-55 q^{12}+56 q^{11}-62 q^{10}+56 q^9-55 q^8+41 q^7-36 q^6+23 q^5-17 q^4+9 q^3-6 q^2 \nonumber\\
 &+2 q-1 ) \lambda ^2 +  (q^{18}-q^{17}+2 q^{16}-2 q^{15}+4 q^{14}-4 q^{13}+6 q^{12}-5 q^{11}+7 q^{10}-5 q^9+6 q^8 \nonumber\\
 &-4 q^7+4 q^6-2 q^5+2 q^4-q^3+q^2 ) \lambda -q^{15}+2 q^{14}-3 q^{13}+4 q^{12}-5 q^{11}+5 q^{10}-5 q^9 \nonumber\\
 &\left. +4 q^8-3 q^7+2 q^6-q^5 \right)
\end{align}

\begin{align}
 \bar{H}&_{\ydiagram{2,1}}(\mathbf{7_2}) = \frac{1}{\lambda ^{12} q^9} 
 \left(  (q^{14}-2 q^{13}+3 q^{12}-4 q^{11}+5 q^{10}-5 q^9+5 q^8-4 q^7+3 q^6-2 q^5 \nonumber\right.\\
 &+q^4)\lambda ^9+ (q^{15}-2 q^{14}+3 q^{13}-5 q^{12}+7 q^{11}-8 q^{10}+8 q^9-8 q^8+7 q^7-5 q^6+3 q^5 \nonumber\\
 &-2 q^4+q^3)\lambda ^8+(q^{16}-2 q^{15}+3 q^{14}-4 q^{13}+7 q^{12}-8 q^{11}+9 q^{10}-9 q^9+9 q^8-8 q^7 \nonumber\\
 &+7 q^6-4 q^5+3 q^4-2 q^3+q^2)\lambda ^7+ (q^{17}-2 q^{16}+3 q^{15}-6 q^{14}+9 q^{13}-12 q^{12}+14 q^{11} \nonumber\\
 & -17 q^{10}+17 q^9-17 q^8+14 q^7-12 q^6+9 q^5-6 q^4+3 q^3-2 q^2+q)\lambda ^6+ (q^{18}-2 q^{17} \nonumber\\
 & +4 q^{16}-7 q^{15}+11 q^{14}-14 q^{13}+19 q^{12}-22 q^{11}+24 q^{10}-25 q^9+24 q^8-22 q^7+19 q^6 \nonumber\\
 &-14 q^5+11 q^4-7 q^3+4 q^2-2 q+1) \lambda ^5+ (-q^{18}+q^{17}-3 q^{16}+4 q^{15}-7 q^{14}+9 q^{13} \nonumber\\
 &-12 q^{12}+14 q^{11}-16 q^{10}+16 q^9-16 q^8+14 q^7-12 q^6+9 q^5-7 q^4+4 q^3-3 q^2 \nonumber\\
 &+q-1)\lambda ^4+ (q^{15}+q^{12}+q^{10}-2 q^9+q^8+q^6+q^3)\lambda ^3+ (-q^{13}+q^{12}-q^{11}+q^{10} \nonumber\\
 &\left. -3 q^9+q^8-q^7+q^6-q^5)\lambda ^2+  (q^{12}+q^{10}-q^9+q^8+q^6)\lambda-q^9 \right)
\end{align}

\begin{align}
 \bar{H}&_{\ydiagram{2,1}}(\mathbf{7_3}) = -\frac{\lambda ^6}{q^{12}} 
 \left( (q^{17}+2 q^{15}-q^{14}+2 q^{13}+2 q^{11}-q^{10}+2 q^9+q^7)\lambda ^6+ (-q^{20}-3 q^{18} \nonumber \right.\\
 &+2 q^{17}-5 q^{16}+3 q^{15}-8 q^{14}+4 q^{13}-8 q^{12}+4 q^{11}-8 q^{10}+3 q^9-5 q^8+2 q^7 \nonumber\\
 &-3 q^6-q^4)\lambda ^5+ (-q^{19}+2 q^{18}-3 q^{17}+8 q^{16}-9 q^{15}+13 q^{14}-13 q^{13}+18 q^{12}-13 q^{11} \nonumber\\
 &+13 q^{10}-9 q^9+8 q^8-3 q^7+2 q^6-q^5)\lambda ^4+ (q^{24}-q^{23}+4 q^{22}-5 q^{21}+9 q^{20}-10 q^{19} \nonumber\\
 &+14 q^{18}-13 q^{17}+14 q^{16}-11 q^{15}+10 q^{14}-7 q^{13}+6 q^{12}-7 q^{11}+10 q^{10}-11 q^9+14 q^8 \nonumber\\
 &-13 q^7+14 q^6-10 q^5+9 q^4-5 q^3+4 q^2-q+1)\lambda ^3+ (-q^{24}+2 q^{23}-5 q^{22}+9 q^{21} \nonumber\\
 & -16 q^{20}+22 q^{19}-28 q^{18}+33 q^{17}-40 q^{16} +41 q^{15}-39 q^{14}+41 q^{13}-44 q^{12}+41 q^{11}\nonumber\\
 &-39 q^{10}+41 q^9-40 q^8+33 q^7 -28 q^6+22 q^5-16 q^4+9 q^3-5 q^2+2 q-1)\lambda ^2 \nonumber\\
 & +  (-q^{23}+2 q^{22}-6 q^{21}+11 q^{20}-17 q^{19}+23 q^{18}-31 q^{17}+35 q^{16}-38 q^{15}+41 q^{14} \nonumber\\
 &-42 q^{13}+40 q^{12}-42 q^{11}+41 q^{10}-38 q^9+35 q^8-31 q^7+23 q^6-17 q^5+11 q^4-6 q^3 \nonumber\\
 &+2 q^2-q)\lambda-q^{22}+2 q^{21}-3 q^{20}+6 q^{19}-10 q^{18}+11 q^{17}-12 q^{16}+15 q^{15}-16 q^{14}  \nonumber\\
 &\left. +15 q^{13}-15 q^{12}+15 q^{11}-16 q^{10}+15 q^9-12 q^8+11 q^7-10 q^6+6 q^5-3 q^4+2 q^3-q^2 \right)
\end{align}

\begin{align}
 \bar{H}&_{\ydiagram{2,1}}(\mathbf{7_4}) = -\frac{\lambda ^3}{q^9} 
 \left( q^9 \lambda ^9+ (-q^{12}+q^{11}-q^{10}+2 q^9-q^8+q^7-q^6)\lambda ^8 + (-q^{14}-3 q^{12}+q^{11} \nonumber\right.\\
 &-2 q^{10}+4 q^9-2 q^8+q^7-3 q^6-q^4)\lambda ^7 + (q^{16}+q^{14}-q^{13}-2 q^{12}-q^{11}+4 q^9-q^7-2 q^6 \nonumber\\
 &-q^5+q^4+q^2)\lambda ^6 + (q^{18}-2 q^{17}+4 q^{16}-6 q^{15}+13 q^{14}-16 q^{13}+21 q^{12}-28 q^{11}+33 q^{10} \nonumber\\
 &-28 q^9+33 q^8-28 q^7+21 q^6-16 q^5+13 q^4-6 q^3+4 q^2-2 q+1)\lambda ^5+ (-q^{18}+4 q^{17} \nonumber\\
 & -8 q^{16}+14 q^{15}-24 q^{14}+37 q^{13}-48 q^{12}+57 q^{11}-66 q^{10}+70 q^9-66 q^8+57 q^7-48 q^6 \nonumber\\
 &+37 q^5-24 q^4+14 q^3-8 q^2+4 q-1)\lambda ^4 + (-2 q^{17}+6 q^{16}-14 q^{15}+21 q^{14}-36 q^{13} \nonumber\\
 &+51 q^{12}-61 q^{11} +70 q^{10}-78 q^9+70 q^8-61 q^7+51 q^6-36 q^5+21 q^4-14 q^3\nonumber\\
 &+6 q^2-2 q)\lambda ^3 + (-3 q^{16}+8 q^{15}-15 q^{14}+22 q^{13}-32 q^{12}+47 q^{11}-51 q^{10}+48 q^9-51 q^8 \nonumber\\
 &+47 q^7-32 q^6+22 q^5-15 q^4+8 q^3-3 q^2)\lambda ^2 +  (-2 q^{15}+6 q^{14}-10 q^{13}+16 q^{12}-22 q^{11} \nonumber\\
 &+26 q^{10}-28 q^9+26 q^8-22 q^7+16 q^6-10 q^5+6 q^4-2 q^3)\lambda -q^{14}+4 q^{13}-6 q^{12}+6 q^{11} \nonumber\\
 & \left. -9 q^{10} +12 q^9-9 q^8+6 q^7-6 q^6+4 q^5-q^4\right)
\end{align}

\begin{align}
 \bar{H}&_{\ydiagram{2,1}}(\mathbf{7_5}) = \frac{1}{\lambda ^{12} q^{12}} 
 \left( (q^{22}-2 q^{21}+5 q^{20}-9 q^{19}+15 q^{18}-20 q^{17}+27 q^{16}-32 q^{15}+38 q^{14} \nonumber \right.\\
 &-40 q^{13}+42 q^{12}-40 q^{11}+38 q^{10}-32 q^9+27 q^8-20 q^7+15 q^6-9 q^5+5 q^4-2 q^3 \nonumber\\
 &+q^2) \lambda ^6+(q^{23}-3 q^{22}+9 q^{21}-19 q^{20}+33 q^{19}-49 q^{18}+69 q^{17}-90 q^{16}+107 q^{15}\nonumber\\
 &-121 q^{14} +130 q^{13}-134 q^{12}+130 q^{11}-121 q^{10}+107 q^9-90 q^8+69 q^7-49 q^6+33 q^5 \nonumber\\
 &-19 q^4+9 q^3-3 q^2+q) \lambda ^5+(q^{24}-3 q^{23}+8 q^{22}-18 q^{21}+30 q^{20}-48 q^{19}+69 q^{18} \nonumber\\
 &-92 q^{17}+111 q^{16}-133 q^{15}+146 q^{14}-156 q^{13}+158 q^{12}-156 q^{11}+146 q^{10}-133 q^9 \nonumber\\
 &+111 q^8-92 q^7+69 q^6-48 q^5+30 q^4-18 q^3+8 q^2-3 q+1) \lambda ^4+(-q^{24}+2 q^{23} \nonumber\\
 &-6 q^{22}+10 q^{21}-17 q^{20}+24 q^{19}-32 q^{18}+41 q^{17}-47 q^{16}+51 q^{15}-55 q^{14}+58 q^{13} \nonumber\\
 &-56 q^{12}+58 q^{11}-55 q^{10}+51 q^9-47 q^8+41 q^7-32 q^6+24 q^5-17 q^4+10 q^3-6 q^2 \nonumber\\
 &+2 q-1) \lambda ^3+(q^{21}+2 q^{19}-4 q^{18}+9 q^{17}-12 q^{16}+17 q^{15}-24 q^{14}+27 q^{13}-26 q^{12} \nonumber\\
 &+27 q^{11}-24 q^{10}+17 q^9-12 q^8+9 q^7-4 q^6+2 q^5+q^3) \lambda ^2+(q^{20}-2 q^{19}+3 q^{18}-6 q^{17} \nonumber\\
 &+9 q^{16}-11 q^{15}+12 q^{14}-14 q^{13}+16 q^{12}-14 q^{11}+12 q^{10}-11 q^9+9 q^8-6 q^7+3 q^6 \nonumber\\
 &\left.-2 q^5+q^4) \lambda -q^{17}+2 q^{16}-3 q^{15}+4 q^{14}-5 q^{13}+5 q^{12}-5 q^{11}+4 q^{10}-3 q^9+2 q^8-q^7 \right)
\end{align}

\begin{align}
 \bar{H}&_{\ydiagram{2,1}}(\mathbf{7_6}) = \frac{1}{\lambda ^9 q^{10}} 
 \left( (q^{15}-2 q^{14}+3 q^{13}-4 q^{12}+5 q^{11}-5 q^{10}+5 q^9-4 q^8+3 q^7  \nonumber \right.\\
 &-2 q^6+q^5) \lambda ^9+(-q^{18}+2 q^{17}-4 q^{16}+7 q^{15}-11 q^{14}+14 q^{13}-18 q^{12}+20 q^{11}-21 q^{10} \nonumber\\
 &+20 q^9-18 q^8+14 q^7-11 q^6+7 q^5-4 q^4+2 q^3-q^2) \lambda ^8+(q^{20}-3 q^{19} +9 q^{18}-19 q^{17} \nonumber\\
 &+34 q^{16}-51 q^{15}+74 q^{14}-95 q^{13}+113 q^{12}-124 q^{11}+131 q^{10}-124 q^9+113 q^8-95 q^7 \nonumber\\
 &+74 q^6-51 q^5+34 q^4-19 q^3+9 q^2-3 q+1) \lambda ^7+(-3 q^{20}+9 q^{19}-22 q^{18}+41 q^{17} \nonumber\\
 &-71 q^{16}+105 q^{15}-142 q^{14}+177 q^{13}-213 q^{12}+229 q^{11}-236 q^{10}+229 q^9-213 q^8 \nonumber\\
 &+177 q^7-142 q^6+105 q^5-71 q^4+41 q^3-22 q^2+9 q-3) \lambda ^6+(3 q^{20}-8 q^{19}+19 q^{18} \nonumber\\
 &-34 q^{17}+58 q^{16}-83 q^{15}+117 q^{14}-143 q^{13}+169 q^{12}-183 q^{11}+194 q^{10}-183 q^9 \nonumber\\
 &+169 q^8-143 q^7+117 q^6-83 q^5+58 q^4-34 q^3+19 q^2-8 q+3) \lambda ^5+(-q^{20}+2 q^{19} \nonumber\\
 &-7 q^{18}+11 q^{17}-20 q^{16}+26 q^{15}-39 q^{14}+44 q^{13}-52 q^{12}+53 q^{11}-61 q^{10}+53 q^9 \nonumber\\
 &-52 q^8+44 q^7-39 q^6+26 q^5-20 q^4+11 q^3-7 q^2+2 q-1) \lambda ^4+(2 q^{18}-q^{17}+3 q^{16} \nonumber\\
 &-2 q^{15}+2 q^{14}+5 q^{13}+6 q^{11}-7 q^{10}+6 q^9+5 q^7+2 q^6-2 q^5+3 q^4-q^3+2 q^2) \lambda ^3 \nonumber\\
 &+(-q^{16}-3 q^{15}+q^{14}-2 q^{13}+4 q^{12}-7 q^{11}+q^{10}-7 q^9+4 q^8-2 q^7+q^6-3 q^5-q^4) \lambda ^2 \nonumber\\
 &\left. +(2 q^{13}+q^{12}+q^{11}-2 q^{10}+q^9+q^8+2 q^7) \lambda  -q^{10} \right)
\end{align}

\begin{align}
 \bar{H}&_{\ydiagram{2,1}}(\mathbf{7_7}) = \frac{1}{\lambda ^3 q^{10}} 
 ( q^{10} \lambda ^9 +(-2 q^{13}-2 q^{11}+2 q^{10}-2 q^9-2 q^7) \lambda ^8+(q^{16}+2 q^{15}+4 q^{13} \nonumber\\
 &-4 q^{12}+8 q^{11}-4 q^{10}+8 q^9-4 q^8+4 q^7+2 q^5+q^4) \lambda ^7+(-2 q^{18}+2 q^{17}-8 q^{16}+8 q^{15} \nonumber\\
 &-15 q^{14}+16 q^{13}-24 q^{12}+21 q^{11}-28 q^{10}+21 q^9-24 q^8+16 q^7-15 q^6+8 q^5-8 q^4 \nonumber\\
 &+2 q^3-2 q^2) \lambda ^6+(q^{20}-2 q^{19}+10 q^{18}-18 q^{17}+36 q^{16}-58 q^{15}+88 q^{14}-110 q^{13} \nonumber\\
 &+145 q^{12}-156 q^{11}+164 q^{10}-156 q^9+145 q^8-110 q^7+88 q^6-58 q^5+36 q^4-18 q^3 \nonumber\\
 &+10 q^2-2 q+1) \lambda ^5+(-3 q^{20}+8 q^{19}-22 q^{18}+46 q^{17}-82 q^{16}+124 q^{15}-186 q^{14} \nonumber\\
 &+240 q^{13}-286 q^{12}+319 q^{11}-340 q^{10}+319 q^9-286 q^8+240 q^7-186 q^6+124 q^5 \nonumber\\
 &-82 q^4+46 q^3-22 q^2+8 q-3) \lambda ^4+(3 q^{20}-10 q^{19}+23 q^{18}-45 q^{17}+83 q^{16} \nonumber\\
 &-127 q^{15}+178 q^{14}-234 q^{13}+284 q^{12}-314 q^{11}+326 q^{10}-314 q^9+284 q^8-234 q^7 \nonumber\\
 &+178 q^6-127 q^5+83 q^4-45 q^3+23 q^2-10 q+3) \lambda ^3+(-q^{20}+4 q^{19}-10 q^{18}+18 q^{17} \nonumber\\
 &-32 q^{16}+54 q^{15}-75 q^{14}+95 q^{13}-122 q^{12}+136 q^{11}-134 q^{10}+136 q^9-122 q^8+95 q^7 \nonumber\\
 &-75 q^6+54 q^5-32 q^4+18 q^3-10 q^2+4 q-1) \lambda ^2+(q^{18}-3 q^{17}+3 q^{16}-2 q^{15}+6 q^{14} \nonumber\\
 &-8 q^{13}+q^{12}-3 q^{11}+10 q^{10}-3 q^9+q^8-8 q^7+6 q^6-2 q^5+3 q^4-3 q^3+q^2) \lambda \nonumber\\
 &-q^{15}+4 q^{14}-6 q^{13}+6 q^{12}-9 q^{11}+12 q^{10}-9 q^9+6 q^8-6 q^7+4 q^6-q^5 )
\end{align}

\begin{align}
 \bar{H}&_{\ydiagram{2,1}}(\mathbf{8_1}) = \frac{1}{\lambda ^3 q^9}
\left( q^9 \lambda ^{12}+(-q^{12}-q^{10}+q^9-q^8-q^6) \lambda ^{11}+(q^{13}-q^{12}+q^{11}-q^{10}+3 q^9 \nonumber\right.\\
&-q^8+q^7-q^6+q^5) \lambda ^{10}+(q^{13}-2 q^{12}+q^{11}-2 q^{10}+3 q^9-2 q^8+q^7-2 q^6+q^5) \lambda ^9 \nonumber\\
&+(q^{18}-2 q^{17}+4 q^{16}-6 q^{15}+10 q^{14}-12 q^{13}+15 q^{12}-18 q^{11}+19 q^{10}-19 q^9+19 q^8 \nonumber\\
&-18 q^7+15 q^6-12 q^5+10 q^4-6 q^3+4 q^2-2 q+1) \lambda ^8+(-2 q^{18}+3 q^{17}-7 q^{16}+10 q^{15} \nonumber\\
&-15 q^{14}+19 q^{13}-23 q^{12}+26 q^{11}-28 q^{10}+28 q^9-28 q^8+26 q^7-23 q^6+19 q^5-15 q^4 \nonumber\\
&+10 q^3-7 q^2+3 q-2) \lambda ^7+(q^{18}-2 q^{17}+4 q^{16}-6 q^{15}+9 q^{14}-12 q^{13}+16 q^{12}-16 q^{11} \nonumber\\
&+18 q^{10}-21 q^9+18 q^8-16 q^7+16 q^6-12 q^5+9 q^4-6 q^3+4 q^2-2 q+1) \lambda ^6 \nonumber\\
&+(q^{17}-2 q^{16}+3 q^{15}-6 q^{14}+7 q^{13}-8 q^{12}+12 q^{11}-12 q^{10}+10 q^9-12 q^8+12 q^7-8 q^6 \nonumber\\
&+7 q^5-6 q^4+3 q^3-2 q^2+q) \lambda ^5+(q^{16}-2 q^{15}+3 q^{14}-5 q^{13}+8 q^{12}-9 q^{11}+11 q^{10} \nonumber\\
&-11 q^9+11 q^8-9 q^7+8 q^6-5 q^5+3 q^4-2 q^3+q^2) \lambda ^4+(q^{15}-2 q^{14}+3 q^{13}-7 q^{12}+7 q^{11} \nonumber\\
&-9 q^{10}+11 q^9-9 q^8+7 q^7-7 q^6+3 q^5-2 q^4+q^3) \lambda ^3+(q^{14}-2 q^{13}+q^{12}-4 q^{11}+5 q^{10} \nonumber\\
&\left.-2 q^9+5 q^8-4 q^7+q^6-2 q^5+q^4) \lambda ^2+(-q^{12}+2 q^9-q^6) \lambda +q^9 \right)
\end{align}

\begin{align}
 \bar{H}&_{\ydiagram{2,1}}(\mathbf{8_2}) = \frac{\lambda ^3}{q^{15}}
( (q^{25}-2 q^{24}+3 q^{23}-6 q^{22} +10 q^{21}-11 q^{20}+12 q^{19}-15 q^{18}+16 q^{17} \nonumber\\
&-15 q^{16}+15 q^{15}-15 q^{14}+16 q^{13} -15 q^{12}+12 q^{11}-11 q^{10}+10 q^9-6 q^8+3 q^7-2 q^6 \nonumber\\
&+q^5) \lambda ^6+(-q^{28}+q^{27} -q^{26}+q^{25}-q^{24}-5 q^{23}+9 q^{22}-13 q^{21}+20 q^{20}-28 q^{19}+29 q^{18} \nonumber\\
&-31 q^{17} +33 q^{16}-35 q^{15}+33 q^{14}-31 q^{13}+29 q^{12}-28 q^{11}+20 q^{10}-13 q^9+9 q^8-5 q^7 \nonumber\\
& -q^6+q^5-q^4+q^3-q^2) \lambda ^5+(q^{30}-2 q^{29}+5 q^{28}-7 q^{27}+14 q^{26}-17 q^{25} +22 q^{24} \nonumber\\
&-22 q^{23}+26 q^{22}-20 q^{21}+19 q^{20}-12 q^{19}+12 q^{18}-7 q^{17}+8 q^{16} -4 q^{15}+8 q^{14}-7 q^{13} \nonumber\\
&+12 q^{12}-12 q^{11}+19 q^{10}-20 q^9+26 q^8-22 q^7+22 q^6 -17 q^5+14 q^4-7 q^3+5 q^2 \nonumber\\
&-2 q+1) \lambda ^4+(-2 q^{30}+3 q^{29}-8 q^{28}+13 q^{27} -25 q^{26}+30 q^{25}-45 q^{24}+50 q^{23}-62 q^{22} \nonumber\\
&+58 q^{21}-69 q^{20}+64 q^{19}-71 q^{18} +62 q^{17}-71 q^{16}+65 q^{15}-71 q^{14}+62 q^{13}-71 q^{12} \nonumber\\
&+64 q^{11}-69 q^{10}+58 q^9 -62 q^8+50 q^7-45 q^6+30 q^5-25 q^4+13 q^3-8 q^2+3 q-2) \lambda ^3 \nonumber\\
&+(q^{30}-q^{29} +5 q^{28}-7 q^{27}+15 q^{26}-18 q^{25}+32 q^{24}-32 q^{23}+46 q^{22}-43 q^{21}+57 q^{20} \nonumber\\
& -48 q^{19}+61 q^{18}-52 q^{17}+64 q^{16}-52 q^{15}+64 q^{14}-52 q^{13}+61 q^{12}-48 q^{11} +57 q^{10} \nonumber\\
&-43 q^9+46 q^8-32 q^7+32 q^6-18 q^5+15 q^4-7 q^3+5 q^2-q+1) \lambda ^2 +(-q^{28}-3 q^{26} \nonumber\\
&+2 q^{25}-7 q^{24}+4 q^{23}-12 q^{22}+7 q^{21}-15 q^{20}+8 q^{19}-18 q^{18} +8 q^{17}-18 q^{16}+9 q^{15} \nonumber\\
&-18 q^{14}+8 q^{13}-18 q^{12}+8 q^{11}-15 q^{10}+7 q^9-12 q^8 +4 q^7-7 q^6+2 q^5-3 q^4-q^2) \lambda \nonumber\\
& + q^{25}+2 q^{23}-q^{22}+4 q^{21}-q^{20}+4 q^{19}-q^{18}+4 q^{17}-q^{16}+5 q^{15}-q^{14} +4 q^{13}-q^{12} \nonumber\\
&+4 q^{11}-q^{10}+4 q^9-q^8+2 q^7+q^5 )
\end{align}

\begin{align}
 \bar{H}&_{\ydiagram{2,1}}(\mathbf{8_3}) =\frac{1}{\lambda ^6 q^9}
 \left(q^9 \lambda ^{12}+(-q^{12}+2 q^9-q^6) \lambda ^{11}+(-2 q^{12}-q^{10}+3 q^9-q^8-2 q^6) \lambda ^{10} \nonumber\right.\\
 &+(q^{16}-2 q^{15}+3 q^{14}-5 q^{13}+6 q^{12}-9 q^{11}+10 q^{10}-8 q^9+10 q^8-9 q^7+6 q^6-5 q^5 \nonumber\\
 &+3 q^4-2 q^3+q^2) \lambda ^9+(q^{17}-2 q^{16}+6 q^{15}-9 q^{14}+14 q^{13}-20 q^{12}+27 q^{11}-29 q^{10} \nonumber\\
 &+30 q^9-29 q^8+27 q^7-20 q^6+14 q^5-9 q^4+6 q^3-2 q^2+q) \lambda ^8+(q^{18}-3 q^{17}+6 q^{16} \nonumber\\
 &-12 q^{15}+21 q^{14}-32 q^{13}+45 q^{12}-55 q^{11}+63 q^{10}-68 q^9+63 q^8-55 q^7+45 q^6-32 q^5 \nonumber\\
 &+21 q^4-12 q^3+6 q^2-3 q+1) \lambda ^7+(-2 q^{18}+4 q^{17}-10 q^{16}+16 q^{15}-30 q^{14}+46 q^{13} \nonumber\\
 &-60 q^{12}+74 q^{11}-86 q^{10}+89 q^9-86 q^8+74 q^7-60 q^6+46 q^5-30 q^4+16 q^3-10 q^2 \nonumber\\
 &+4 q-2) \lambda ^6+(q^{18}-3 q^{17}+6 q^{16}-12 q^{15}+21 q^{14}-32 q^{13}+45 q^{12}-55 q^{11}+63 q^{10} \nonumber\\
 &-68 q^9+63 q^8-55 q^7+45 q^6-32 q^5+21 q^4-12 q^3+6 q^2-3 q+1) \lambda ^5+(q^{17}-2 q^{16} \nonumber\\
 &+6 q^{15}-9 q^{14}+14 q^{13}-20 q^{12}+27 q^{11}-29 q^{10}+30 q^9-29 q^8+27 q^7-20 q^6+14 q^5 \nonumber\\
 &-9 q^4+6 q^3-2 q^2+q) \lambda ^4+(q^{16}-2 q^{15}+3 q^{14}-5 q^{13}+6 q^{12}-9 q^{11}+10 q^{10}-8 q^9 \nonumber\\
 &+10 q^8-9 q^7+6 q^6-5 q^5+3 q^4-2 q^3+q^2) \lambda ^3+(-2 q^{12}-q^{10}+3 q^9-q^8-2 q^6) \lambda ^2 \nonumber\\
 &\left.+(-q^{12}+2 q^9-q^6) \lambda +q^9 \right)
\end{align}

\begin{align}
 \bar{H}&_{\ydiagram{2,1}}(\mathbf{8_4}) = \frac{1}{\lambda ^6 q^{12}}
\left( (q^{17}+2 q^{15}-q^{14}+2 q^{13}+2 q^{11}-q^{10}+2 q^9+q^7) \lambda ^9+(-q^{20}-3 q^{18} \nonumber\right.\\
&+2 q^{17}-5 q^{16}+3 q^{15}-8 q^{14}+4 q^{13}-8 q^{12}+4 q^{11}-8 q^{10}+3 q^9-5 q^8+2 q^7-3 q^6 \nonumber\\
&-q^4) \lambda ^8+(q^{21}-q^{20}+3 q^{19}-4 q^{18}+8 q^{17}-6 q^{16}+12 q^{15}-11 q^{14}+15 q^{13}-10 q^{12} \nonumber\\
&+15 q^{11}-11 q^{10}+12 q^9-6 q^8+8 q^7-4 q^6+3 q^5-q^4+q^3) \lambda ^7+(q^{24}-2 q^{23}+6 q^{22} \nonumber\\
&-11 q^{21}+19 q^{20}-27 q^{19}+38 q^{18}-48 q^{17}+59 q^{16}-67 q^{15}+72 q^{14}-77 q^{13}+78 q^{12} \nonumber\\
&-77 q^{11}+72 q^{10}-67 q^9+59 q^8-48 q^7+38 q^6-27 q^5+19 q^4-11 q^3+6 q^2-2 q+1) \lambda ^6 \nonumber\\
&+(-2 q^{24}+4 q^{23}-10 q^{22}+17 q^{21}-30 q^{20}+39 q^{19}-54 q^{18}+65 q^{17}-79 q^{16}+85 q^{15} \nonumber\\
&-93 q^{14}+96 q^{13}-100 q^{12}+96 q^{11}-93 q^{10}+85 q^9-79 q^8+65 q^7-54 q^6+39 q^5-30 q^4 \nonumber\\
&+17 q^3-10 q^2+4 q-2) \lambda ^5+(q^{24}-3 q^{23}+6 q^{22}-10 q^{21}+18 q^{20}-24 q^{19}+31 q^{18}-36 q^{17} \nonumber\\
&+42 q^{16}-42 q^{15}+45 q^{14}-44 q^{13}+44 q^{12}-44 q^{11}+45 q^{10}-42 q^9+42 q^8-36 q^7+31 q^6 \nonumber\\
&-24 q^5+18 q^4-10 q^3+6 q^2-3 q+1) \lambda ^4+(q^{23}-3 q^{22}+6 q^{21}-8 q^{20}+14 q^{19}-18 q^{18} \nonumber\\
&+18 q^{17}-18 q^{16}+20 q^{15}-16 q^{14}+14 q^{13}-14 q^{12}+14 q^{11}-16 q^{10}+20 q^9-18 q^8+18 q^7 \nonumber\\
&-18 q^6+14 q^5-8 q^4+6 q^3-3 q^2+q) \lambda ^3+(q^{22}-3 q^{21}+4 q^{20}-6 q^{19}+8 q^{18}-11 q^{17}+9 q^{16} \nonumber\\
&-9 q^{15}+11 q^{14}-10 q^{13}+6 q^{12}-10 q^{11}+11 q^{10}-9 q^9+9 q^8-11 q^7+8 q^6-6 q^5+4 q^4 \nonumber\\
&-3 q^3+q^2) \lambda ^2+(-q^{20}+q^{19}-2 q^{15}+5 q^{14}-5 q^{13}+4 q^{12}-5 q^{11}+5 q^{10}-2 q^9+q^5-q^4) \lambda \nonumber\\
&\left. +q^{17}-2 q^{16}+3 q^{15}-4 q^{14}+5 q^{13}-5 q^{12} +5 q^{11}-4 q^{10}+3 q^9-2 q^8+q^7\right)
\end{align}

\begin{align}
 \bar{H}&_{\ydiagram{2,1}}(\mathbf{8_6}) = \frac{1}{q^{12}}
 \left(  (q^{17}-2 q^{16}+3 q^{15}-4 q^{14}+5 q^{13}-5 q^{12}+5 q^{11}-4 q^{10}+3 q^9-2 q^8+q^7) \lambda ^9 \nonumber\right.\\
 &+(-q^{20}+2 q^{19}-4 q^{18}+7 q^{17}-11 q^{16}+14 q^{15}-18 q^{14}+20 q^{13}-21 q^{12}+20 q^{11}-18 q^{10} \nonumber\\
 &+14 q^9-11 q^8+7 q^7-4 q^6+2 q^5-q^4) \lambda ^8+(-q^{19}+3 q^{18}-7 q^{17}+12 q^{16}-18 q^{15}+24 q^{14} \nonumber\\
 &-28 q^{13}+30 q^{12}-28 q^{11}+24 q^{10}-18 q^9+12 q^8-7 q^7+3 q^6-q^5) \lambda ^7+(q^{24}-3 q^{23}+8 q^{22} \nonumber\\
 &-14 q^{21}+25 q^{20}-36 q^{19}+52 q^{18}-66 q^{17}+80 q^{16}-92 q^{15}+104 q^{14}-108 q^{13}+109 q^{12} \nonumber\\
 &-108 q^{11}+104 q^{10}-92 q^9+80 q^8-66 q^7+52 q^6-36 q^5+25 q^4-14 q^3+8 q^2-3 q+1) \lambda ^6 \nonumber\\
 &+(-2 q^{24}+5 q^{23}-14 q^{22}+27 q^{21}-49 q^{20}+77 q^{19}-113 q^{18}+150 q^{17}-193 q^{16}+230 q^{15} \nonumber\\
 &-258 q^{14}+278 q^{13}-288 q^{12}+278 q^{11}-258 q^{10}+230 q^9-193 q^8+150 q^7-113 q^6+77 q^5 \nonumber\\
 &-49 q^4+27 q^3-14 q^2+5 q-2) \lambda ^5+(q^{24}-3 q^{23}+8 q^{22}-19 q^{21}+37 q^{20}-61 q^{19}+93 q^{18} \nonumber\\
 &-135 q^{17}+175 q^{16}-214 q^{15}+247 q^{14}-271 q^{13}+275 q^{12}-271 q^{11}+247 q^{10}-214 q^9 \nonumber\\
 &+175 q^8-135 q^7+93 q^6-61 q^5+37 q^4-19 q^3+8 q^2-3 q+1) \lambda ^4+(q^{23}-3 q^{22}+8 q^{21} \nonumber\\
 &-15 q^{20}+28 q^{19}-43 q^{18}+64 q^{17}-82 q^{16}+107 q^{15}-120 q^{14}+135 q^{13}-137 q^{12}+135 q^{11} \nonumber\\
 &-120 q^{10}+107 q^9-82 q^8+64 q^7-43 q^6+28 q^5-15 q^4+8 q^3-3 q^2+q) \lambda ^3+(q^{22}-2 q^{21} \nonumber\\
 &+4 q^{20}-9 q^{19}+12 q^{18}-17 q^{17}+23 q^{16}-28 q^{15}+31 q^{14}-35 q^{13}+34 q^{12}-35 q^{11}+31 q^{10} \nonumber\\
 &-28 q^9+23 q^8-17 q^7+12 q^6-9 q^5+4 q^4-2 q^3+q^2) \lambda ^2+(-q^{20}-2 q^{18}+2 q^{17}-2 q^{16} \nonumber\\
 &+2 q^{15}-5 q^{14}+2 q^{13}-4 q^{12}+2 q^{11}-5 q^{10}+2 q^9-2 q^8+2 q^7-2 q^6-q^4) \lambda+q^{17}+2 q^{15} \nonumber\\
 &\left.-q^{14}+2 q^{13}+2 q^{11}-q^{10}+2 q^9+q^7  \right)
\end{align}

\begin{align}
 \bar{H}&_{\ydiagram{2,1}}(\mathbf{8_7}) = -\frac{1}{\lambda ^6 q^{15}}
 \left( (q^{25}-2 q^{24}+3 q^{23}-6 q^{22}+10 q^{21}-11 q^{20}+12 q^{19}-15 q^{18}+16 q^{17} \nonumber\right.\\
 &-15 q^{16}+15 q^{15}-15 q^{14}+16 q^{13}-15 q^{12}+12 q^{11}-11 q^{10}+10 q^9-6 q^8+3 q^7-2 q^6 \nonumber\\
 &+q^5) \lambda ^6+(-q^{28}+q^{27}-2 q^{26}+4 q^{25}-7 q^{24}+6 q^{23}-11 q^{22}+17 q^{21}-19 q^{20}+21 q^{19} \nonumber\\
 &-29 q^{18}+30 q^{17}-32 q^{16}+32 q^{15}-32 q^{14}+30 q^{13}-29 q^{12}+21 q^{11}-19 q^{10}+17 q^9 \nonumber\\
 &-11 q^8+6 q^7-7 q^6+4 q^5-2 q^4+q^3-q^2) \lambda ^5+(q^{30}-2 q^{29}+6 q^{28}-11 q^{27}+23 q^{26} \nonumber\\
 &-34 q^{25}+55 q^{24}-75 q^{23}+104 q^{22}-124 q^{21}+155 q^{20}-171 q^{19}+190 q^{18}-198 q^{17} \nonumber\\
 &+210 q^{16}-204 q^{15}+210 q^{14}-198 q^{13}+190 q^{12}-171 q^{11}+155 q^{10}-124 q^9+104 q^8 \nonumber\\
 &-75 q^7+55 q^6-34 q^5+23 q^4-11 q^3+6 q^2-2 q+1) \lambda ^4+(-2 q^{30}+4 q^{29}-11 q^{28} \nonumber\\
 &+20 q^{27}-40 q^{26}+57 q^{25}-91 q^{24}+118 q^{23}-159 q^{22}+181 q^{21}-220 q^{20}+233 q^{19} \nonumber\\
 &-260 q^{18}+255 q^{17}-273 q^{16}+264 q^{15}-273 q^{14}+255 q^{13}-260 q^{12}+233 q^{11}-220 q^{10} \nonumber\\
 &+181 q^9-159 q^8+118 q^7-91 q^6+57 q^5-40 q^4+20 q^3-11 q^2+4 q-2) \lambda ^3 \nonumber\\
 &+(q^{30}-2 q^{29}+7 q^{28}-11 q^{27}+22 q^{26}-31 q^{25}+49 q^{24}-55 q^{23}+76 q^{22}-78 q^{21}+90 q^{20} \nonumber\\
 &-77 q^{19}+85 q^{18}-66 q^{17}+72 q^{16}-56 q^{15}+72 q^{14}-66 q^{13}+85 q^{12}-77 q^{11}+90 q^{10} \nonumber\\
 &-78 q^9+76 q^8-55 q^7+49 q^6-31 q^5+22 q^4-11 q^3+7 q^2-2 q+1) \lambda ^2+(-q^{28}+q^{27} \nonumber\\
 &-3 q^{26}+2 q^{25}-3 q^{24}-2 q^{23}+5 q^{22}-18 q^{21}+25 q^{20}-45 q^{19}+56 q^{18}-75 q^{17}+78 q^{16} \nonumber\\
 &-88 q^{15}+78 q^{14}-75 q^{13}+56 q^{12}-45 q^{11}+25 q^{10}-18 q^9+5 q^8-2 q^7-3 q^6+2 q^5-3 q^4 \nonumber\\
 &+q^3-q^2) \lambda+q^{25}-2 q^{24}+5 q^{23}-9 q^{22}+15 q^{21}-20 q^{20}+27 q^{19}-32 q^{18}+38 q^{17}-40 q^{16} \nonumber\\
 &\left.+42 q^{15}-40 q^{14}+38 q^{13}-32 q^{12}+27 q^{11}-20 q^{10}+15 q^9-9 q^8+5 q^7-2 q^6+q^5 \right)
\end{align}

\begin{align}
 \bar{H}&_{\ydiagram{2,1}}(\mathbf{8_8}) = -\frac{1}{\lambda ^6 q^{12}}
 \left( (q^{17}-2 q^{16}+3 q^{15}-4 q^{14}+5 q^{13}-5 q^{12}+5 q^{11}-4 q^{10}+3 q^9-2 q^8 \nonumber\right.\\
 &+q^7) \lambda ^9+(-q^{20}+q^{19}-q^{18}+q^{17}-2 q^{16}-q^{13}-q^{11}-2 q^8+q^7-q^6+q^5-q^4) \lambda ^8 \nonumber\\
 &+(q^{22}-3 q^{21}+6 q^{20}-9 q^{19}+15 q^{18}-21 q^{17}+26 q^{16}-28 q^{15}+35 q^{14}-35 q^{13}+35 q^{12} \nonumber\\
 &-35 q^{11}+35 q^{10}-28 q^9+26 q^8-21 q^7+15 q^6-9 q^5+6 q^4-3 q^3+q^2) \lambda ^7+(q^{23}-4 q^{22} \nonumber\\
 &+9 q^{21}-16 q^{20}+30 q^{19}-45 q^{18}+60 q^{17}-78 q^{16}+97 q^{15}-103 q^{14}+113 q^{13}-121 q^{12} \nonumber\\
 &+113 q^{11}-103 q^{10}+97 q^9-78 q^8+60 q^7-45 q^6+30 q^5-16 q^4+9 q^3-4 q^2+q) \lambda ^6 \nonumber\\
 &+(q^{24}-4 q^{23}+9 q^{22}-20 q^{21}+38 q^{20}-64 q^{19}+94 q^{18}-137 q^{17}+175 q^{16}-214 q^{15} \nonumber\\
 &+247 q^{14}-271 q^{13}+271 q^{12}-271 q^{11}+247 q^{10}-214 q^9+175 q^8-137 q^7+94 q^6-64 q^5 \nonumber\\
 &+38 q^4-20 q^3+9 q^2-4 q+1) \lambda ^5+(-2 q^{24}+6 q^{23}-15 q^{22}+31 q^{21}-56 q^{20}+91 q^{19} \nonumber\\
 &-138 q^{18}+190 q^{17}-242 q^{16}+296 q^{15}-339 q^{14}+365 q^{13}-374 q^{12}+365 q^{11}-339 q^{10} \nonumber\\
 &+296 q^9-242 q^8+190 q^7-138 q^6+91 q^5-56 q^4+31 q^3-15 q^2+6 q-2) \lambda ^4 \nonumber\\
 &+(q^{24}-3 q^{23}+9 q^{22}-17 q^{21}+31 q^{20}-52 q^{19}+79 q^{18}-108 q^{17}+145 q^{16}-176 q^{15} \nonumber\\
 &+202 q^{14}-219 q^{13}+230 q^{12}-219 q^{11}+202 q^{10}-176 q^9+145 q^8-108 q^7+79 q^6-52 q^5 \nonumber\\
 &+31 q^4-17 q^3+9 q^2-3 q+1) \lambda ^3+(-q^{20}+q^{19}-2 q^{18}+6 q^{17}-9 q^{16}+11 q^{15}-16 q^{14} \nonumber\\
 &+18 q^{13}-19 q^{12}+18 q^{11}-16 q^{10}+11 q^9-9 q^8+6 q^7-2 q^6+q^5-q^4) \lambda ^2+(-q^{20}+2 q^{19} \nonumber\\
 &-4 q^{18}+7 q^{17}-11 q^{16}+14 q^{15}-18 q^{14}+20 q^{13}-21 q^{12}+20 q^{11}-18 q^{10}+14 q^9-11 q^8 \nonumber\\
 &+7 q^7-4 q^6+2 q^5-q^4) \lambda+q^{17}-2 q^{16}+3 q^{15}-4 q^{14}+5 q^{13}-5 q^{12}+5 q^{11}-4 q^{10}+3 q^9 \nonumber\\
 &\left.-2 q^8+q^7 \right)
\end{align}

\begin{align}
 \bar{H}&_{\ydiagram{2,1}}(\mathbf{8_9}) = \frac{1}{\lambda ^3 q^{15}}
 \left( (q^{25}-2 q^{24}+5 q^{23}-9 q^{22}+15 q^{21}-20 q^{20}+27 q^{19}-32 q^{18}+38 q^{17} \nonumber \right.\\
& -40 q^{16}+42 q^{15}-40 q^{14}+38 q^{13}-32 q^{12}+27 q^{11}-20 q^{10}+15 q^9-9 q^8+5 q^7-2 q^6 \nonumber\\
&+q^5) \lambda ^6+(-q^{28}+q^{27}-3 q^{26}+3 q^{25}-5 q^{24}+q^{23}-q^{22}-5 q^{21}+8 q^{20}-18 q^{19}+16 q^{18} \nonumber\\
&-21 q^{17}+20 q^{16}-26 q^{15}+20 q^{14}-21 q^{13}+16 q^{12}-18 q^{11}+8 q^{10}-5 q^9-q^8+q^7-5 q^6 \nonumber\\
&+3 q^5-3 q^4+q^3-q^2) \lambda ^5+(q^{30}-2 q^{29}+7 q^{28}-12 q^{27}+25 q^{26}-36 q^{25}+58 q^{24}-74 q^{23} \nonumber\\
&+103 q^{22}-121 q^{21}+155 q^{20}-175 q^{19}+208 q^{18}-224 q^{17}+247 q^{16}-242 q^{15}+247 q^{14} \nonumber\\
&-224 q^{13}+208 q^{12}-175 q^{11}+155 q^{10}-121 q^9+103 q^8-74 q^7+58 q^6-36 q^5+25 q^4 \nonumber\\
&-12 q^3+7 q^2-2 q+1) \lambda ^4+(-2 q^{30}+4 q^{29}-12 q^{28}+22 q^{27}-44 q^{26}+64 q^{25}-103 q^{24} \nonumber\\
&+136 q^{23}-186 q^{22}+225 q^{21}-286 q^{20}+332 q^{19}-389 q^{18}+414 q^{17}-454 q^{16}+459 q^{15} \nonumber\\
&-454 q^{14}+414 q^{13}-389 q^{12}+332 q^{11}-286 q^{10}+225 q^9-186 q^8+136 q^7-103 q^6 \nonumber\\
&+64 q^5-44 q^4+22 q^3-12 q^2+4 q-2) \lambda ^3+(q^{30}-2 q^{29}+7 q^{28}-12 q^{27}+25 q^{26}-36 q^{25} \nonumber\\
&+58 q^{24}-74 q^{23}+103 q^{22} -121 q^{21}+155 q^{20}-175 q^{19}+208 q^{18}-224 q^{17}+247 q^{16} \nonumber\\
&-242 q^{15}+247 q^{14}-224 q^{13}+208 q^{12}-175 q^{11}+155 q^{10}-121 q^9+103 q^8-74 q^7 \nonumber\\
&+58 q^6-36 q^5+25 q^4-12 q^3+7 q^2-2 q+1) \lambda ^2+(-q^{28}+q^{27}-3 q^{26}+3 q^{25}-5 q^{24} \nonumber\\
&+q^{23}-q^{22}-5 q^{21}+8 q^{20}-18 q^{19}+16 q^{18}-21 q^{17}+20 q^{16}-26 q^{15}+20 q^{14}-21 q^{13} \nonumber\\
&+16 q^{12}-18 q^{11}+8 q^{10}-5 q^9-q^8+q^7-5 q^6+3 q^5-3 q^4+q^3-q^2) \lambda+q^{25}-2 q^{24} \nonumber\\
&+5 q^{23}-9 q^{22}+15 q^{21}-20 q^{20}+27 q^{19}-32 q^{18}+38 q^{17}-40 q^{16}+42 q^{15}-40 q^{14}+38 q^{13} \nonumber\\
&\left.-32 q^{12}+27 q^{11}-20 q^{10}+15 q^9-9 q^8+5 q^7-2 q^6+q^5 \right)
\end{align}

\begin{align}
 \bar{H}&_{\ydiagram{2,1}}(\mathbf{8_{11}}) =\frac{1}{q^{12}}
 \left( (q^{17}-2 q^{16}+3 q^{15}-4 q^{14}+5 q^{13}-5 q^{12}+5 q^{11}-4 q^{10}+3 q^9-2 q^8+q^7) \lambda ^9 \nonumber\right.\\
 &+(-q^{20}+2 q^{19}-4 q^{18}+6 q^{17}-10 q^{16}+11 q^{15}-14 q^{14}+15 q^{13}-16 q^{12}+15 q^{11}-14 q^{10} \nonumber\\
 &+11 q^9-10 q^8+6 q^7-4 q^6+2 q^5-q^4) \lambda ^8+(q^{20}-q^{19}+6 q^{18}-11 q^{17}+19 q^{16}-25 q^{15} \nonumber\\
 &+37 q^{14}-40 q^{13}+43 q^{12}-40 q^{11}+37 q^{10}-25 q^9+19 q^8-11 q^7+6 q^6-q^5+q^4) \lambda ^7 \nonumber\\
 &+(q^{24}-3 q^{23}+7 q^{22}-14 q^{21}+24 q^{20}-35 q^{19}+49 q^{18}-67 q^{17}+72 q^{16}-83 q^{15}+96 q^{14} \nonumber\\
 &-99 q^{13}+87 q^{12}-99 q^{11}+96 q^{10}-83 q^9+72 q^8-67 q^7+49 q^6-35 q^5+24 q^4-14 q^3 \nonumber\\
 &+7 q^2-3 q+1) \lambda ^6+(-2 q^{24}+6 q^{23}-17 q^{22}+34 q^{21}-63 q^{20}+103 q^{19}-151 q^{18}+203 q^{17} \nonumber\\
 &-267 q^{16}+325 q^{15}-356 q^{14}+392 q^{13}-411 q^{12}+392 q^{11}-356 q^{10}+325 q^9-267 q^8 \nonumber\\
 &+203 q^7 -151 q^6+103 q^5-63 q^4+34 q^3-17 q^2+6 q-2) \lambda ^5+(q^{24}-4 q^{23}+14 q^{22} \nonumber\\
 &-32 q^{21}+67 q^{20}-114 q^{19}+179 q^{18}-258 q^{17}+346 q^{16}-423 q^{15}+492 q^{14}-539 q^{13} \nonumber\\
 &+554 q^{12}-539 q^{11}+492 q^{10}-423 q^9+346 q^8-258 q^7+179 q^6-114 q^5+67 q^4-32 q^3 \nonumber\\
 &+14 q^2-4 q+1) \lambda ^4+(q^{23}-5 q^{22}+16 q^{21}-36 q^{20}+64 q^{19}-113 q^{18}+170 q^{17}-226 q^{16} \nonumber\\
 &+288 q^{15}-349 q^{14}+378 q^{13}-384 q^{12}+378 q^{11}-349 q^{10}+288 q^9-226 q^8+170 q^7 \nonumber\\
 &-113 q^6+64 q^5-36 q^4+16 q^3-5 q^2+q) \lambda ^3+(q^{22}-4 q^{21}+9 q^{20}-21 q^{19}+35 q^{18} \nonumber\\
 &-51 q^{17}+77 q^{16}-101 q^{15}+114 q^{14}-130 q^{13}+139 q^{12}-130 q^{11}+114 q^{10}-101 q^9 \nonumber\\
 &+77 q^8-51 q^7+35 q^6-21 q^5+9 q^4-4 q^3+q^2) \lambda ^2+(-q^{20}+2 q^{19}-3 q^{18}+6 q^{17} \nonumber\\
 &-7 q^{16}+9 q^{15}-12 q^{14}+13 q^{13}-11 q^{12}+13 q^{11}-12 q^{10}+9 q^9-7 q^8+6 q^7-3 q^6 \nonumber\\
 &\left.+2 q^5-q^4) \lambda+q^{17}-2 q^{16}+3 q^{15}-4 q^{14}+5 q^{13}-5 q^{12}+5 q^{11}-4 q^{10}+3 q^9-2 q^8+q^7 \right)
\end{align}

\begin{align}
 \bar{H}&_{\ydiagram{2,1}}(\mathbf{8_{12}}) = \frac{1}{\lambda ^6 q^{10}}
 \left( q^{10} \lambda ^{12}+(-2 q^{13}-q^{11}+3 q^{10}-q^9-2 q^7) \lambda ^{11}+(q^{16}+2 q^{15}-2 q^{14} \nonumber\right.\\
 &+2 q^{13}-6 q^{12}+8 q^{11}-4 q^{10}+8 q^9-6 q^8+2 q^7-2 q^6+2 q^5+q^4) \lambda ^{10}+(-2 q^{18}+3 q^{17} \nonumber\\
 &-6 q^{16}+8 q^{15}-9 q^{14}+7 q^{13}-13 q^{12}+10 q^{11}-6 q^{10}+10 q^9-13 q^8+7 q^7-9 q^6+8 q^5 \nonumber\\
 &-6 q^4+3 q^3-2 q^2) \lambda ^9+(q^{20}-3 q^{19}+11 q^{18}-23 q^{17}+43 q^{16}-68 q^{15}+104 q^{14}-137 q^{13} \nonumber\\
 &+167 q^{12}-187 q^{11}+199 q^{10}-187 q^9+167 q^8-137 q^7+104 q^6-68 q^5+43 q^4-23 q^3 \nonumber\\
 &+11 q^2-3 q+1) \lambda ^8+(-4 q^{20}+14 q^{19}-37 q^{18}+76 q^{17}-139 q^{16}+219 q^{15}-322 q^{14} \nonumber\\
 &+424 q^{13}-513 q^{12}+575 q^{11}-604 q^{10}+575 q^9-513 q^8+424 q^7-322 q^6+219 q^5 \nonumber\\
 &-139 q^4+76 q^3-37 q^2+14 q-4) \lambda ^7+(6 q^{20}-22 q^{19}+56 q^{18}-112 q^{17}+203 q^{16} \nonumber\\
 &-322 q^{15}+458 q^{14}-595 q^{13}+730 q^{12}-810 q^{11}+835 q^{10}-810 q^9+730 q^8-595 q^7 \nonumber\\
 &+458 q^6-322 q^5+203 q^4-112 q^3+56 q^2-22 q+6) \lambda ^6+(-4 q^{20}+14 q^{19}-37 q^{18} \nonumber\\
 &+76 q^{17}-139 q^{16}+219 q^{15}-322 q^{14}+424 q^{13}-513 q^{12}+575 q^{11}-604 q^{10}+575 q^9 \nonumber\\
 &-513 q^8+424 q^7-322 q^6+219 q^5-139 q^4+76 q^3-37 q^2+14 q-4) \lambda ^5+(q^{20}-3 q^{19} \nonumber\\
 &+11 q^{18}-23 q^{17}+43 q^{16}-68 q^{15}+104 q^{14}-137 q^{13}+167 q^{12}-187 q^{11}+199 q^{10}-187 q^9 \nonumber\\
 &+167 q^8-137 q^7+104 q^6-68 q^5+43 q^4-23 q^3+11 q^2-3 q+1) \lambda ^4+(-2 q^{18}+3 q^{17} \nonumber\\
 &-6 q^{16}+8 q^{15}-9 q^{14}+7 q^{13}-13 q^{12}+10 q^{11}-6 q^{10}+10 q^9-13 q^8+7 q^7-9 q^6+8 q^5 \nonumber\\
 &-6 q^4+3 q^3-2 q^2) \lambda ^3+(q^{16}+2 q^{15}-2 q^{14}+2 q^{13}-6 q^{12}+8 q^{11}-4 q^{10}+8 q^9-6 q^8 \nonumber\\
 &\left.+2 q^7-2 q^6+2 q^5+q^4) \lambda ^2+(-2 q^{13}-q^{11}+3 q^{10}-q^9-2 q^7) \lambda +q^{10} \right)
\end{align}

\begin{align}
 \bar{H}&_{\ydiagram{2,1}}(\mathbf{8_{13}}) = -\frac{1}{\lambda ^6 q^{12}}
\left( (q^{17}-4 q^{16}+6 q^{15}-6 q^{14}+9 q^{13}-12 q^{12}+9 q^{11}-6 q^{10}+6 q^9 \nonumber\right.\\
&-4 q^8+q^7) \lambda ^9 +(-q^{20}+3 q^{19}-2 q^{18}-3 q^{16}+2 q^{15}+6 q^{14}-5 q^{13}-5 q^{11}+6 q^{10}+2 q^9 \nonumber\\
&-3 q^8-2 q^6 +3 q^5-q^4) \lambda ^8+(q^{22}-5 q^{21}+11 q^{20}-18 q^{19}+30 q^{18}-47 q^{17}+62 q^{16} \nonumber\\
&-74 q^{15}+88 q^{14} -97 q^{13}+98 q^{12}-97 q^{11}+88 q^{10}-74 q^9+62 q^8-47 q^7+30 q^6-18 q^5 \nonumber\\
&+11 q^4-5 q^3 +q^2) \lambda ^7+(q^{23}-6 q^{22}+17 q^{21}-34 q^{20}+62 q^{19}-107 q^{18}+158 q^{17}-209 q^{16} \nonumber\\
&+267 q^{15} -320 q^{14}+348 q^{13}-354 q^{12}+348 q^{11}-320 q^{10}+267 q^9-209 q^8+158 q^7 \nonumber\\
&-107 q^6 +62 q^5-34 q^4+17 q^3-6 q^2+q) \lambda ^6+(q^{24}-5 q^{23}+15 q^{22}-33 q^{21}+68 q^{20} \nonumber\\
&-123 q^{19} +194 q^{18}-280 q^{17}+383 q^{16}-480 q^{15}+557 q^{14}-615 q^{13}+636 q^{12}-615 q^{11} \nonumber\\
&+557 q^{10} -480 q^9+383 q^8-280 q^7+194 q^6-123 q^5+68 q^4-33 q^3+15 q^2-5 q+1) \lambda ^5 \nonumber\\
& +(-2 q^{24}+7 q^{23}-18 q^{22}+38 q^{21}-75 q^{20}+127 q^{19}-198 q^{18}+284 q^{17}-379 q^{16} +471 q^{15} \nonumber\\
&-545 q^{14}+599 q^{13}-618 q^{12}+599 q^{11}-545 q^{10}+471 q^9-379 q^8+284 q^7 -198 q^6 \nonumber\\
&+127 q^5-75 q^4+38 q^3-18 q^2+7 q-2) \lambda ^4+(q^{24}-3 q^{23}+8 q^{22}-17 q^{21} +32 q^{20} \nonumber\\
&-54 q^{19}+86 q^{18}-126 q^{17}+166 q^{16}-209 q^{15}+247 q^{14}-272 q^{13}+274 q^{12} -272 q^{11} \nonumber\\
&+247 q^{10}-209 q^9+166 q^8-126 q^7+86 q^6-54 q^5+32 q^4-17 q^3+8 q^2 -3 q+1) \lambda ^3 \nonumber\\
&+(q^{19}-q^{18}+3 q^{17}-4 q^{16}+8 q^{15}-6 q^{14}+10 q^{13}-10 q^{12}+10 q^{11}-6 q^{10} +8 q^9-4 q^8 \nonumber\\
&+3 q^7-q^6+q^5) \lambda ^2+(-q^{20}+2 q^{19}-4 q^{18}+6 q^{17}-10 q^{16}+13 q^{15}-17 q^{14} +18 q^{13} \nonumber\\
&-20 q^{12}+18 q^{11}-17 q^{10}+13 q^9-10 q^8+6 q^7-4 q^6+2 q^5-q^4) \lambda+ q^{17}-2 q^{16}+3 q^{15} \nonumber\\
&\left.-4 q^{14}+5 q^{13}-5 q^{12}+5 q^{11}-4 q^{10}+3 q^9-2 q^8 +q^7 \right)
\end{align}

\begin{align}
 \bar{H}&_{\ydiagram{2,1}}(\mathbf{8_{14}}) = \frac{1}{q^{12}}
 \left( (q^{17}-4 q^{16}+6 q^{15}-6 q^{14}+9 q^{13}-12 q^{12}+9 q^{11}-6 q^{10}+6 q^9 \nonumber\right.\\
 &-4 q^8+q^7) \lambda ^9+(-q^{20}+4 q^{19}-6 q^{18}+8 q^{17}-17 q^{16}+24 q^{15}-22 q^{14}+28 q^{13}-36 q^{12} \nonumber\\
 &+28 q^{11}-22 q^{10}+24 q^9-17 q^8+8 q^7-6 q^6+4 q^5-q^4) \lambda ^8+(-q^{21}+2 q^{20}-q^{19}+5 q^{18} \nonumber\\
 &-14 q^{17}+15 q^{16}-19 q^{15}+36 q^{14}-38 q^{13}+30 q^{12}-38 q^{11}+36 q^{10}-19 q^9+15 q^8-14 q^7 \nonumber\\
 &+5 q^6-q^5+2 q^4-q^3) \lambda ^7+(q^{24}-4 q^{23}+10 q^{22}-21 q^{21}+39 q^{20}-65 q^{19}+100 q^{18} \nonumber\\
 &-145 q^{17}+189 q^{16}-233 q^{15}+279 q^{14}-303 q^{13}+306 q^{12}-303 q^{11}+279 q^{10}-233 q^9 \nonumber\\
 &+189 q^8-145 q^7+100 q^6-65 q^5+39 q^4-21 q^3+10 q^2-4 q+1) \lambda ^6+(-2 q^{24}+8 q^{23} \nonumber\\
 &-22 q^{22}+51 q^{21}-98 q^{20}+169 q^{19}-268 q^{18}+388 q^{17}-516 q^{16}+647 q^{15}-760 q^{14} \nonumber\\
 &+831 q^{13}-856 q^{12}+831 q^{11}-760 q^{10}+647 q^9-516 q^8+388 q^7-268 q^6+169 q^5 \nonumber\\
 &-98 q^4+51 q^3-22 q^2+8 q-2) \lambda ^5+(q^{24}-5 q^{23}+17 q^{22}-43 q^{21}+91 q^{20}-169 q^{19} \nonumber\\
 &+274 q^{18}-408 q^{17}+566 q^{16}-717 q^{15}+840 q^{14}-932 q^{13}+970 q^{12}-932 q^{11}+840 q^{10} \nonumber\\
 &-717 q^9+566 q^8-408 q^7+274 q^6-169 q^5+91 q^4-43 q^3+17 q^2-5 q+1) \lambda ^4 \nonumber\\
 &+(q^{23}-6 q^{22}+18 q^{21}-42 q^{20}+81 q^{19}-141 q^{18}+221 q^{17}-308 q^{16}+399 q^{15}-483 q^{14} \nonumber\\
 &+536 q^{13}-552 q^{12}+536 q^{11}-483 q^{10}+399 q^9-308 q^8+221 q^7-141 q^6+81 q^5-42 q^4 \nonumber\\
 &+18 q^3-6 q^2+q) \lambda ^3+(q^{22}-4 q^{21}+10 q^{20}-21 q^{19}+37 q^{18}-57 q^{17}+83 q^{16}-109 q^{15} \nonumber\\
 &+129 q^{14}-145 q^{13}+152 q^{12}-145 q^{11}+129 q^{10}-109 q^9+83 q^8-57 q^7+37 q^6-21 q^5 \nonumber\\
 &+10 q^4-4 q^3+q^2) \lambda ^2+(-q^{20}+2 q^{19}-3 q^{18}+5 q^{17}-6 q^{16}+7 q^{15}-9 q^{14}+9 q^{13}-8 q^{12} \nonumber\\
 &+9 q^{11}-9 q^{10}+7 q^9-6 q^8+5 q^7-3 q^6+2 q^5-q^4) \lambda+q^{17}-2 q^{16}+3 q^{15}-4 q^{14}+5 q^{13} \nonumber\\
 &\left.-5 q^{12}+5 q^{11}-4 q^{10}+3 q^9-2 q^8+q^7 \right)
\end{align}

These normalized HOMFLY invariants enjoy some symmetry properties. The colored HOMFLY invariants of a knot $\mathcal{K}$ and its mirror image $\mathcal{K}^*$ are related by\cite{MR2177747}
\begin{equation}
	\mathcal{W}_R(\mathcal{K}^*)(q,\lambda) = \mathcal{W}_R(\mathcal{K})(q^{-1},\lambda^{-1}) \ .
\end{equation}
So the colored HOMFLY invariants of an amphichiral knot should be invariant under the transformation $q \mapsto q^{-1}, \lambda \mapsto \lambda^{-1}$. Indeed the HOMFLY invariants of $\mathbf{4_1}$, $\mathbf{6_3}$, $\mathbf{8_3}$, $\mathbf{8_9}$, and $\mathbf{8_{12}}$ colored by $\ydiagram{2,1}$ respect this symmetry.

On the other hand, the colored HOMFLY invariants colored with $R$ and its transpose $\tilde{R}$ of the same knot are related by\cite{Itoyama:2012fq,Zodinmawia:2011oya}
\begin{equation}
	\mathcal{W}_{\tilde{R}}(\mathcal{K})(q,\lambda) = \mathcal{W}_R(\mathcal{K})(q^{-1},\lambda) \ .
\end{equation}
So the HOMFLY invariants colored with transpose-symmetric young diagrams should be invariant under the transformation $q\mapsto q^{-1}$. All the stated normalized HOMFLY invariants colored by $\ydiagram{2,1}$ do indeed respect this symmetry. 

For the knots with braid index three --- namely the knots $\mathbf{4_1}$, $\mathbf{5_2}$, $\mathbf{6_2}$, $\mathbf{6_3}$, $\mathbf{7_3}$, $\mathbf{7_5}$, $\mathbf{8_2}$, $\mathbf{8_7}$, $\mathbf{8_9}$ --- using a complementary method these HOMFLY polynomials have also been computed in refs.~\cite{Anokhina:2012rm,Anokhina:2013ica}. Their findings are in perfect agreement with our results, which thus serves as a non--trivial check on both approaches.

%%%%%%%%%%%%%%%%%%%%%%%%%%%%%%
\section{Conclusions and prospects}\label{sec:conclusion}
%%%%%%%%%%%%%%%%%%%%%%%%%%%%%%
Building on the interesting works~\cite{Ramadevi:2000gq,Borhade:2003cu, Zodinmawia:2011oya}, we have assembled the necessary tools to determine colored HOMFLY polynomials for two-bridge hyperbolic knots. Our findings generalize these previous results by systematically implementing non--trivial multiplicities for primaries, which arise in the employed conformal field theory approach. For two--bridge hyperbolic knots with up to eight crossings, we have provided for the general formulae --- consistently including non--trivial multiplicities --- in terms of crossing matrices of the underlying WZW conformal field theory. 

In order to arrive at polynomial expressions for colored HOMFLY invariants, it is therefore necessary to explicitly evaluate the crossing matrices required for a particular coloring with a representation of $SU(N)$. In this work, we have determined the relevant crossing matrices required for coloring with $\ydiagram{2,1}$, with which we have computed the HOMFLY polynomials $\bar H_{\ydiagram{2,1}}$ for all two--bridge hyperbolic knots with up to eight crossings. Using a different approach, for those knots with braid index three the discussed HOMFLY invariants have previously been determined in the interesting works~\cite{Anokhina:2012rm,Anokhina:2013ica}. We find agreement with these results, while the HOMFLY invariants for the other analyzed knots are new (at least to our knowledge). 

The problem of determining the crossing matrices for representations of $SU(N)$ is non--trivial and corresponds to calculating the associated quantum 6j--symbol of the quantum group $\mathcal{U}_qsu(N)$. Formulating the problem in terms of quantum 6j--symbols has the advantage that one can exploit their symmetries. We use these symmetries to iteratively construct quantum 6j--symbols from known and simple ones. This procedure is potentially suitable to determine any $\mathcal{U}_qsu(N)$ quantum 6j--symbol \cite{Butler:1981}. However, an implementation of a fully--automated algorithm to determine a particular quantum 6j--symbols remains challenging. Guided by crossing symmetries among conformal blocks in the s--, t-- and u--channels in conformal field theory, we propose the eigenvector method to further reduce the number of quantum 6j--symbols that need to be computed for the crossing matrices of interest. Although, we have concentrated on the quantum group $\mathcal{U}_qsu(N)$ the discussed techniques generalize to the calculation of quantum 6j--symbols for other quantum groups as well.

Finally, we develop the projector method to compute classical 6j--symbols of $SU(N)$ as well, which is inspired by computations of classical 6j--symbols of $U(N)$ in refs.~\cite{MR2418111,Elvang:2003ue}. The generalization from $U(N)$ to $SU(N)$ enables us to realize both representations and their conjugate representations of $SU(N)$ (for general $N$) simultaneously. As outlined, the projector method can be implemented as a fully--automated algorithm to calculate classical 6j--symbols. For us the computation of classical 6j--symbols is yet another method to get a handle on $\mathcal{U}_qsu(N)$ quantum 6j--symbols, as a quantum 6j--symbol of $\mathcal{U}_qsu(N)$ reduces in the limit $q\to1$ to its classical counterpart. Thus, exhibiting consistency among the discussed mutually independent approaches serves for us as highly non--trivial check on the presented computations and techniques. More generally, we believe that the projector method for representations of $SU(N)$ may prove useful in other contexts as well, such as the calculation of color factors in $SU(N)$ Yang Mills amplitudes for general~$N$. 

In the projector method for $SU(N)$, the number of terms in the graphical representation for $SU(N)$ grows factorially with the number of boxes in the associated composite Young tableau. Therefore it would be desirable to improve on the graphical representation of $SU(N)$ projectors in order to arrive at a less expensive growth behavior. In addition, it could be interesting to develop graphical representations of projector for other Lie groups, too. A generalization of the projector method to quantum groups would be desirable as well.

In this work we have focussed on the computation of knot invariants using the correspondence between Chern--Simons theory and WZW models \cite{Witten:1988hf}. However, more generally the computation of quantum 6j--symbols --- combined with bootstrap methods --- has immediate applications for calculating correlation functions in WZW models. From this perspective, it would indeed be interesting to employ our approaches to quantum 6j--symbols for other quantum groups, so as to also describe correlators in WZW models bases on other affine Lie groups than $\widehat{su(N)}_k$.  As discussed in refs.~\cite{Behrend:1999bn,Felder:1999ka,Felder:1999mq}, quantum 6j--symbols arise naturally in boundary correlation functions in WZW models with branes as well. Thus, this is yet another context, where our methods may become useful. Alternatively, it would be interesting to see, if --- analogously to the presented eigenvector method originating from recoupling relations in the bulk sector --- the boundary sector of WZW models can yield additional constraints on the structure quantum 6j--symbols to be implemented in our computational approaches. 

The fruitful interplay between correlators in WZW models and their connection to Wilson loop expectation values in Chern--Simons theory can even be further exploited in the context of knot theory. In this work, we have focussed on two--bridge knots. We required that their (quasi--)plat representations have four strands so that the quantum states on the boundaries $\Sigma_1,\Sigma_2$ of the cut--open three manifolds $M_1,M_2$ are in one--to--one correspondence with four point functions of the associated WZW model. On the one hand, for some higher--bridge knots --- with (quasi--)plat representations with more than four strands --- it is possible to cut $S^3$ into three manifolds with more than one boundary such that each boundary is still punctured by only four strands. Then a similar procedure can be applied to derive colored HOMFLY invariants for these knots~\cite{Kaul:1991np, Kaul:1992rs, Nawata:2013qpa}. On the other hand, from the conformal field theory point of view the crossing matrices are the fundamental data. Once the crossing matrices are determined, the structure constants in the operator algebra can be derived via bootstrap techniques, and in principal all the correlation functions can be computed. In particular, one can decompose an $n$--point function via the operator algebra into a product of three--point functions, represented by a trivalent tree diagram. Distinct decomposition into different trivalent tree diagrams furnish distinct bases of the boundary Hilbert spaces. These Hilbert bases are again related among each other by suitable crossing matrices. To arrive at formulae for colored HOMFLY invariants in terms of the crossing matrices, transformations among suitable bases must be performed such that the action of the braid operators in the (quasi--)plat representation of the knot can consecutively be performed on adopted eigenstate bases. We plan to come back to the outlined approach to calculate colored HOMFLY invariants for higher--bridge knots in the future.

A computational challenge is that the method used in this paper is of relatively low efficiency. Let us consider a two--bridge knot and split a (quasi--)plat representation of the knot from top to bottom to different layers, where each layer consisting of only central braidings or only side braidings. For instance, the (quasi--)plat representations of the knot $\mathbf{5_2}$ and the knot $\mathbf{6_2}$ in Figs.~\ref{fig:4s1And6s1} have two and four layers, respectively. Suppose a (quasi--)plat representation has $k$ layers, and the Hilbert space has dimension $D$, then the memory consumed and the time spent to compute the HOMFLY invariant are both proportional to $D^k$. In other words, both space and time complexity of this algorithm follows power laws with high exponents as the dimension of the Hilbert space grows. It even suffers from exponential growth as the number of layers in the (quasi--)plat representations of knots increases. Thus, the calculations can quickly become computational expensive in memory and time for (quasi-)plat representations with many levels and colored with high--dimensional representations. Therefore, the lesson is that for a given knot one should use the (quasi--)plat representations with as few levels as possible. The minimal number of levels for a knot can be used to define the computational complexity of the knot.

%%%%%%%%%%%%%%%%%%
\bigskip
\subsection*{Acknowledgments}
%%%%%%%%%%%%%%%%%%
%%
We would like to thank
Philip Butler,
Andrey Morozov,
Masoud Soroush,
and
Christoph Schweigert
for useful correspondence.
J.G. is supported by the BCGS program, and H.J.~is supported by the DFG grant KL 2271/1-1.

%%%%%%%%%%%%%%%%%%%
%\bigskip
\newpage
\appendix
%%%%%%%%%%%%%%%%%%%

%%%%%%%%%%%%%%%%%%%%%%%%%%%%%%%%%%%%%
\section{Examples of bootstrap computation}\label{sec:BootstrapExample}
We give a concrete example here of the bootstrap method in action. Let us compute the value of the following quantum 6j--symbol
\begin{equation} \label{equ:target6j}
	\begin{Bmatrix}
		21;0 & 0;21 & 1;1\\
		21;0 & 21;0 & 1;1
	\end{Bmatrix}_{0000} \ .
\end{equation}
We first need to convert it to primitive 6j--symbols. In order to do so, we apply the pentagon relation with $\nu_1 = \nu_2 = [1;0]$. On the right hand of the pentagon relation there will be only primitive 6j--symbols. We collect them as follows
\[\begin{aligned}
	&\begin{Bmatrix}
		21;1 & 1;21 & 1;1\\
		21;0 & 21;0 & 1;0
	\end{Bmatrix}_{000r_4} , \quad r_4 = 0,1,2 \\
	&\begin{Bmatrix}
		0;21 & 21;0 & 1;1\\
		1;21 & 0;2 & 1;0
	\end{Bmatrix}_{0000} \quad
	\begin{Bmatrix}
		0;21 & 21;0 & 1;1\\
		1;21 & 0;11 & 1;0
	\end{Bmatrix}_{0000} \\
	&\begin{Bmatrix}
		0;21 & 21;0 & 1;1\\
		0;2 & 0;2 & 1;0
	\end{Bmatrix}_{0000} \quad
	\begin{Bmatrix}
		0;21 & 21;0 & 1;1\\
		0;11 & 0;2 & 1;0
	\end{Bmatrix}_{0000} \quad
	\begin{Bmatrix}
		0;21 & 21;0 & 1;1\\
		0;11 & 0;11 & 1;0
	\end{Bmatrix}_{0000}\\
	&\begin{Bmatrix}
		0;21 & 21;0 & 1;1\\
		0;1 & 0;1 & 21;1
	\end{Bmatrix}_{0000} \quad
	\begin{Bmatrix}
		0;21 & 21;0 & 1;1\\
		0;1 & 0;1 & 2;0
	\end{Bmatrix}_{0000}
\end{aligned}\]
The 6j--symbols in the last row are already type II 6j--symbols. All the rest 6j--symbols belong to type III, and can be converted to type IIs and type IVs by applying the pentagon relation with $\nu_2=[1;0]$ and some appropriate $\nu_1$ (here we choose $\nu_1 = [1;0]$). We collect all the type IV 6j--symbols

\vskip2ex
\noindent
\begin{tabular}{c c c}
$\begin{Bmatrix}
	1;21 & 21;1^2 & 1;0 \\
	1;21 & 0;21 & 0;1
\end{Bmatrix}_{0000}$
&$\begin{Bmatrix}
	1;21 & 21;2 & 1;0 \\
	1;21 & 0;21 & 0;1
\end{Bmatrix}_{0000}$
& \\
$\begin{Bmatrix}
	21;1 & 0;21 & 1;0 \\
	21;1 & 1^2;1 & 0;1
\end{Bmatrix}_{0000}$
&$\begin{Bmatrix}
	21;1 & 0;21 & 1;0 \\
	21;1 & 2;1 & 0;1
\end{Bmatrix}_{0000}$
& \\
$\begin{Bmatrix}
	1^2;0 & 0;21 & 1;0 \\
	1^2;0 & 1^2;1 & 0;1
\end{Bmatrix}_{0000}$
&$\begin{Bmatrix}
	1^2;0 & 0;21 & 1;0 \\
	1^2;0 & 1;0 & 0;1
\end{Bmatrix}_{0000}$
& \\
$\begin{Bmatrix}
	2;0 & 0;21 & 1;0 \\
	2;0 & 2;1 & 0;1
\end{Bmatrix}_{0000}$
&$\begin{Bmatrix}
	2;0 & 0;21 & 1;0 \\
	2;0 & 1;0 & 0;1
\end{Bmatrix}_{0000}$
& \\
$\begin{Bmatrix}
	21;1 & 0;21 & 1;0 \\
	1^2;0 & 1^2;1 & 0;1
\end{Bmatrix}_{0000}$
& $\begin{Bmatrix}
	21;1 & 0;21 & 1;0 \\
	2;0 & 2;1 & 0;1
\end{Bmatrix}_{0000}$ 
& $\begin{Bmatrix}
	1^2;0 & 0;21 & 1;0 \\
	2;0 & 1;0 & 0;1
\end{Bmatrix}_{0000}$
\end{tabular}
\vskip2ex

\noindent as well as all the type II 6j--symbols

\vskip2ex
\noindent
\begin{tabular}{ccl}
$\begin{Bmatrix}
21;1 & 1;21 & 1;1\\
1;0 & 1;0 & 1^2;21
\end{Bmatrix}_{000r_4}$
&$\begin{Bmatrix}
21;1 & 1;21 & 1;1\\
1;0 & 1;0 & 2;21
\end{Bmatrix}_{000r_4}$& \\
$\begin{Bmatrix}
21;1 & 1;21 & 1;1\\
1;0 & 1;0 & 1;1^2
\end{Bmatrix}_{000r_4}$
&$\begin{Bmatrix}
21;1 & 1;21 & 1;1\\
1;0 & 1;0 & 1;2
\end{Bmatrix}_{000r_4}$\\
$\begin{Bmatrix}
21;0 & 0;21 & 1;1\\
1;0 & 1;0 & 1;21
\end{Bmatrix}_{000r'_4}$
&$\begin{Bmatrix}
21;0 & 0;21 & 1;1\\
1;0 & 1;0 & 0;1^2
\end{Bmatrix}_{000r'_4}$
&$\begin{Bmatrix}
21;0 & 0;21 & 1;1\\
1;0 & 1;0 & 0;2
\end{Bmatrix}_{000r'_4}$ \\
$\begin{Bmatrix}
21;0 & 0;1^2 & 1;1\\
1;0 & 1;0 & 1;1^2
\end{Bmatrix}_{0000}$
&$\begin{Bmatrix}
21;0 & 0;2 & 1;1\\
1;0 & 1;0 & 1;2
\end{Bmatrix}_{0000} 
$& \\
$\begin{Bmatrix}
1^2;0 & 0;1^2 & 1;1\\
1;0 & 1;0 & 1;1^2
\end{Bmatrix}_{0000}$
&$\begin{Bmatrix}
2;0 & 0;2 & 1;1\\
1;0 & 1;0 & 1;2
\end{Bmatrix}_{0000}$& \\
$\begin{Bmatrix}
1^2;0 & 0;1^2 & 1;1\\
1;0 & 1;0 & 0;1
\end{Bmatrix}_{0000}$
&$\begin{Bmatrix}
1^2;0 & 0;2 & 1;1\\
1;0 & 1;0 & 0;1
\end{Bmatrix}_{0000}$
&$\begin{Bmatrix}
2;0 & 0;2 & 1;1\\
1;0 & 1;0 & 0;1
\end{Bmatrix}_{0000}$
\end{tabular}
\vskip1ex
\noindent with $r_4 = 0,1,2$ and $r'_4=0,1$. 
\vskip2ex

In the type IV 6j--symbols we have used the tetrahedral symmetry to make sure that the two representations in the last column are conjugate to each other. Among all the type IVs, except for those in the last row all other 6j--symbols satisfy that the two representations in the first column are identical. So we can apply the generalized Racah backcoupling rule to convert them to trivial 6j--symbols only (we also need to use the normalization and $q\rightarrow q^{-1}$ tricks discussed in Sec.~\ref{sec:Cores}). The type IVs in the last row can be related to those above them. For example, in the 6j--symbol
\[
\begin{Bmatrix}
	21;1 & 0;21 & 1;0\\
	1^2;0 & 1^2;1 & 0;1
\end{Bmatrix}_{0000}
\]
the representation at position (2,1) can be either $[1^2;0]$ or $[21;1]$. So its absolute value is related to that of 
\[
\begin{Bmatrix}
	21;1 & 0;21 & 1;0\\
	21;1 & 1^2;1 & 0;1
\end{Bmatrix}_{0000}
\]
in the second row via the unitarity relation. Its phase can be arbitrarily chosen. In this way we can solve for all these type IVs.

Next let us determine the type II 6j--symbols. Since all these type IIs are descendable at position (1,3) --- the representation at this position can either be $[0;0]$ or $[1;1]$ --- their absolute values are determined. The only things we need to take care of are the separation of multiplicities and the choice of phase. For the former issue, we give the example of solving the 6j--symbols in the third row, which only differ by the representation at position (2,3). We use the following multiplicity separation scheme in accordance with the order of the three families of 6j--symbols:
\vskip2ex
\begin{center}
\begin{tabular}{c|c:c:c}
$r_4\backslash \mu_3$ & $[1;21]$ & $[0;1^2]$ & $[0;2]$ \\\hline
0 & $*$ & $-$ & $-$  \\
1 & 0 & $*$ & $-$ \\
\end{tabular}
\end{center}
\vskip2ex

\noindent In other words we set 
\[
\begin{Bmatrix}
	21;0 & 0;21 & 1;1\\
	1;0 & 1;0 & 1;21
\end{Bmatrix}_{0001} = 0  \ .
\]
Then by applying the unitarity relation
\begin{align*}
\dim_q[0;0]\dim_q[1;21]
\Big| &\begin{Bmatrix}
	21;0 & 0;21 & 0;0\\
	1;0 & 1;0 & 1;21
\end{Bmatrix}_{0000}\Big|^2 \\
&+ \dim_q[1;1]\dim_q[1;21]
\Big|\begin{Bmatrix}
	21;0 & 0;21 & 1;1\\
	1;0 & 1;0 & 1;21
\end{Bmatrix}_{0000}\Big|^2 = 1
\end{align*}
we can solve for
\[
\Big|\begin{Bmatrix}
	21;0 & 0;21 & 1;1\\
	1;0 & 1;0 & 1;21
\end{Bmatrix}_{0000} \Big|= \frac{[3]}{[N-1][N][N+1]\sqrt{[N-2][N+2]}} \ ,
\]
and we assign the phase $+1$. For the next two 6j--symbols,
\[
\begin{Bmatrix}
	21;0 & 0;21 & 1;1\\
	1;0 & 1;0 & 0;1^2
\end{Bmatrix}_{0000}\ , \quad
\begin{Bmatrix}
	21;0 & 0;21 & 1;1\\
	1;0 & 1;0 & 0;1^2
\end{Bmatrix}_{0001} \ ,
\] 
the value of the first 6j--symbol is uniquely fixed by the orthogonality relation
\begin{align*}
\dim_q[0;0]&\begin{Bmatrix}
	21;0 & 0;21 & 0;0\\
	1;0 & 1;0 & 1;21
\end{Bmatrix}_{0000}^*
\begin{Bmatrix}
	21;0 & 0;21 & 0;0\\
	1;0 & 1;0 & 0;1^2
\end{Bmatrix}_{0000} \\
&+\dim_q[1;1]
\begin{Bmatrix}
	21;0 & 0;21 & 1;1\\
	1;0 & 1;0 & 1;21
\end{Bmatrix}_{0000}^*
\begin{Bmatrix}
	21;0 & 0;21 & 1;1\\
	1;0 & 1;0 & 0;1^2
\end{Bmatrix}_{0000} = 0 \ ,
\end{align*}
while the absolute value of the second 6j--symbol is obtained by again relating to trivial 6j--symbol using unitarity. We assign the phase $i$ to the second 6j--symbol as its multiplicity sum is odd. The values of the last two 6j--symbols are now both fixed by orthogonality.

As for the choice of phase, we illustrate it by the simple example of
\[
	\begin{Bmatrix}
		1^2;0 & 0;1^2 & 1;1\\
		1;0 & 1;0 & 1;1^2
	\end{Bmatrix}_{0000} \textrm{and }\,
	\begin{Bmatrix}
		1^2;0 & 0;1^2 & 1;1\\
		1;0 & 1;0 & 0;1
	\end{Bmatrix}_{0000} \ .
\]
Clearly we can only assign a phase of our choice to one of the two 6j--symbols, while the phase of the other is determined by the orthogonality relation
\begin{align*}
	\dim_q[0;0]&\begin{Bmatrix}
		1^2;0 & 0;1^2 & 0;0\\
		1;0 & 1;0 & 1;1^2
	\end{Bmatrix}_{0000}
	\begin{Bmatrix}
		1^2;0 & 0;1^2 & 0;0\\
		1;0 & 1;0 & 0;1
	\end{Bmatrix}_{0000}\\
	&+\dim_q[1;1]\begin{Bmatrix}
		1^2;0 & 0;1^2 & 1;1\\
		1;0 & 1;0 & 1;1^2
	\end{Bmatrix}_{0000}	
	\begin{Bmatrix}
		1^2;0 & 0;1^2 & 1;1\\
		1;0 & 1;0 & 0;1
	\end{Bmatrix}_{0000}  = 0 \ .
\end{align*}
It is easy to see in fact this is a special case of the multiplicity separation scheme with $r_4=0$.

After we have solved all the type IV and type II 6j--symbols, we can plug them in to obtain the values of the primitive 6j--symbols, which in turn can be used to solve for the value of \eqref{equ:target6j} that we started with.

%%%%%%%%%%%%%%%%%%%%%%%%%%%%%%%%%%%%%
\section{HOMFLY invariants for the computed knots}\label{sec:HOMFLYFormulae}
\begin{figure}[htbp]
\center
\subfloat[$\mathbf{8_1}$]{\includegraphics[width=0.2\linewidth]{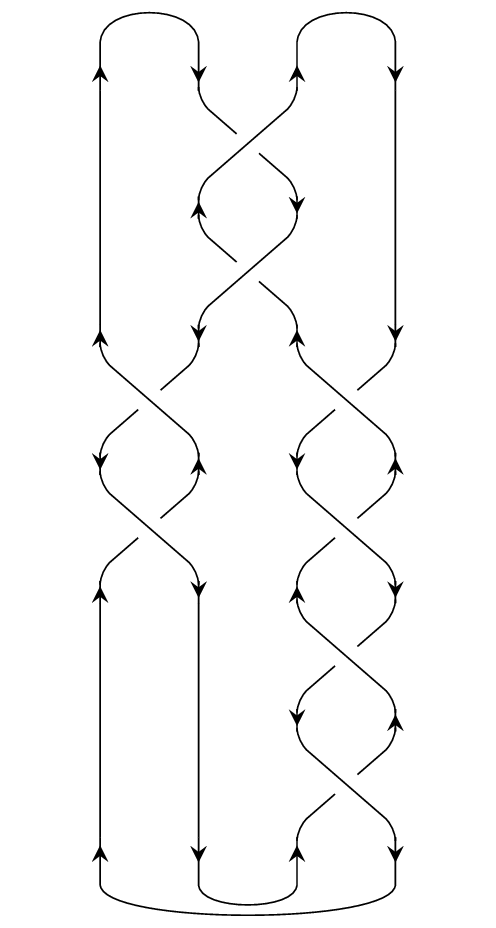}}
\subfloat[$\mathbf{8_2}$]{\includegraphics[width=0.2\linewidth]{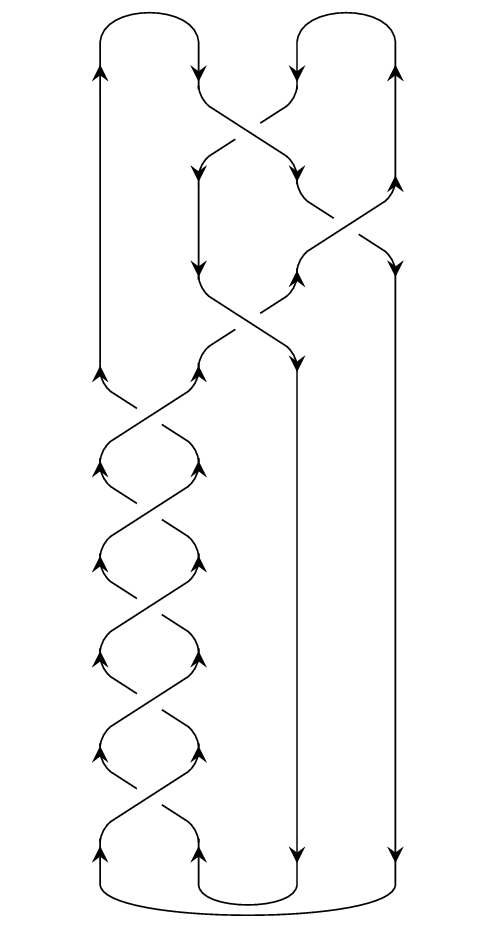}}
\subfloat[$\mathbf{8_3}$]{\includegraphics[width=0.2\linewidth]{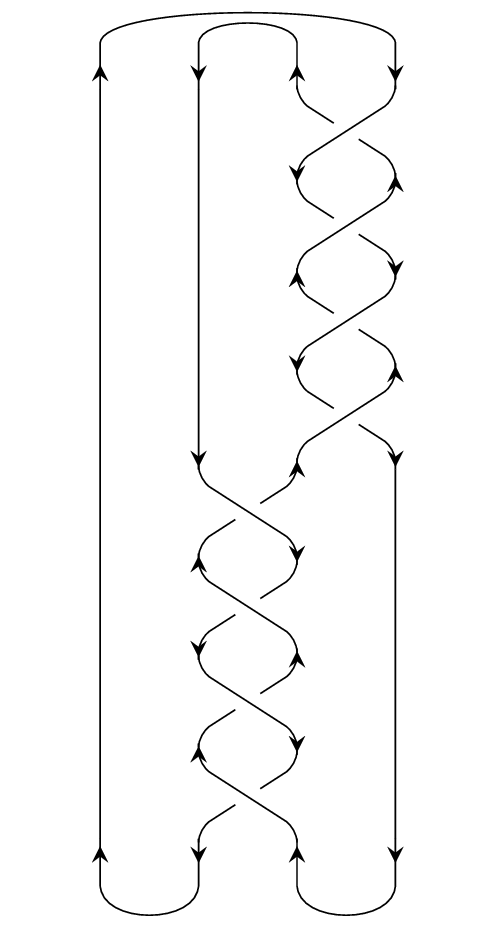}}
\subfloat[$\mathbf{8_4}$]{\includegraphics[width=0.2\linewidth]{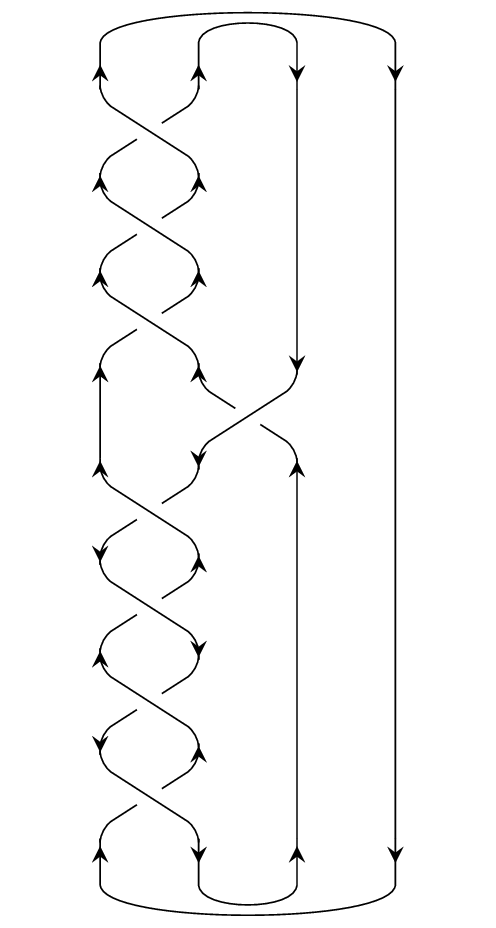}} \\
\subfloat[$\mathbf{8_6}$]{\includegraphics[width=0.2\linewidth]{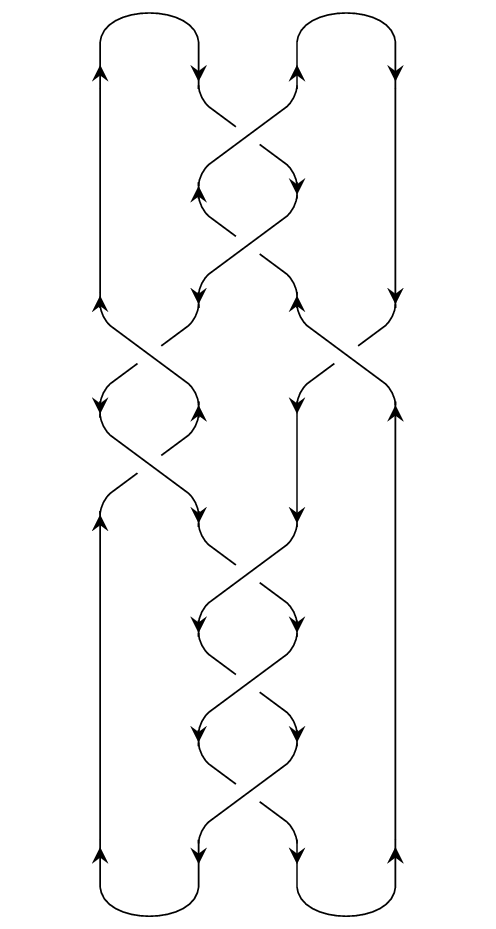}}
\subfloat[$\mathbf{8_7}$]{\includegraphics[width=0.2\linewidth]{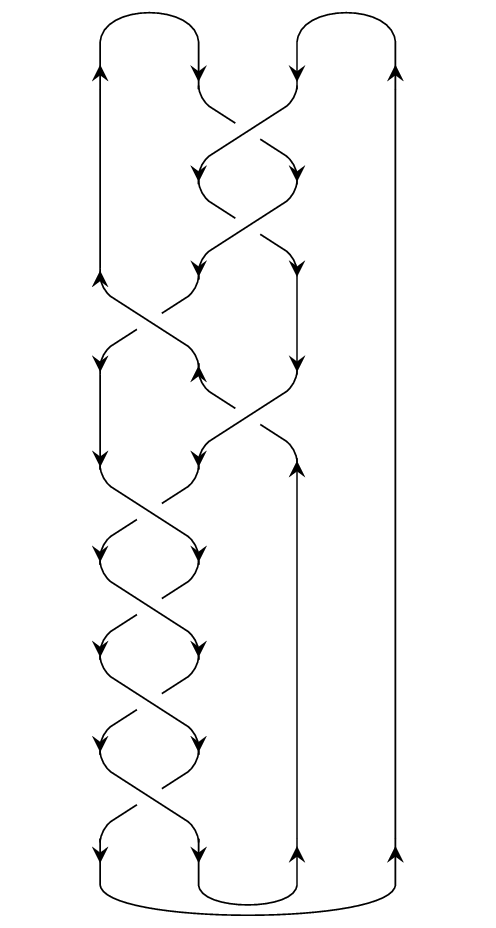}}
\subfloat[$\mathbf{8_8}$]{\includegraphics[width=0.2\linewidth]{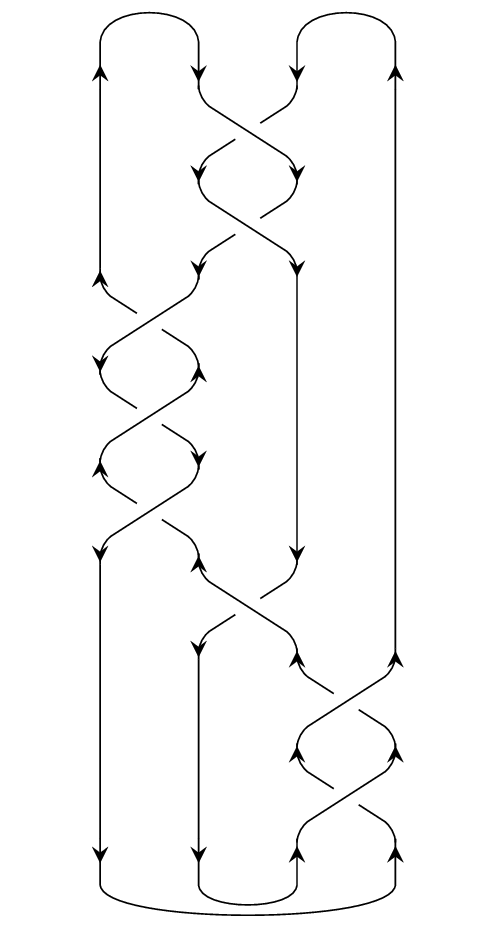}}
\subfloat[$\mathbf{8_9}$]{\includegraphics[width=0.2\linewidth]{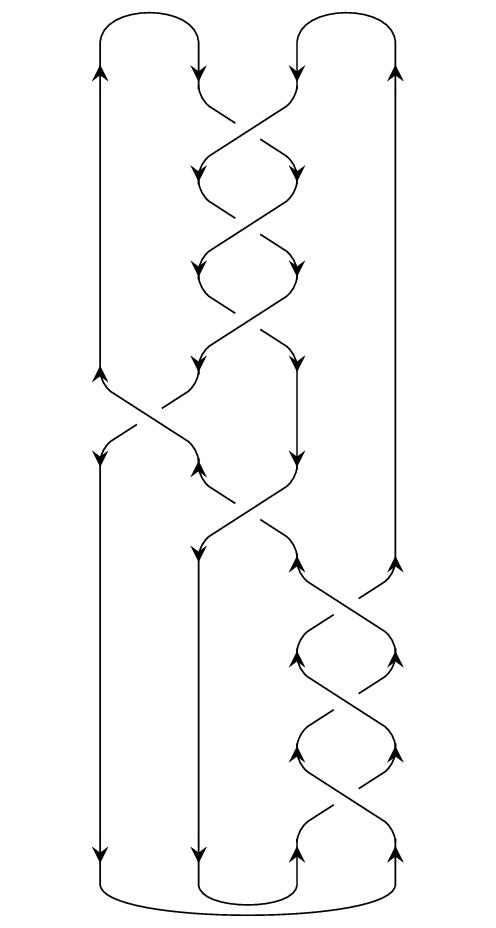}} \\
\subfloat[$\mathbf{8_{11}}$]{\includegraphics[width=0.2\linewidth]{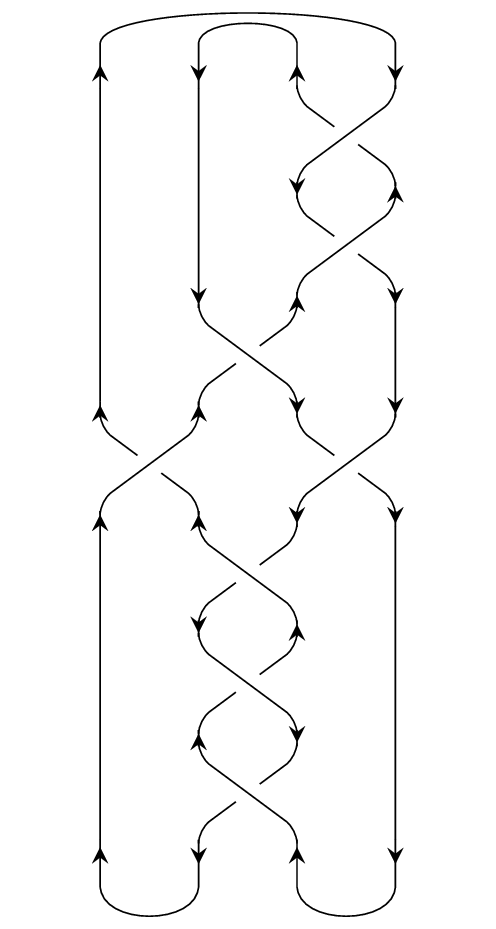}}
\subfloat[$\mathbf{8_{12}}$]{\includegraphics[width=0.2\linewidth]{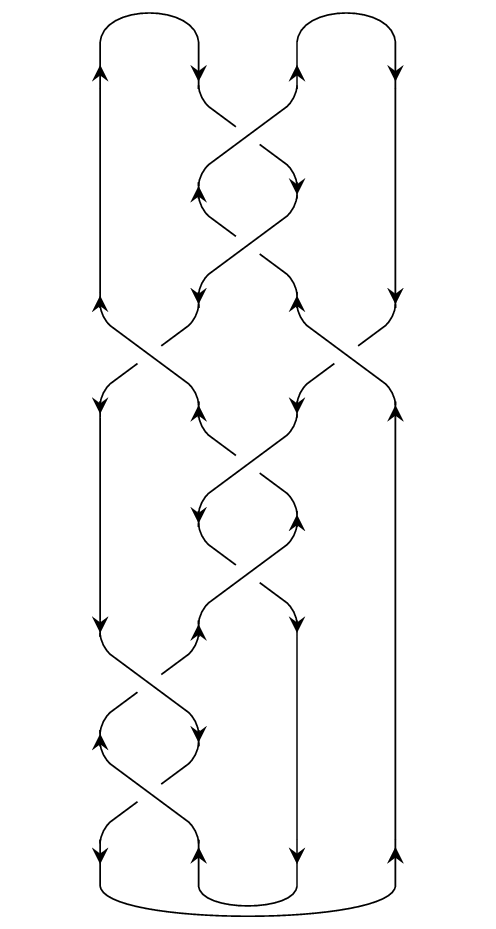}}
\subfloat[$\mathbf{8_{13}}$]{\includegraphics[width=0.2\linewidth]{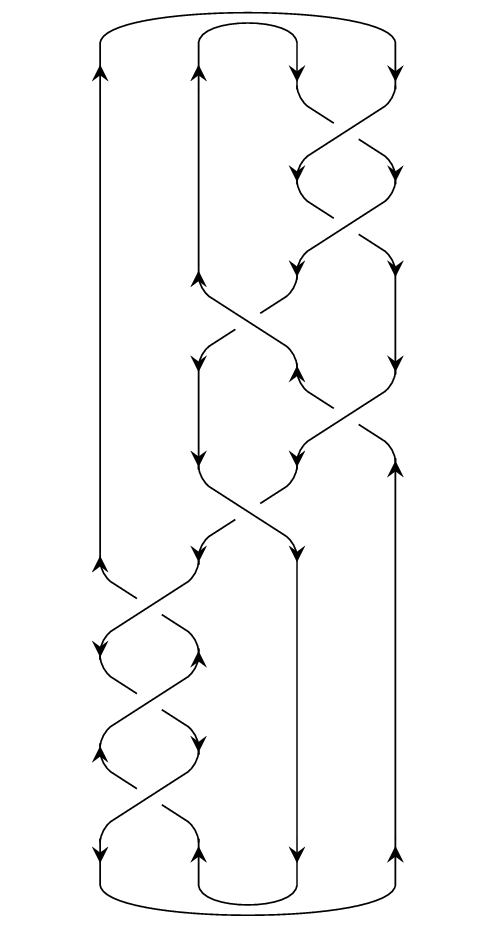}}
\subfloat[$\mathbf{8_{14}}$]{\includegraphics[width=0.2\linewidth]{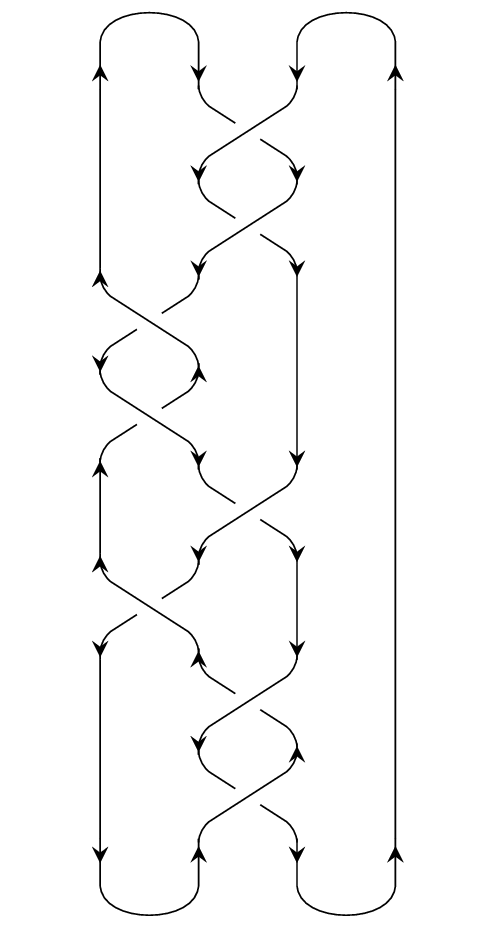}}
\caption{Quasi--plat representations of hyperbolic knots with eight crossings.}\label{fig:8sx}
\end{figure}

In the following, the formulae for computing the quantum knot invariants are written down according to the (quasi--)plat representations in Figs.~\ref{fig:4s1And6s1} \footnote{These formulae differ from those in the Appendix A of ref.~\cite{Zodinmawia:2011oya} by the proper inclusion of multiplicity labels. Besides they are based on simplified (quasi-)plat representations to reduce computation time; c.f., with the discussion on computational complexities in Sec.~\ref{sec:conclusion}.} and Figs.~\ref{fig:8sx}. In each formula, the integer on the left hand side next to the knot symbol is the framing of the (quasi--)plat representation. In order to get the (unnormalized) HOMFLY invariant with framing 0, one has multiply it with the appropriate $U(1)$ factor and then perform the framing transformation. Besides, each formula is summed over all the representation labels $s^{(i)},t^{(j)}$ and multiplicity labels $r^{(k)}_l$ appearing in the summand. We use $|s^{(i)}|$ and $|t^{(j)}|$ as shorthand for quantum dimensions. The choices of 2j-- and 2j--phases are given in Sec.~\ref{sec:3jPhases}, while the braiding eigenvalues are given in eqs.~\eqref{equ:BraidingEigenvalues}. Finally, the listed crossing matrices are linked to the quantum 6j--symbols according to eq.~\eqref{equ:Cross6j}.

\begin{align*}
\mathcal{W}_R(\mathbf{4_1},0) = &\{R\} \sum_{\ldots} \sqrt{|s^{(1)}|} \{ R,\bar{R},s^{(1)},r^{(1)}_2\} \delta_{r^{(1)}_1,r^{(1)}_2} \left(\lambda^{(-)}_{R\bar{R};s^{(1)}r^{(1)}_2}\right)^2 
a^{t^{(2)}, r^{(2)}_3r^{(2)}_4}_{s^{(1)},r^{(1)}_1r^{(1)}_2} \begin{bmatrix}
R & \bar{R} \\
R & \bar{R}
\end{bmatrix}^*\\
& \left( \lambda^{(-)}_{R\bar{R};t^{(2)}r^{(2)}_4} \right)^{-2} \sqrt{|t^{(2)}|} \{R,\bar{R},t^{(2)},r^{(2)}_4\} \delta_{r^{(2)}_3,r^{(2)}_4} \ .
\end{align*}

\begin{align*}
\mathcal{W}_R(\mathbf{5_2},5) = &\sum_{\ldots} \sqrt{|s^{(1)}|} \{\bar{R},\bar{R},\bar{s}^{(1)},r^{(1)}_2\} \delta_{r^{(1)}_1,r^{(1)}_2} \left( \lambda^{(+)}_{\bar{R}\bar{R};s^{(1)}r^{(1)}_2} \right)^{-2}
a^{t^{(2)}, r^{(2)}_3r^{(2)}_4}_{s^{(1)},r^{(1)}_1r^{(1)}_2} \begin{bmatrix}
R & \bar{R} \\
\bar{R} & R
\end{bmatrix}^*\\
& \left( \lambda^{(-)}_{R\bar{R};t^{(2)}r^{(2)}_4} \right)^{-2} \left( \lambda^{(-)}_{R\bar{R};t^{(2)}r^{(2)}_3} \right)^{-1} \sqrt{|t^{(2)}|} \{R,\bar{R},t^{(2)},r^{(2)}_4\} \delta_{r^{(2)}_3,r^{(2)}_4} \ .
\end{align*}

\begin{align*}
\mathcal{W}_R(\mathbf{6_1},2) = &\{R\} \sum_{\ldots} \sqrt{|s^{(1)}|} \{ R,\bar{R},s^{(1)},r^{(1)}_2\} \delta_{r^{(1)}_1,r^{(1)}_2} \left(\lambda^{(-)}_{R\bar{R};s^{(1)}r^{(1)}_2}\right)^2 
a^{t^{(2)}, r^{(2)}_3r^{(2)}_4}_{s^{(1)},r^{(1)}_1r^{(1)}_2} \begin{bmatrix}
R & \bar{R} \\
R & \bar{R}
\end{bmatrix}^*\\
& \left( \lambda^{(-)}_{R\bar{R};t^{(2)}r^{(2)}_4} \right)^{-4} \sqrt{|t^{(2)}|} \{R,\bar{R},t^{(2)},r^{(2)}_4\} \delta_{r^{(2)}_3,r^{(2)}_4} \ .
\end{align*}

\begin{align*}
\mathcal{W}_R(\mathbf{6_2},2) =& \{R\} \sum_{\ldots} \sqrt{|s^{(1)}|}\{R,R,\bar{s}^{(1)},r^{(1)}_2\} \delta_{r^{(1)}_1,r^{(1)}_2} 
\lambda^{(+)}_{RR;s^{(1)}r^{(1)}_2} 
a^{t^{(2)}, r^{(2)}_3r^{(2)}_4}_{s^{(1)},r^{(1)}_1r^{(1)}_2} \begin{bmatrix}
\bar{R} & R \\
R & \bar{R}
\end{bmatrix}^*\\
&\lambda^{(-)}_{R\bar{R};t^{(2)}r^{(2)}_3} 
a^{t^{(2)}, r^{(2)}_3r^{(2)}_4}_{s^{(3)},r^{(3)}_1r^{(3)}_2} \begin{bmatrix}
\bar{R} & R \\
\bar{R} & R
\end{bmatrix}
\left( \lambda^{(-)}_{R\bar{R};s^{(3)}r^{(3)}_2} \right)^{-1}
a^{t^{(4)}, r^{(4)}_3r^{(4)}_4}_{s^{(3)},r^{(3)}_1r^{(3)}_2} \begin{bmatrix}
\bar{R} & \bar{R} \\
R & R
\end{bmatrix}^* \\
&\left(\lambda^{(+)}_{\bar{R}\bar{R};t^{(4)}r^{(4)}_4}\right)^{-2} \left( \lambda^{(+)}_{RR;\bar{t}^{(4)}r^{(4)}_3} \right)^{-1} \sqrt{|t^{(4)}|} \{R,R,t^{(4)},r^{(4)}_4\} \delta_{r^{(4)}_3,r^{(4)}_4} \ .
\end{align*}

\begin{align*}
\mathcal{W}_R(\mathbf{6_3},0) = &\{R\} \sum_{\ldots} \sqrt{|s^{(1)}|} \{R,R,\bar{s}^{(1)},r^{(1)}_2\} \delta_{r^{(1)}_1,r^{(1)}_2}
\left(\lambda^{(+)}_{RR;s^{(1)}r^{(1)}_2}\right)^{-2}
a^{t^{(2)}, r^{(2)}_3r^{(2)}_4}_{s^{(1)},r^{(1)}_1r^{(1)}_2} \begin{bmatrix}
\bar{R} & R \\
R & \bar{R}
\end{bmatrix}^* \\
&\left(\lambda^{(-)}_{R\bar{R};t^{(2)}r^{(2)}_3}\right)^{-1}
a^{t^{(2)}, r^{(2)}_3r^{(2)}_4}_{s^{(3)},r^{(3)}_1r^{(3)}_2} \begin{bmatrix}
\bar{R} & R \\
\bar{R} & R
\end{bmatrix}
\lambda^{(-)}_{R\bar{R};s^{(3)}r^{(3)}_2}
a^{t^{(4)}, r^{(4)}_3r^{(4)}_4}_{s^{(3)},r^{(3)}_1r^{(3)}_2} \begin{bmatrix}
\bar{R} & \bar{R} \\
R & R
\end{bmatrix}^*\\
&\left(\lambda^{(+)}_{\bar{R}\bar{R};t^{(4)}r^{(4)}_4}\right)^{2}
\sqrt{|t^{(4)}|} \{R,R,t^{(4)},r^{(4)}_4\} \delta_{r^{(4)}_3,r^{(4)}_4} \ .
\end{align*}

\begin{align*}
\mathcal{W}_R(\mathbf{7_2},7) = & \sum_{\ldots} \sqrt{|s^{(1)}|} \{R,R,\bar{s}^{(1)},r^{(1)}_2\} \delta_{r^{(1)}_1,r^{(1)}_2}
\left( \lambda^{(+)}_{RR;s^{(1)}r^{(1)}_2} \right)^{-2}
a^{t^{(2)}, r^{(2)}_3r^{(2)}_4}_{s^{(1)},r^{(1)}_1r^{(1)}_2} \begin{bmatrix}
\bar{R} & R \\
R & \bar{R}
\end{bmatrix}^*\\
&\left( \lambda^{(-)}_{R\bar{R};t^{(2)}r^{(2)}_4} \right)^{-3} \left( \lambda^{(-)}_{R\bar{R};t^{(2)}r^{(2)}_3} \right)^{-2}
\sqrt{|t^{(2)}|} \{R,\bar{R},t^{(2)},r^{(2)}_4\} \delta_{r^{(2)}_3,r^{(2)}_4} \ .
\end{align*}

\begin{align*}
\mathcal{W}_R(\mathbf{7_3},-7) = &\sum_{\ldots} \sqrt{|s^{(1)}|} \{R,\bar{R},s^{(1)},r^{(1)}_2\} \delta_{r^{(1)}_1,r^{(1)}_2}
\left( \lambda^{(-)}_{R\bar{R};s^{(1)}r^{(1)}_2} \right)^{3}
a^{t^{(2)}, r^{(2)}_3r^{(2)}_4}_{s^{(1)},r^{(1)}_1r^{(1)}_2} \begin{bmatrix}
\bar{R} & \bar{R} \\
R & R
\end{bmatrix}^*\\
& \left( \lambda^{(+)}_{\bar{R}\bar{R};t^{(2)}r^{(2)}_4} \right)^{3} \lambda^{(+)}_{RR;\bar{t}^{(2)}r^{(2)}_3}
\sqrt{|t^{(2)}|} \{R,R,t^{(2)},r^{(2)}_4\} \delta_{r^{(2)}_3,r^{(2)}_4} \ .
\end{align*}

\begin{align*}
\mathcal{W}_R(\mathbf{7_4},-7) =&\sum_{\ldots} \sqrt{|s^{(1)}|} \{R,R,\bar{s}^{(1)},r^{(1)}_2\} \delta_{r^{(1)}_1,r^{(1)}_2}
\lambda^{(+)}_{RR;s^{(1)}r^{(1)}_2}
a^{t^{(2)}, r^{(2)}_3r^{(2)}_4}_{s^{(1)},r^{(1)}_1r^{(1)}_2} \begin{bmatrix}
\bar{R} & R \\
R & \bar{R}
\end{bmatrix}^*\\
& \lambda^{(-)}_{R\bar{R};t^{(2)}r^{(2)}_4} \lambda^{(-)}_{R\bar{R};t^{(2)}r^{(2)}_3}
a^{t^{(2)}, r^{(2)}_3r^{(2)}_4}_{s^{(3)},r^{(3)}_1r^{(3)}_2} \begin{bmatrix}
R & \bar{R} \\
\bar{R} & R
\end{bmatrix}
\lambda^{(+)}_{\bar{R}\bar{R};s^{(3)}r^{(3)}_2}
a^{t^{(4)}, r^{(4)}_3r^{(4)}_4}_{s^{(3)},r^{(3)}_1r^{(3)}_2} \begin{bmatrix}
R & \bar{R} \\
\bar{R} & R
\end{bmatrix}^*\\
&\left(\lambda^{(-)}_{R\bar{R};t^{(4)}r^{(4)}_4} \right)^2 \lambda^{(-)}_{R\bar{R};t^{(4)}r^{(4)}_3}
\sqrt{|t^{(4)}|} \{R,\bar{R},t^{(4)},r^{(4)}_4\} \delta_{r^{(4)}_3,r^{(4)}_4} \ .
\end{align*}

\begin{align*}
\mathcal{W}_R(\mathbf{7_5},7) =& \sum_{\ldots} \sqrt{|s^{(1)}|} \{R,\bar{R},s^{(1)},r^{(1)}_2\} \delta_{r^{(1)}_1,r^{(1)}_2}
\left(\lambda^{(-)}_{R\bar{R};s^{(1)}r^{(1)}_2} \right)^{-1}
a^{t^{(2)}, r^{(2)}_3r^{(2)}_4}_{s^{(1)},r^{(1)}_1r^{(1)}_2} \begin{bmatrix}
\bar{R} & \bar{R} \\
R & R
\end{bmatrix}^*\\
&\left(\lambda^{(+)}_{RR;\bar{t}^{(2)}r^{(2)}_3} \right)^{-1}
a^{t^{(2)}, r^{(2)}_3r^{(2)}_4}_{s^{(3)},r^{(3)}_1r^{(3)}_2} \begin{bmatrix}
\bar{R} & \bar{R} \\
R & R
\end{bmatrix}
\left(\lambda^{(-)}_{R\bar{R};s^{(3)}r^{(3)}_2} \right)^{-2}
a^{t^{(4)}, r^{(4)}_3r^{(4)}_4}_{s^{(3)},r^{(3)}_1r^{(3)}_2} \begin{bmatrix}
\bar{R} & \bar{R} \\
R & R
\end{bmatrix}^*\\
&\left(\lambda^{(+)}_{\bar{R}\bar{R};t^{(4)}r^{(4)}_4} \right)^{-2}\left(\lambda^{(+)}_{RR;\bar{t}^{(4)}r^{(4)}_3} \right)^{-1}
\sqrt{|t^{(4)}|}\{R,R,t^{(4)},r^{(4)}_4\} \delta_{t^{(4)}_3,t^{(4)}_4} \ .
\end{align*}

\begin{align*}
\mathcal{W}_R(\mathbf{7_6},3) =& \sum_{\ldots} \sqrt{|s^{(1)}|} \{R,\bar{R},s^{(1)},r^{(1)}_2\} \delta_{r^{(1)}_1,r^{(1)}_2}
\left(\lambda^{(-)}_{R\bar{R};s^{(1)}r^{(1)}_2} \right)^{-2}
a^{t^{(2)}, r^{(2)}_3r^{(2)}_4}_{s^{(1)},r^{(1)}_1r^{(1)}_2} \begin{bmatrix}
\bar{R} & R \\
\bar{R} & R
\end{bmatrix}^*\\
&\left(\lambda^{(-)}_{R\bar{R};t^{(2)}r^{(2)}_3} \right)^{2}
a^{t^{(2)}, r^{(2)}_3r^{(2)}_4}_{s^{(3)},r^{(3)}_1r^{(3)}_2} \begin{bmatrix}
\bar{R} & R \\
\bar{R} & R
\end{bmatrix}
\left(\lambda^{(-)}_{R\bar{R};s^{(3)}r^{(3)}_2} \right)^{-1}
a^{t^{(4)}, r^{(4)}_3r^{(4)}_4}_{s^{(3)},r^{(3)}_1r^{(3)}_2} \begin{bmatrix}
\bar{R} & \bar{R} \\
R & R
\end{bmatrix}^*\\
&\left(\lambda^{(+)}_{\bar{R}\bar{R};t^{(4)}r^{(4)}_4} \right)^{-2}
\sqrt{|t^{(4)}|}\{R,R,t^{(4)},r^{(4)}_4\} \delta_{t^{(4)}_3,t^{(4)}_4} \ .
\end{align*}

\begin{align*}
\mathcal{W}_R(\mathbf{7_7},-1) = &\sum_{\ldots} \sqrt{|s^{(1)}|} \{R,R,\bar{s}^{(1)},r^{(1)}_2\} \delta_{r^{(1)}_1,r^{(1)}_2}
\lambda^{(+)}_{RR;s^{(1)}r^{(1)}_2}
a^{t^{(2)}, r^{(2)}_3r^{(2)}_4}_{s^{(1)},r^{(1)}_1r^{(1)}_2} \begin{bmatrix}
\bar{R} & R \\
R & \bar{R}
\end{bmatrix}^*\\
&\lambda^{(-)}_{R\bar{R};t^{(2)}r^{(2)}_4}
a^{t^{(2)}, r^{(2)}_3r^{(2)}_4}_{s^{(3)},r^{(3)}_1r^{(3)}_2} \begin{bmatrix}
R & \bar{R} \\
R & \bar{R}
\end{bmatrix}
\left( \lambda^{(-)}_{R\bar{R};s^{(3)}r^{(3)}_2} \right)^{-1}
a^{t^{(4)}, r^{(4)}_3r^{(4)}_4}_{s^{(3)},r^{(3)}_1r^{(3)}_2} \begin{bmatrix}
R & R \\
\bar{R} & \bar{R}
\end{bmatrix}^*\\
&\left( \lambda^{(+)}_{\bar{R}\bar{R};\bar{t}^{(4)}r^{(4)}_3} \right)^{-1}
a^{t^{(4)}, r^{(4)}_3r^{(4)}_4}_{s^{(5)},r^{(5)}_1r^{(5)}_2} \begin{bmatrix}
R & R \\
\bar{R} & \bar{R}
\end{bmatrix}
\left( \lambda^{(-)}_{R\bar{R};s^{(5)}r^{(5)}_2} \right)^{-1}
a^{t^{(6)}, r^{(6)}_3r^{(6)}_4}_{s^{(5)},r^{(5)}_1r^{(5)}_2} \begin{bmatrix}
R & \bar{R} \\
R & \bar{R}
\end{bmatrix}^*\\
&\left( \lambda^{(-)}_{R\bar{R};t^{(6)}r^{(6)}_4} \right)^{2} \sqrt{|t^{(6)}|} \{R,\bar{R},t^{(6)},r^{(6)}_4\} \delta_{r^{(6)}_3,r^{(6)}_4} \ .
\end{align*}

\begin{align*}
\mathcal{W}_R(\mathbf{8_1},-4) = &\sum_{\ldots} \sqrt{|s^{(1)}|} \{\bar{R},R,\bar{s}^{(1)},r^{(1)}_2\} \delta_{r^{(1)}_1,r^{(1)}_2} \left( \lambda^{(-)}_{\bar{R}R;s^{(1)}r^{(1)}_2} \right)^{-2}
a^{t^{(2)},r^{(2)}_3r^{(2)}_4}_{s^{(1)},r^{(1)}_1r^{(1)}_2}\begin{bmatrix}
R & \bar{R} \\
R & \bar{R} 
\end{bmatrix}^*\\
&\left( \lambda^{(-)}_{R\bar{R};t^{(2)}r^{(2)}_4}  \right)^2 \left(  \lambda^{(-)}_{R\bar{R};t^{(2)}r^{(2)}_3} \right)^4
\sqrt{|t^{(2)}|}\{R,\bar{R},t^{(2)},r^{(2)}_3\} \delta_{r^{(1)}_3, r^{(1)}_4}	\ .
\end{align*}

\begin{align*}
\mathcal{W}_R(\mathbf{8_2},-4) =& \sum_{\ldots} \sqrt{|s^{(1)}|} \{ \bar{R},\bar{R},\bar{s}^{(1)},r^{(1)}_2\} \delta_{r^{(1)}_1,r^{(1)}_2}
\left( \lambda^{(+)}_{\bar{R}\bar{R};s^{(1)}r^{(1)}_2} \right)^{-1}
a^{t^{(2)},r^{(2)}_3r^{(2)}_4}_{s^{(1)},r^{(1)}_1r^{(1)}_2}\begin{bmatrix}
R & \bar{R} \\
\bar{R} & R 
\end{bmatrix}^*\\
&\left( \lambda^{(-)}_{\bar{R}R;\bar{t}^{(2)}r^{(2)}_3} \right)^{-1}
a^{t^{(2)},r^{(2)}_3r^{(2)}_4}_{s^{(3)},r^{(3)}_1r^{(3)}_2}\begin{bmatrix}
R & \bar{R} \\
R & \bar{R} 
\end{bmatrix}
 \lambda^{(-)}_{\bar{R}R;s^{(3)}r^{(3)}_2} \;
a^{t^{(4)},r^{(4)}_3r^{(4)}_4}_{s^{(3)},r^{(3)}_1r^{(3)}_2}\begin{bmatrix}
R & R \\
\bar{R} & \bar{R} 
\end{bmatrix}^*\\
&\left( \lambda^{(+)}_{RR;t^{(4)}r^{(4)}_4} \right)^{5}
\sqrt{|t^{(4)}|} \{R,R,\bar{t}^{(4)},r^{(4)}_4 \} \delta_{r^{(4)}_3,r^{(4)}_4} \ .
\end{align*}

\begin{align*}
\mathcal{W}_R(\mathbf{8_3},0) =& \sum_{\ldots} \sqrt{|t^{(1)}|} \{ R,\bar{R},t^{(1)},r^{(1)}_3\} \delta_{r^{(1)}_3,r^{(1)}_4}
\left( \lambda^{(-)}_{R\bar{R};\bar{t}^{(1)}r^{(1)}_3} \right)^{-4}
a^{t^{(1)},r^{(1)}_3r^{(1)}_4}_{s^{(2)},r^{(2)}_1r^{(2)}_2}\begin{bmatrix}
R & \bar{R} \\
R & \bar{R} 
\end{bmatrix}\\
&\left( \lambda^{(-)}_{\bar{R}R;s^{(2)}r^{(2)}_2} \right)^{4} 
\sqrt{|s^{(2)}|} \{\bar{R},R,\bar{s}^{(2)},r^{(2)}_2\} \delta_{r^{(2)}_1,r^{(2)}_2} \ .
\end{align*}

\begin{align*}
\mathcal{W}_R(\mathbf{8_4},0) =& \{R\} \sum_{\ldots} \sqrt{|t^{(1)}|} \{ R,R,\bar{t}^{(1)},r^{(1)}_4\} \delta_{r^{(1)}_3,r^{(1)}_4}
\left( \lambda^{(+)}_{RR;t^{(1)}r^{(1)}_4} \right)^{-3}
a^{t^{(1)},r^{(1)}_3r^{(1)}_4}_{s^{(2)},r^{(2)}_1r^{(2)}_2}\begin{bmatrix}
R & R \\
\bar{R} & \bar{R} 
\end{bmatrix}\\
&\left( \lambda^{(-)}_{R\bar{R};s^{(2)}r^{(2)}_2} \right)^{-1}
a^{t^{(3)},r^{(3)}_3r^{(3)}_4}_{s^{(2)},r^{(2)}_1r^{(2)}_2}\begin{bmatrix}
R & \bar{R} \\
R & \bar{R} 
\end{bmatrix}^*
\left( \lambda^{(-)}_{R\bar{R};t^{(3)}r^{(3)}_4} \right)^{4}\\
&\sqrt{|t^{(3)}|} \{R,\bar{R},\bar{t}^{(3)}, r^{(3)}_4\} \; \delta_{r^{(3)}_3,r^{(3)}_4}	\ .
\end{align*}

\begin{align*}
\mathcal{W}_R(\mathbf{8_6},-4) =& \{R\} \sum_{\ldots} \sqrt{|s^{(1)}|} \{ \bar{R},R,\bar{s}^{(1)},r^{(1)}_2\} \delta_{r^{(1)}_1,r^{(1)}_2}
\left( \lambda^{(-)}_{\bar{R}R;s^{(1)}r^{(1)}_2} \right)^{-2}
a^{t^{(2)},r^{(2)}_3r^{(2)}_4}_{s^{(1)},r^{(1)}_1r^{(1)}_2}\begin{bmatrix}
R & \bar{R} \\
R & \bar{R} 
\end{bmatrix}^*\\
&\left( \lambda^{(-)}_{R\bar{R};t^{(2)}r^{(2)}_4} \right)^{2} \lambda^{(-)}_{R\bar{R};\bar{t}^{(2)}r^{(2)}_3}\;
a^{t^{(2)},r^{(2)}_3r^{(2)}_4}_{s^{(3)},r^{(3)}_1r^{(3)}_2}\begin{bmatrix}
R & \bar{R} \\
\bar{R} & R 
\end{bmatrix}
\left( \lambda^{(+)}_{\bar{R}\bar{R};s^{(3)}r^{(3)}_2} \right)^{3} \\
&\sqrt{|s^{(3)}|} \{ \bar{R},\bar{R},\bar{s}^{(3)},r^{(3)}_2\} \; \delta_{r^{(3)}_1,r^{(3)}_2} \ .
\end{align*}

\begin{align*}
\mathcal{W}_R(\mathbf{8_7},2) =& \sum_{\ldots} \sqrt{|s^{(1)}|} \{ \bar{R},\bar{R},\bar{s}^{(1)},r^{(1)}_2\} \delta_{r^{(1)}_1,r^{(1)}_2}
\left( \lambda^{(+)}_{\bar{R}\bar{R};s^{(1)}r^{(1)}_2} \right)^{2}
a^{t^{(2)},r^{(2)}_3r^{(2)}_4}_{s^{(1)},r^{(1)}_1r^{(1)}_2}\begin{bmatrix}
R & \bar{R} \\
\bar{R} & R
\end{bmatrix}^*\\
& \lambda^{(-)}_{R\bar{R};t^{(2)}r^{(2)}_4} \;
a^{t^{(2)},r^{(2)}_3r^{(2)}_4}_{s^{(3)},r^{(3)}_1r^{(3)}_2}\begin{bmatrix}
\bar{R} & R \\
\bar{R} & R
\end{bmatrix}
\left( \lambda^{(-)}_{R\bar{R};s^{(3)}r^{(3)}_2} \right)^{-1}
a^{t^{(4)},r^{(4)}_3r^{(4)}_4}_{s^{(3)},r^{(3)}_1r^{(3)}_2}\begin{bmatrix}
\bar{R} & \bar{R} \\
R & R
\end{bmatrix}^*\\
&\left( \lambda^{(+)}_{\bar{R}\bar{R};t^{(4)}r^{(4)}_4} \right)^{-4}
\sqrt{|t^{(4)}|} \{ \bar{R},\bar{R},\bar{t}^{(4)}, r^{(4)}_4 \}\; \delta_{r^{(4)}_3,r^{(4)}_4} \ .
\end{align*}

\begin{align*}
\mathcal{W}_R(\mathbf{8_8},2) =& \sum_{\ldots} \sqrt{|s^{(1)}|} \{ \bar{R},\bar{R},\bar{s}^{(1)},r^{(1)}_2\} \;\delta_{r^{(1)}_1,r^{(1)}_2}
\left( \lambda^{(+)}_{\bar{R}\bar{R};s^{(1)}r^{(1)}_2} \right)^{-2}
a^{t^{(2)},r^{(2)}_3r^{(2)}_4}_{s^{(1)},r^{(1)}_1r^{(1)}_2}\begin{bmatrix}
R & \bar{R} \\
\bar{R} & R
\end{bmatrix}^*\\
&\left( \lambda^{(-)}_{R\bar{R};t^{(2)}r^{(2)}_4} \right)^{-3}
a^{t^{(2)},r^{(2)}_3r^{(2)}_4}_{s^{(3)},r^{(3)}_1r^{(3)}_2}\begin{bmatrix}
\bar{R} & R \\
\bar{R} & R
\end{bmatrix}
\lambda^{(-)}_{R\bar{R};s^{(3)}r^{(3)}_2} \;
a^{t^{(4)},r^{(4)}_3r^{(4)}_4}_{s^{(3)},r^{(3)}_1r^{(3)}_2}\begin{bmatrix}
\bar{R} & \bar{R} \\
R & R
\end{bmatrix}^*\\
&\left( \lambda^{(+)}_{RR;\bar{t}^{(4)}r^{(4)}_3} \right)^{2}
\sqrt{|t^{(4)}|} \{R,R,t^{(4)},r^{(4)}_3\} \; \delta_{r^{(4)}_3,r^{(4)}_4} \ .
\end{align*}

\begin{align*}
\mathcal{W}_R(\mathbf{8_9},0) =& \sum_{\ldots} \sqrt{|s^{(1)}|} \{ \bar{R},\bar{R},\bar{s}^{(1)},r^{(1)}_2\} \;\delta_{r^{(1)}_1,r^{(1)}_2}
\left( \lambda^{(+)}_{\bar{R}\bar{R};s^{(1)}r^{(1)}_2} \right)^{3}
a^{t^{(2)},r^{(2)}_3r^{(2)}_4}_{s^{(1)},r^{(1)}_1r^{(1)}_2}\begin{bmatrix}
R & \bar{R} \\
\bar{R} & R
\end{bmatrix}^*\\
& \lambda^{(-)}_{R\bar{R};t^{(2)}r^{(2)}_4}  \;
a^{t^{(2)},r^{(2)}_3r^{(2)}_4}_{s^{(3)},r^{(3)}_1r^{(3)}_2}\begin{bmatrix}
\bar{R} & R \\
\bar{R} & R
\end{bmatrix}
\left( \lambda^{(-)}_{R\bar{R};s^{(3)}r^{(3)}_2} \right)^{-1}
a^{t^{(4)},r^{(4)}_3r^{(4)}_4}_{s^{(3)},r^{(3)}_1r^{(3)}_2}\begin{bmatrix}
\bar{R} & \bar{R} \\
R & R
\end{bmatrix}^*\\
&\left( \lambda^{(+)}_{RR;\bar{t}^{(4)}r^{(4)}_3} \right)^{-3}
\sqrt{|t^{(4)}|} \{R,R,t^{(4)},r^{(4)}_3\} \; \delta_{r^{(4)}_3,r^{(4)}_4} \ .
\end{align*}

\begin{align*}
\mathcal{W}_R(\mathbf{8_{11}},-4) =& \sum_{\ldots} \sqrt{|t^{(1)}|} \{ R,\bar{R},t^{(1)},r^{(1)}_3\}\; \delta_{r^{(1)}_3,r^{(1)}_4}
\left( \lambda^{(-)}_{R\bar{R};\bar{t}^{(1)}r^{(1)}_3} \right)^{-2}
a^{t^{(1)},r^{(1)}_3r^{(1)}_4}_{s^{(2)},r^{(2)}_1r^{(2)}_2}\begin{bmatrix}
R & \bar{R} \\
R & \bar{R} 
\end{bmatrix}\\
& \lambda^{(-)}_{\bar{R}R;s^{(2)}r^{(2)}_2} \;
a^{t^{(3)},r^{(3)}_3r^{(3)}_4}_{s^{(2)},r^{(2)}_1r^{(2)}_2}\begin{bmatrix}
R & R \\
\bar{R} & \bar{R} 
\end{bmatrix}^*
\lambda^{(+)}_{RR;t^{(3)}r^{(3)}_4} \; \lambda^{(+)}_{\bar{R}\bar{R};\bar{t}^{(3)}r^{(3)}_3} \;
a^{t^{(3)},r^{(3)}_3r^{(3)}_4}_{s^{(4)},r^{(4)}_1r^{(4)}_2}\begin{bmatrix}
R & R \\
\bar{R} & \bar{R} 
\end{bmatrix}\\
&\left( \lambda^{(-)}_{R\bar{R};s^{(4)}r^{(4)}_2} \right)^{3}
\sqrt{|s^{(4)}|} \{R,\bar{R},\bar{s}^{(4)},r^{(4)}_2\} \; \delta_{r^{(4)}_1,r^{(4)}_2} \ .
\end{align*}

\begin{align*}
\mathcal{W}_R(\mathbf{8_{12}},0) =& \sum_{\ldots} \sqrt{|s^{(1)}|} \{ \bar{R},R,\bar{s}^{(1)},r^{(1)}_2\} \;\delta_{r^{(1)}_1,r^{(1)}_2}
\left( \lambda^{(-)}_{\bar{R} R;s^{(1)}r^{(1)}_2} \right)^{-2}
a^{t^{(2)},r^{(2)}_3r^{(2)}_4}_{s^{(1)},r^{(1)}_1r^{(1)}_2}\begin{bmatrix}
R & \bar{R} \\
R & \bar{R}
\end{bmatrix}^*\\
& \lambda^{(-)}_{R \bar{R};t^{(2)}r^{(2)}_4} \; \lambda^{(-)}_{R \bar{R};\bar{t}^{(2)}r^{(2)}_3} \;
a^{t^{(2)},r^{(2)}_3r^{(2)}_4}_{s^{(3)},r^{(3)}_1r^{(3)}_2}\begin{bmatrix}
\bar{R} & R \\
\bar{R} & R
\end{bmatrix}
\left( \lambda^{(-)}_{R \bar{R};s^{(3)}r^{(3)}_2} \right)^{-2}
a^{t^{(4)},r^{(4)}_3r^{(4)}_4}_{s^{(3)},r^{(3)}_1r^{(3)}_2}\begin{bmatrix}
\bar{R} & R \\
\bar{R} & R
\end{bmatrix}^*\\
&\left( \lambda^{(-)}_{\bar{R} R;t^{(4)}r^{(4)}_4} \right)^{2} 
\sqrt{|t^{(4)}|} \{\bar{R},R,\bar{t}^{(4)},r^{(4)}_4\} \; \delta_{r^{(4)}_3,r^{(4)}_4} \ .
\end{align*}

\begin{align*}
\mathcal{W}_R(\mathbf{8_{13}},2) =&\{R\} \sum_{\ldots} \sqrt{|t^{(1)}|} \{ \bar{R},\bar{R},t^{(1)},r^{(1)}_3\}\; \delta_{r^{(1)}_3,r^{(1)}_4}
\left( \lambda^{(+)}_{\bar{R}\bar{R};\bar{t}^{(1)}r^{(1)}_3} \right)^{2}
a^{t^{(1)},r^{(1)}_3r^{(1)}_4}_{s^{(2)},r^{(2)}_1r^{(2)}_2}\begin{bmatrix}
R & R \\
\bar{R} & \bar{R} 
\end{bmatrix}\\
& \lambda^{(-)}_{R\bar{R};s^{(2)}r^{(2)}_2} \;
a^{t^{(3)},r^{(3)}_3r^{(3)}_4}_{s^{(2)},r^{(2)}_1r^{(2)}_2}\begin{bmatrix}
R & \bar{R} \\
R & \bar{R} 
\end{bmatrix}^*
\left( \lambda^{(-)}_{R\bar{R};\bar{t}^{(3)}r^{(3)}_3} \right)^{-1}
a^{t^{(3)},r^{(3)}_3r^{(3)}_4}_{s^{(4)},r^{(4)}_1r^{(4)}_2}\begin{bmatrix}
R & \bar{R} \\
\bar{R} & R
\end{bmatrix}\\
&\left( \lambda^{(+)}_{\bar{R}\bar{R};s^{(4)}r^{(4)}_2} \right)^{-1}
a^{t^{(5)},r^{(5)}_3r^{(5)}_4}_{s^{(4)},r^{(4)}_1r^{(4)}_2}\begin{bmatrix}
R & \bar{R} \\
\bar{R} & R
\end{bmatrix}^*
\left( \lambda^{(-)}_{R\bar{R};t^{(5)}r^{(5)}_4} \right)^{-3} \\
&\sqrt{|t^{(5)}|} \{\bar{R},R,\bar{t}^{(5)},r^{(5)}_4\} \; \delta_{r^{(5)}_3,r^{(5)}_4} \ .
\end{align*}

\begin{align*}
\mathcal{W}_R(\mathbf{8_{14}},-4) =&\{R\} \sum_{\ldots} \sqrt{|s^{(1)}|} \{ \bar{R},\bar{R},\bar{s}^{(1)},r^{(1)}_2\} \;\delta_{r^{(1)}_1,r^{(1)}_2}
\left( \lambda^{(+)}_{\bar{R} \bar{R};s^{(1)}r^{(1)}_2} \right)^{2}
a^{t^{(2)},r^{(2)}_3r^{(2)}_4}_{s^{(1)},r^{(1)}_1r^{(1)}_2}\begin{bmatrix}
R & \bar{R} \\
\bar{R} & R
\end{bmatrix}^*\\
&\left( \lambda^{(-)}_{R \bar{R};t^{(2)}r^{(2)}_4} \right)^{2}
a^{t^{(2)},r^{(2)}_3r^{(2)}_4}_{s^{(3)},r^{(3)}_1r^{(3)}_2}\begin{bmatrix}
R & \bar{R} \\
\bar{R} & R
\end{bmatrix}
\lambda^{(+)}_{\bar{R} \bar{R};s^{(3)}r^{(3)}_2} \;
a^{t^{(4)},r^{(4)}_3r^{(4)}_4}_{s^{(3)},r^{(3)}_1r^{(3)}_2}\begin{bmatrix}
R & \bar{R} \\
\bar{R} & R
\end{bmatrix}^*\\
&\lambda^{(-)}_{R \bar{R};t^{(4)}r^{(4)}_4} \;
a^{t^{(4)},r^{(4)}_3r^{(4)}_4}_{s^{(5)},r^{(5)}_1r^{(5)}_2}\begin{bmatrix}
\bar{R} & R \\
\bar{R} & R
\end{bmatrix}
\left( \lambda^{(-)}_{R \bar{R};s^{(5)}r^{(5)}_2} \right)^{-2} \\
&\sqrt{|s^{(5)}|} \{ R,\bar{R},\bar{s}^{(5)}, r^{(5)}_2\} \; \delta_{r^{(5)}_1,r^{(5)}_2} \ .
\end{align*}

%%%%%%%%%%%%%%%%%%%%%%%%%%%%%%%%%%%%%

\begin{figure}[htbp]
\center
\subfloat[$\mathbf{4_1}$]{\includegraphics[width=0.2\linewidth]{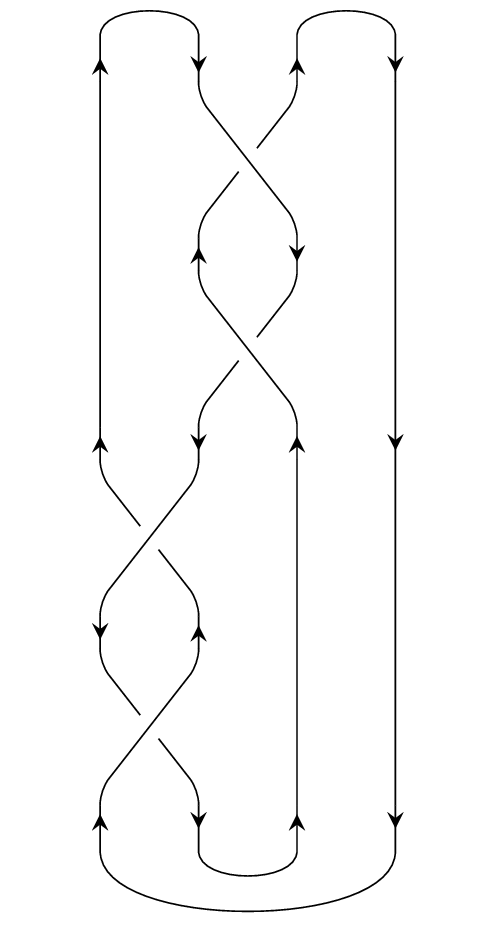}}
\subfloat[$\mathbf{5_2}$]{\includegraphics[width=0.2\linewidth]{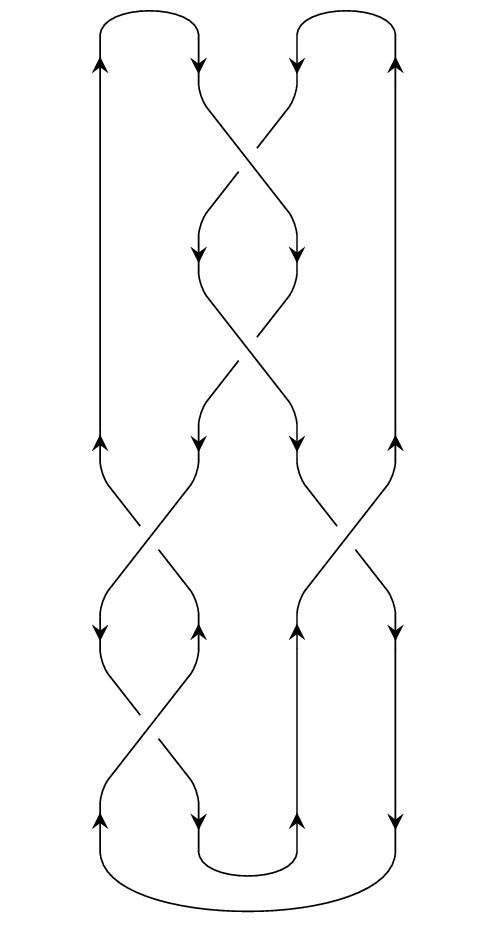}}
\subfloat[$\mathbf{6_1}$]{\includegraphics[width=0.2\linewidth]{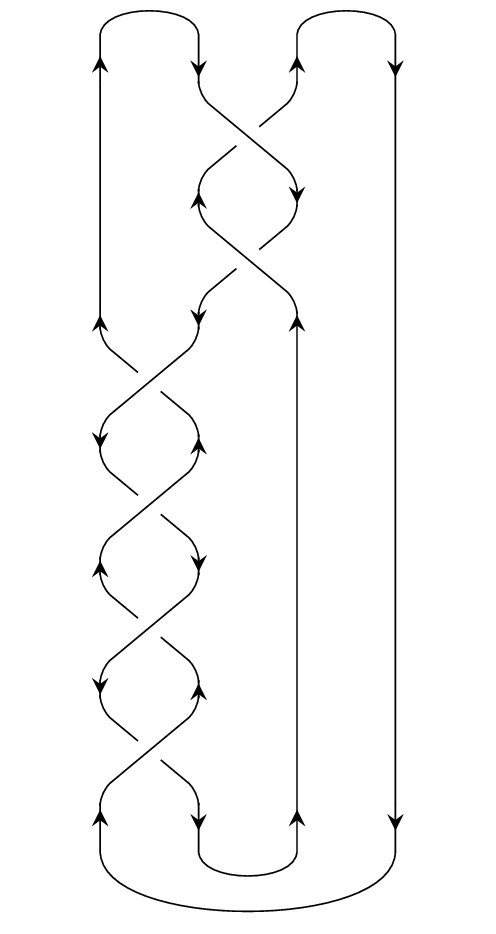}}
\subfloat[$\mathbf{6_2}$]{\includegraphics[width=0.2\linewidth]{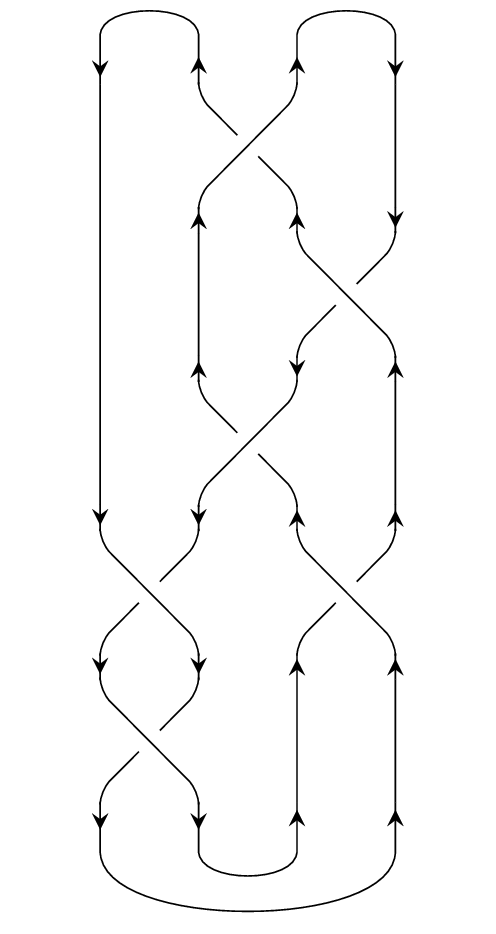}} \\
\subfloat[$\mathbf{6_3}$]{\includegraphics[width=0.2\linewidth]{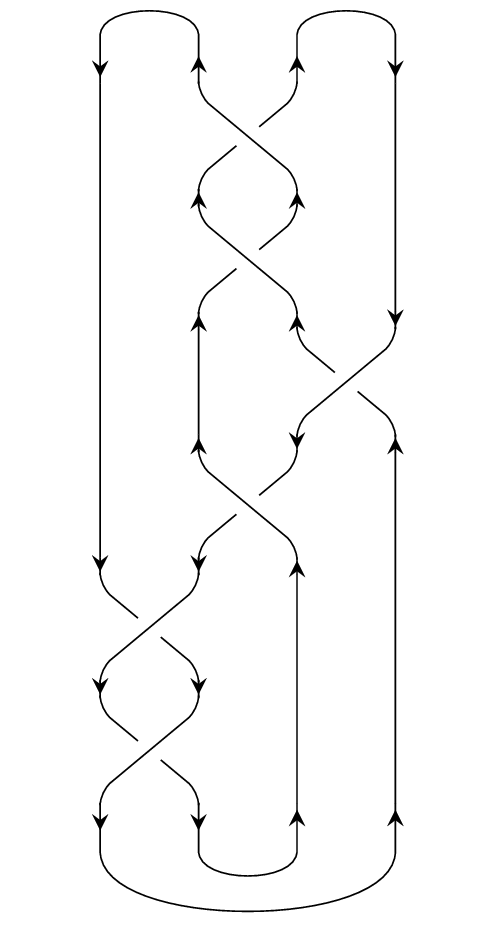}}
\subfloat[$\mathbf{7_2}$]{\includegraphics[width=0.2\linewidth]{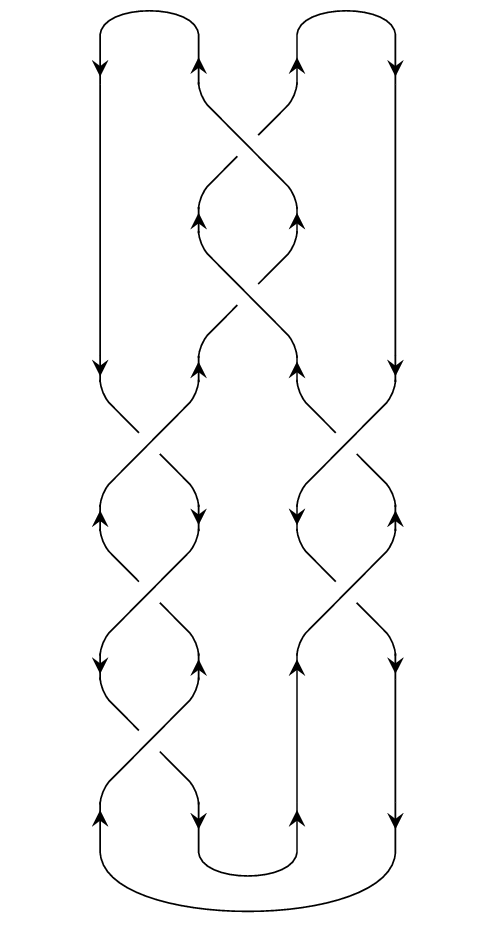}}
\subfloat[$\mathbf{7_3}$]{\includegraphics[width=0.2\linewidth]{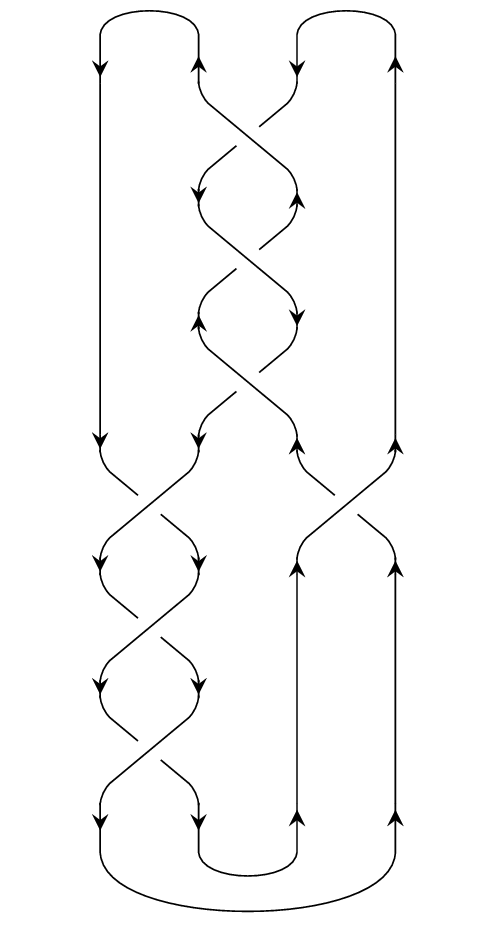}}
\subfloat[$\mathbf{7_4}$]{\includegraphics[width=0.2\linewidth]{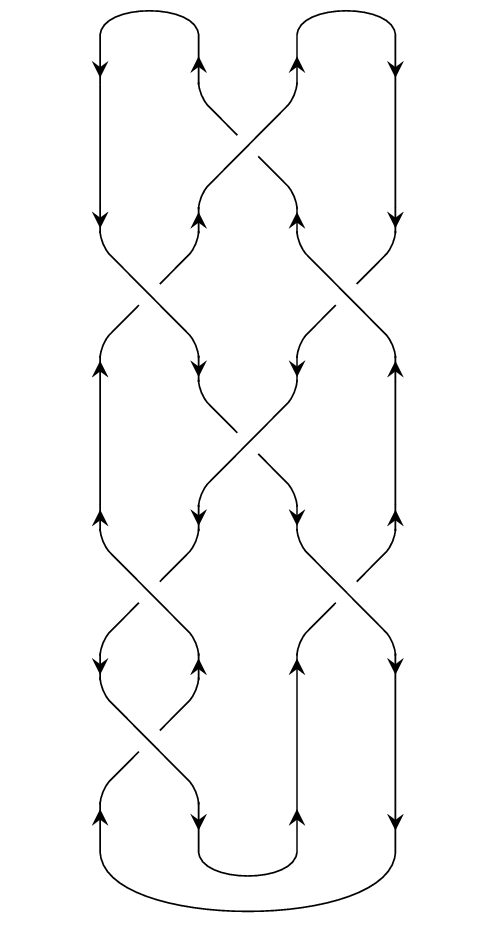}} \\
\subfloat[$\mathbf{7_5}$]{\includegraphics[width=0.2\linewidth]{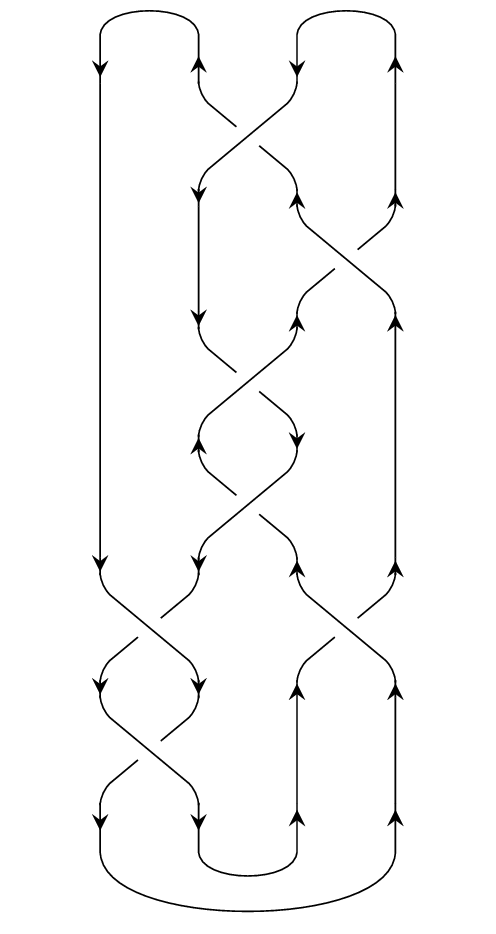}}
\subfloat[$\mathbf{7_6}$]{\includegraphics[width=0.2\linewidth]{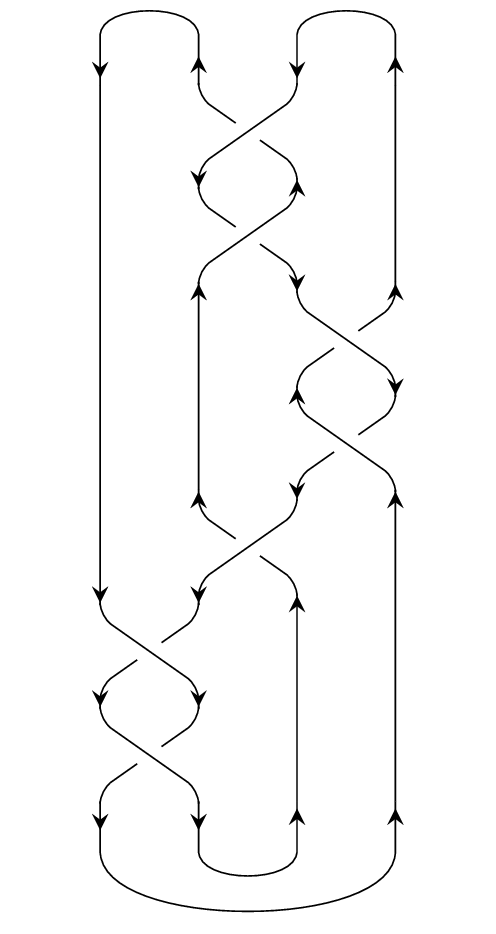}}
\subfloat[$\mathbf{7_7}$]{\includegraphics[width=0.2\linewidth]{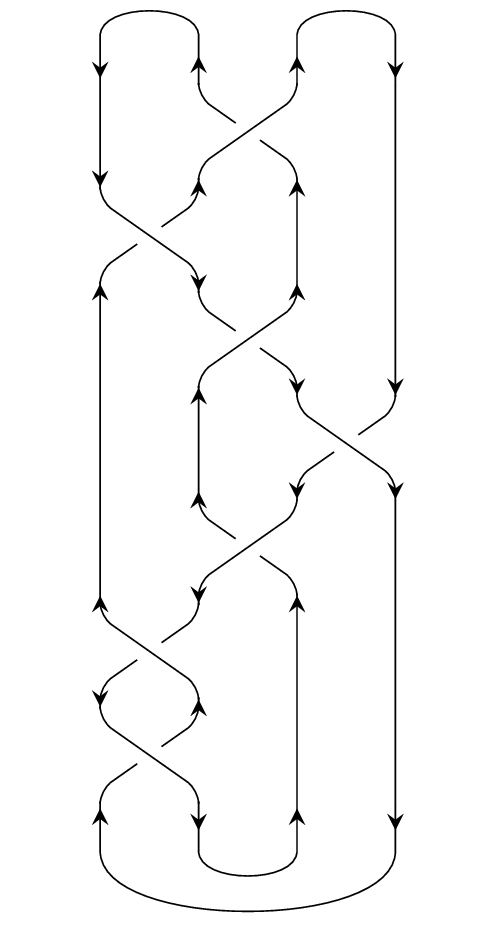}}
\caption{Quasi--plat representations of hyperbolic knots with up to seven crossings.}\label{fig:4s1And6s1}
\end{figure}

%%%%%%%%%%%%%%%%%%%%%%%%%%%%%%%%%%%%%

\newpage
\bibliographystyle{amsmod}
\bibliography{GJ}
\end{document}